\documentclass[twocolumn]{aastex631}
\usepackage{xfrac,graphicx,color,float,hyperref}
\usepackage{soul} 
\usepackage{seqsplit}
\usepackage{amsmath}
\usepackage{multirow}

\newcommand{\bagpipes}{\textsc{bagpipes}}
\newcommand{\sersic}{S\'ersic}

\newcommand{\lephare}{\textsc{LePHARE}}
\newcommand{\cigale}{\textsc{Cigale}}
\newcommand{\flares}{\textsc{Flares}}

\received{August 6, 2025}
\submitjournal{ApJ}

\begin{document}

\title{Physical properties of galaxies and the UV Luminosity Function from $z\sim6$ to $z\sim14$ in COSMOS-Web}

\shortauthors{Franco et al.}
\correspondingauthor{M. Franco}
\suppressAffiliations

\email{maximilien.franco@cea.fr}

%regarder la liste des architectes

\author[0000-0002-3560-8599]{Maximilien Franco}
\affiliation{Université Paris-Saclay, Université Paris Cité, CEA, CNRS, AIM, 91191 Gif-sur-Yvette, France}
\affiliation{The University of Texas at Austin, 2515 Speedway Blvd Stop C1400, Austin, TX 78712, USA}

\author[0000-0002-0930-6466]{Caitlin M. Casey}
\affiliation{Department of Physics, University of California, Santa Barbara, Santa Barbara, CA 93106, USA}
\affiliation{The University of Texas at Austin, 2515 Speedway Blvd Stop C1400, Austin, TX 78712, USA}
\affiliation{Cosmic Dawn Center (DAWN), Denmark}

\author[0000-0003-3596-8794]{Hollis B. Akins}
\affiliation{The University of Texas at Austin, 2515 Speedway Blvd Stop C1400, Austin, TX 78712, USA}

\author[0000-0002-7303-4397]{Olivier Ilbert}
\affiliation{Aix Marseille Univ, CNRS, CNES, LAM, Marseille, France }

\author[0000-0002-7087-0701]{Marko Shuntov}
\affiliation{Cosmic Dawn Center (DAWN), Denmark} 
\affiliation{Niels Bohr Institute, University of Copenhagen, Jagtvej 128, DK-2200, Copenhagen, Denmark}
\affiliation{University of Geneva, 24 rue du Général-Dufour, 1211 Genève 4, Switzerland}

\author[0000-0001-8519-1130]{Steven L. Finkelstein}
\affiliation{Department of Astronomy, The University of Texas at Austin, Austin, TX, USA}

\author[0000-0003-2397-0360]{Louise Paquereau} 
\affiliation{Institut d’Astrophysique de Paris, UMR 7095, CNRS, and Sorbonne Université, 98 bis boulevard Arago, F-75014 Paris, France}

\author[0000-0002-9382-9832]{Andreas L. Faisst}
\affiliation{Caltech/IPAC, MS 314-6, 1200 E. California Blvd. Pasadena, CA 91125, USA}

\author[0000-0002-6610-2048]{Anton M. Koekemoer}
\affiliation{Space Telescope Science Institute, 3700 San Martin Dr., Baltimore, MD 21218, USA} 

\author[0000-0002-3301-3321]{Michaela Hirschmann}
\affiliation{Institute of Physics, GalSpec, Ecole Polytechnique Federale de Lausanne, Observatoire de Sauverny, Chemin Pegasi 51, 1290 Versoix, Switzerland}
\affiliation{INAF, Astronomical Observatory of Trieste, Via Tiepolo 11, 34131 Trieste, Italy}

\author[0009-0000-2800-6402]{Sebastiano Cantarella}
\affiliation{Astronomy Section, Department of Physics, University of Trieste, via G.B. Tiepolo 11, I-34143, Trieste, Italy}
\affiliation{INAF, Astronomical Observatory of Trieste, Via Tiepolo 11, 34131 Trieste, Italy}

\author[0000-0003-4761-2197]{Nicole E. Drakos}
\affiliation{Department of Physics and Astronomy, University of Hawaii, Hilo, 200 W Kawili St, Hilo, HI 96720, USA}

\author[0000-0003-3903-6935]{Stephen M.~Wilkins} 
\affiliation{Astronomy Centre, University of Sussex, Falmer, Brighton BN1 9QH, UK}
\affiliation{Institute of Space Sciences and Astronomy, University of Malta, Msida MSD 2080, Malta}

\author[0000-0002-9489-7765]{Henry Joy McCracken}
\affiliation{Institut d’Astrophysique de Paris, UMR 7095, CNRS, and Sorbonne Université, 98 bis boulevard Arago, F-75014 Paris, France}

\author[0000-0001-9187-3605]{Jeyhan S. Kartaltepe}
\affiliation{Laboratory for Multiwavelength Astrophysics, School of Physics and Astronomy, Rochester Institute of Technology, 84 Lomb Memorial Drive, Rochester, NY 14623, USA}

\author[0000-0001-7711-3677]{Claudia Maraston}
\affiliation{Institute of Cosmology and Gravitation, University of Portsmouth, Dennis Sciama Building, Burnaby Road, Portsmouth PO13FX, UK}

\author[0009-0000-5827-5435]{Fatemeh Abedini}
\affiliation{Institute for Advanced Studies in Basic Sciences (IASBS), 444 Prof. Yousef Sobouti Blvd., Zanjan 45137-66731, Iran}

\author[0009-0000-7385-3539]{Mark J. Achenbach}
\affiliation{Department of Physics and Astronomy, University of Hawaii at Manoa, 2505 Correa Rd, Honolulu, HI 96822, USA}

\author[0000-0002-0569-5222]{Rafael C. Arango-Toro}
\affiliation{Aix Marseille Univ, CNRS, CNES, LAM, Marseille, France }

\author[0000-0002-8008-9871]{Fabrizio Gentile}
\affiliation{Université Paris-Saclay, Université Paris Cité, CEA, CNRS, AIM, 91191 Gif-sur-Yvette, France}
\affiliation{INAF- Osservatorio di Astrofisica e Scienza dello Spazio, Via Gobetti 93/3, I-40129, Bologna, Italy}

\author[0000-0002-0236-919X]{Ghassem Gozaliasl}
\affiliation{Department of Computer Science, Aalto University, P.O. Box 15400, FI-00076 Espoo, Finland}
\affiliation{Department of Physics, University of, P.O. Box 64, FI-00014 Helsinki, Finland}

\author[0000-0001-9840-4959]{Kohei Inayoshi}
\affiliation{Kavli Institute for Astronomy and Astrophysics, Peking University, Beijing 100871, China}

\author[0000-0002-2603-2639]{Darshan Kakkad}
\affiliation{Centre for Astrophysics Research, University of Hertfordshire, Hatfield, AL10 9AB, UK}
\affiliation{Space Telescope Science Institute, 3700 San Martin Drive, Baltimore, 21218,USA}

\author[0009-0006-2285-6792]{Atousa Kalantari}
\affiliation{Institute for Advanced Studies in Basic Sciences (IASBS), 444 Prof. Yousef Sobouti Blvd., Zanjan 45137-66731, Iran}

\author[0000-0002-0101-336X]{Ali Ahmad Khostovan}
\affiliation{Department of Physics and Astronomy, University of Kentucky, 505 Rose Street, Lexington, KY 40506, USA}
\affiliation{Laboratory for Multiwavelength Astrophysics, School of Physics and Astronomy, Rochester Institute of Technology, 84 Lomb Memorial Drive, Rochester, NY 14623, USA}

\author[0000-0002-5588-9156]{Vasily Kokorev}
\affiliation{Kapteyn Astronomical Institute, University of Groningen, PO Box 800, 9700 AV Groningen, The Netherlands}

\author[0000-0003-3216-7190]{Erini Lambrides}
\affiliation{NASA-Goddard Space Flight Center, Code 662, Greenbelt, MD, 20771, USA}

\author[0009-0004-2523-4425]{Gavin Leroy}
\affiliation{Institute for Computational Cosmology, Department of Physics, Durham University, South Road, Durham DH1 3LE, UK}

\author[0000-0002-6085-3780]{Richard Massey}
\affiliation{Institute for Computational Cosmology, Department of Physics, Durham University, South Road, Durham DH1 3LE, UK}

\author[0000-0001-5846-4404]{Bahram Mobasher}
\affiliation{Department of Physics and Astronomy, University of California, Riverside, 900 University Avenue, Riverside, CA 92521, USA}

\author[0009-0001-3422-3048]{Sophie L. Newman}
\affiliation{Institute of Cosmology and Gravitation, University of Portsmouth, Dennis Sciama Building, Burnaby Road, Portsmouth PO13FX, UK}

\author[0000-0002-4485-8549]{Jason Rhodes}
\affiliation{Jet Propulsion Laboratory, California Institute of Technology, 4800 Oak Grove Drive, Pasadena, CA 91001, USA}

\author[0000-0003-0427-8387]{R. Michael Rich}
\affiliation{Department of Physics and Astronomy, UCLA, PAB 430 Portola Plaza, Box 951547, Los Angeles, CA 90095-1547}

\author[0000-0002-4271-0364]{Brant E. Robertson}
\affiliation{Department of Astronomy and Astrophysics, University of California, Santa Cruz, 1156 High Street, Santa Cruz, CA 95064, USA}

\author[0000-0002-1233-9998]{David B. Sanders}
\affiliation{Institute for Astronomy, University of Hawaii, 2680 Woodlawn Drive, Honolulu, HI 96822, USA}

\author[0009-0003-4742-7060]{Takumi S. Tanaka}
\affiliation{Department of Astronomy, Graduate School of Science, The University of Tokyo, 7-3-1 Hongo, Bunkyo-ku, Tokyo, 113-0033, Japan}
\affiliation{Kavli Institute for the Physics and Mathematics of the Universe (WPI), The University of Tokyo Institutes for Advanced Study, The University of Tokyo, Kashiwa, Chiba 277-8583, Japan}
\affiliation{Center for Data-Driven Discovery, Kavli IPMU (WPI), UTIAS, The University of Tokyo, Kashiwa, Chiba 277-8583, Japan}

\author[0000-0002-1905-4194]{Aswin P. Vijayan}
\affiliation{Astronomy Centre, University of Sussex, Falmer, Brighton BN1 9QH, UK} 

\author[0000-0003-1614-196X]{John R. Weaver}
\affiliation{MIT Kavli Institute for Astrophysics and Space Research, 77 Massachusetts Ave., Cambridge, MA 02139, USA}
% \affil{Department of Astronomy, University of Massachusetts, Amherst, MA 01003, USA}

\author[0000-0002-8434-880X]{Lilan Yang}
\affiliation{Laboratory for Multiwavelength Astrophysics, School of Physics and Astronomy, Rochester Institute of Technology, 84 Lomb Memorial Drive, Rochester, NY 14623, USA}

\author[0000-0002-3462-4175]{Si-Yue Yu}
\affiliation{Kavli Institute for the Physics and Mathematics of the Universe (WPI), The University of Tokyo, Kashiwa, Chiba 277-8583, Japan}
\affiliation{Department of Astronomy, School of Science, The University of Tokyo, 7-3-1 Hongo, Bunkyo, Tokyo 113-0033, Japan}

\collaboration{40}{\vspace{-20pt}}

\begin{abstract} 

We present measurements of the rest-frame ultraviolet luminosity function (UVLF) in three redshift bins over $z\sim5.5$–14 from the \textit{JWST} COSMOS-Web survey. Our samples, selected using the dropout technique in the HST/ACS F814W, \textit{JWST}/NIRCam F115W, and F150W filters, contain a total of 3099 galaxies spanning a wide luminosity range from faint ($M_{\rm UV}\sim-19$ mag) to bright ($M_{\rm UV}\sim-22.5$ mag). The galaxies are undergoing rapid star formation, with blue stellar populations. Surprisingly, their median UV spectral slope $\beta$ does not evolve at $z>8$, suggesting minimal dust, or physical separation of dust and star formation at early epochs. The measured UVLF exhibits an excess at the bright-end ($M_{\rm UV}<-21$ mag) compared to pre-\textit{JWST} empirical results and theoretical predictions of an evolving Schechter function, with the excess beginning at $z\sim9$ and becoming increasingly prominent toward $z\sim12$. Our analysis suggests that reproducing the observed abundance of UV-bright galaxies at high redshift requires a combination of physical processes, including elevated star formation efficiencies, moderate levels of stochasticity in galaxy luminosities, and minimal dust attenuation.
\end{abstract}


\section{Introduction}

The Epoch of Reionization (EoR) represents a pivotal chapter in cosmic history when the intergalactic medium (IGM) transitioned from a mostly neutral to a nearly fully ionized state due to the emergence of the first sources of light. This process marked the end of the so-called ``dark ages" and established the conditions for the transparent Universe observed today \citep{Stark2016, Finkelstein2016, Dayal2018, Robertson2022}.
Observations of Gunn-Peterson absorption in the spectra of the $z>6$ quasars indicate that reionization was essentially complete by $z \sim 6$ \citep{Becker2001}. However, the exact onset and mechanism of reionization following the formation of the first astrophysical objects during the ``cosmic dawn" remains poorly constrained, with significant uncertainties surrounding the timing, the nature of the first ionizing sources, and the extent of their contribution to the early stages of this transformative epoch \citep[e.g.,][]{Robertson2015, Munoz2024, Witstok2025}. Understanding the nature and properties of the most distant galaxies provides crucial insights into the physics of star formation in extreme environments \citep{Alvarez-Marquez2019,Boylan-Kolchin2023, Andalman2025, Conselice2025} feedback mechanisms \citep{Kimm2017, Ma2020, Dekel2023, Li2024}, and the growth of structure in the early Universe.

Recent evidence suggests that low-mass galaxies dominated the ionizing photon budget during the EoR, owing to their high specific star formation rates and relatively low metallicities \citep{Bouwens2012, Finkelstein2012, Robertson2015, Livermore2017, finkelstein2019, Dayal2020, Atek2024}. However, alternative contributions from more massive galaxies, active galactic nuclei (AGN), and even dense stellar systems such as globular clusters remain plausible \citep{ Ricotti2002, Volonteri2009, Chardin2015, Grazian2018, Grazian2024}. Bright galaxies, although fewer in number than faint galaxies, could contribute significantly to the reionization of the Universe due to their high star formation rates and the larger number of ionizing photons per galaxy \citep{Finkelstein2012, Naidu2022, Harikane2023}. 

The advent of the \textit{James Webb Space Telescope} (\textit{JWST}) with unprecedented sensitivity and resolution in the near-infrared (NIR) has revolutionized our ability to probe the EoR, enabling the detection of galaxies at $z > 9$. Since the launch of \textit{JWST}, the number of $z\ge9$ candidates has surged dramatically \citep[e.g.,][]{Pontoppidan2022, Castellano2022, Naidu2022, finkelstein2022, Leung2023, Whitler2023, Adams2023, Finkelstein2023, Adams2024, Franco2024, Harikane2023, Harikane2024, Casey2024, Finkelstein2024, Donnan2023, Donnan2024,  Whitler2025} with some spectroscopically confirmed systems reaching up to z$\sim$14.4 \citep{Carniani2024,Schouws2024,Naidu2025}. A remarkable result from early data was the detection of an unexpected abundance of bright galaxies, characterized by ultraviolet (UV) magnitudes brighter than $M_{\rm UV} \sim -20$ \citep{Finkelstein2023, Donnan2023, Castellano2022, Harikane2023, Whitler2025}. 

The apparent overabundance of luminous galaxies at $z > 9$ raises fundamental questions about the physical processes that govern galaxy formation at these epochs. Several hypotheses have been proposed to explain this phenomenon.  Stochastic, ``bursty" star formation histories may play a role, particularly in environments where feedback cycles drive intermittent star formation \citep{Shen2023, Mason2023, Mirocha2023, Ciesla2024, Kravtsov2024, Rojas-Ruiz2024, Cole2025, Kokorev2025b}. Enhanced star formation efficiency in massive halos could account for the rapid build-up of stellar mass in these systems \citep{Inayoshi2022c,Harikane2023, Harikane2024, Somerville2025} or inefficient feedback in early galaxies \citep{Dekel2023, Li2024, Somerville2025}. This excess could also reflect contributions from metal-poor stellar populations, or Population III stars, which are expected to have extremely high UV luminosities relative to their stellar mass \citep{ Mason2023, Yung2024, Jeong2025}. A decrease in the mass--to--UV luminosity ratio induced by a top-heavy initial mass function (IMF) could potentially explain the excess of bright galaxies \citep{Haslbauer2022, Trinca2024, Hutter2024}. This excess could also be due to a very low dust attenuation or a very low dust content (possibly due to radiation pressure expelling dust from galaxies) \citep{Ferrara2023, Ferrara2025, Ziparo2023}. Alternatively, this could result from changes in the $\Lambda$CDM paradigm \citep{Menci2024}. Yet, there is no clear evidence for necessary changes in our cosmological model, given that baryon physics is still not well understood.

However, reconciling these observations with theoretical models requires precise measurements of the UV luminosity function (UVLF) across a wide dynamic range in luminosity and redshift. The UVLF is a cornerstone in the study of galaxy evolution during the EoR. It quantifies the co-moving number density of galaxies as a function of their UV luminosity, providing insights into the cosmic star formation rate density, the ionizing photon budget, and the role of bright versus faint galaxies in driving reionization \citep{Bouwens2015, finkelstein2022b, Harikane2023}. Constraints on the UVLF at $z > 8$ have been limited by small survey areas and shallow depths prior to \textit{JWST}, leading to large uncertainties, particularly at the bright end \citep{Bowler2020, Stefanon2021, Bouwens2021, Varadaraj2023}. The bright end of the UVLF is especially sensitive to cosmic variance and requires large-area surveys to accurately measure the abundance of rare luminous systems \citep{Trenti2008, Oesch2016, Bowler2020, Bhowmick2020}. Testing these hypotheses, and in particular the contribution of bright galaxies to reionization, requires statistically robust samples of galaxies spanning a wide range of luminosities and environments. Until recently, the small areas probed by deep surveys have limited such studies, leaving significant gaps in our observational constraints on reionization.

\textit{JWST} has already extended measurements of the UVLF to unprecedented redshifts, unveiling the bright end as a critical diagnostic for testing galaxy formation models. Recent studies have reported a higher abundance of M$_{\rm UV} < -20$ galaxies at $z>8$ relative to the predictions of an evolving Schechter (or even a double power-law) function \citep{Perez-Gonzalez2023, Donnan2023, Finkelstein2023, Finkelstein2024}. These findings underscore the need for comprehensive surveys that combine depth and wide-area coverage to overcome cosmic variance and statistically probe the brightest galaxies.

The COSMOS-Web treasury program (\citealt{Casey2023}, PIs: Kartaltepe \& Casey, ID=1727) addresses these challenges by offering a contiguous survey area of 0.54 deg$^2$ observed with NIRCam --- the largest contiguous area observed to date with \textit{JWST} --- as well as a 0.20 deg$^2$ (non-contiguous) observation with MIRI. COSMOS-Web reduces cosmic variance to less than 10\% at $6 < z < 12$ for sources with M$_{\rm UV} > -21.25$ AB \citep{Trapp2020} while capturing a statistically significant sample of UV-bright galaxies at $z > 8$. Observations in four NIRCam filters (F115W, F150W, F277W, and F444W) provide photometric redshifts and precise UV luminosity measurements, enabling robust constraints on the UVLF at $z > 8$ not dominated by cosmic variance. Furthermore, the survey’s contiguous field allows for the study of galaxy clustering and large-scale environments, offering unique insights into the role of environment in early galaxy formation and reionization \citep{Behroozi2019, Paquereau2025}.

In this paper, with the addition of archival \textit{HST}/ACS F814W data, we search for galaxies at $z > 5.5$ in the COSMOS-Web survey and subsequently construct the UVLF over the redshift range $z \sim 6-14$, in three redshift bins, $5.5 < z < 8.5$, $8.5 < z < 12$, $12 < z < 15$, covering the full 0.54\,deg$^2$. Section~\ref{sec:data} describes the dataset, data reduction, and photometric catalog creation. Section~\ref{sec:selection} details the selection criteria, purity, completeness, and redshift estimation techniques.  Section~\ref{sec:final_sample} presents the sample. Section~\ref{sec:properties} describes the physical properties of this sample, and we then present the UVLF of these observations in Section~\ref{sec:UVLF}. Finally, Section~\ref{sec:discussion} discusses the implications of our findings for galaxy formation during the EoR and their role in cosmic reionization.  Throughout this paper, we adopt a spatially flat $\Lambda$CDM cosmological model with H$_0$\,=\,70 km\,s$^{-1}$Mpc$^{-1}$, $\Omega_m$\,=\,0.3, and $\Omega_{\Lambda}$\,=\,0.7. We assume a Chabrier \citep{Chabrier2003} IMF. All magnitudes are quoted in the AB system \citep{Oke1983}.

\section{Data}\label{sec:data}

The COSMOS-Web survey (GO \# 1727) is a 270-hour \textit{JWST} treasury program in \textit{JWST} Cycle 1 (PIs: Kartaltepe \& Casey; see overview in \citealt{Casey2023}). The survey is designed to map a contiguous area of 0.54 deg$^2$ with the Near-Infrared Camera (NIRCam; \citealt{rieke03,rieke05,beichman12,rieke23}), complemented by a non-contiguous 0.20 deg$^2$ area observed with the Mid-Infrared Imager (MIRI) in the COSMOS field \citep{scoville07a,capak07a,Koekemoer2007}. Measuring 41.5 arcmin $\times$ 46.6 arcmin, COSMOS-Web covers the largest contiguous area yet observed by \textit{JWST} in a single program, with 152 (19 $\times$ 8) NIRCam pointings.

NIRCam observations have been done through four filters: F115W, F150W, F277W, and F444W.  The achieved depths vary slightly across the mosaic, reaching 5$\sigma$ point-source sensitivities of 26.7--27.4 AB (F115W), 27.0--27.7 AB (F150W), 27.7--28.3 AB (F277W), and 27.6--28.2 AB (F444W) within $0.15\arcsec$ radius apertures (without aperture correction). These filters enable the detection of high-redshift galaxies by sampling their rest-frame ultraviolet and optical emission. The contiguous NIRCam coverage is crucial for ensuring uniform photometry and seamless analysis across the field.

MIRI observations, conducted in the F770W filter, provide mid-infrared imaging at a wavelength of 7.7 microns. MIRI observations achieve depths of 25.2--25.9 AB for point sources within $0.27\arcsec$ radius apertures (with aperture corrections; \citealt{Harish2025}). This variable depth arises from differences in exposure coverage across the mosaic. These observations, which are non-contiguous, lie within the NIRCam footprint \citep{Casey2023}.

\subsection{Data Reduction}

The COSMOS-Web NIRCam data reduction was carried out using the \textit{JWST} Calibration Pipeline \citep{Bushouse2023}, supplemented by several custom modifications to optimize image quality and astrometric precision.  A complete description of the COSMOS-Web data processing and validation is presented in \cite{Franco2025} for NIRCam imaging and \cite{Harish2025} for MIRI imaging, while a brief summary is provided here. All uncalibrated NIRCam images were retrieved from the Mikulski Archive for Space Telescopes (MAST) and processed with pipeline version 1.14.0, incorporating additional corrections similar to those applied in other \textit{JWST} deep-field studies \citep[e.g.,][]{Bagley2023}. These include mitigation of 1/f noise, background subtraction, correction of the wisps and identification of bad pixels. The Calibration Reference Data System (CRDS) mapping pmap-1223, corresponding to the NIRCam instrument mapping imap-0285, was adopted to apply the most up-to-date in-flight calibration files. Final science mosaics were generated at a pixel scale of $0.03\arcsec$/pixel, preserving the spatial resolution necessary for robust photometry. The astrometric solution was refined using the \textit{JWST}/\textit{HST} Alignment Tool (JHAT; \citealt{Rest2023}), aligning the NIRCam images to a reference catalog constructed from an updated and recalibrated version of the \textit{HST}/ACS F814W mosaic \citep{Koekemoer2007} with astrometry tied to Gaia-EDR3 \citep{Gaia_Collaboration2023}. This alignment yields a median offset below 5 mas and a median absolute deviation (MAD) of less than 13 mas across all filters. The MIRI/F770W imaging was reduced using \textit{JWST} pipeline version 1.8.4, applying additional background subtraction steps to correct for instrumental effects. The resulting mosaics were aligned with the NIRCam and \textit{HST}/ACS images at a consistent $0.03\arcsec$/pixel scale.

\subsection{Photometry from the COSMOS-Web Catalog}\label{subsec::photometry}

We base the selection of galaxies in the EoR on the COSMOS-Web catalog described in \citet{Shuntov2025}.  We refer the reader to that work for more detailed information, but provide a brief summary of catalog construction here as it pertains to this sample.  Detection of objects is performed on a $\chi^2_{+}$ image, which is constructed as the quadrature sum of noise-equalized, PSF-homogenized S/N maps for the four NIRCam filters.  The positive subscript means that each S/N distribution has been truncated in the positive direction only, such that when the four filters are added in quadrature, only sources with positive S/N are retained.  Detection is performed using \texttt{SEP} \citep{Barbary2016}, a python implementation of \texttt{SExtractor} \citep{Bertin1996} with a hot/cold approach to capture blended bright sources as well as isolated, low-S/N sources.  Over 780K objects are detected in COSMOS-Web imaging.  We then use \texttt{SourceXtractor++} \citep{Bertin2020, Kummel2020, Kummel2022} to construct \sersic\ models (based only on \textit{JWST} NIRCam imaging) for each of the $\sim$780,000 sources and then extract photometry using those models in over 37 filters (see \citealt{Shuntov2025}) taken at their native resolution.  The uncertainties on the flux densities are then added in quadrature with another term that accounts for Poisson noise from the background, and careful comparisons are made between model-based photometric extractions and aperture-based extractions.  Throughout this work, unless explicitly stated otherwise, we use the \citet{Shuntov2025} model-based photometry, representative of sources' total flux.

\subsection{Relevant Ancillary Data}

The \citet{Shuntov2025} catalog provides photometric extraction in 37 bands across the optical and near-infrared.  Many of these filters are not sufficiently deep to adequately constrain galaxies beyond $z>6$, though others are relevant in this regime, which we briefly highlight here.

For example, the  \textit{HST} Advanced Camera for Surveys (ACS) F814W data in COSMOS reaches a 5$\sigma$ point-source depth of 27.2 AB mag in a 0.24" diameter aperture \citep{Koekemoer2007}. This imaging is particularly valuable for identifying lower-redshift interlopers and confirming the absence of flux for high-redshift galaxy candidates.  The spatial resolution of the \textit{HST} imaging is well matched to the NIRCam imaging (0.09" for \textit{HST}/F814W compared to 0.05"-0.14" for \textit{JWST}).

Ground-based datasets that are of particular use include Subaru Hyper Suprime-Cam (HSC) imaging \citep{Aihara2022} as well as UltraVISTA near-infrared imaging data from Data Release 6 (DR6), with improved depth and coverage compared to earlier releases \citep{McCracken2012,Weaver2022}. The depth of this imaging is sufficient to provide meaningful upper limits (in the case of Subaru HSC imaging in the optical) and marginal detections (in the case of UltraVISTA near-infrared images) for EoR samples (see Table~1 in \citealt{Shuntov2025}).  

\section{Sample Selection}\label{sec:selection}

\subsection{Color Selection Criteria}\label{sec::color_criteria}

To identify high-redshift galaxies, we employ a color selection technique based on the Lyman-$\alpha$ break method \citep{Steidel1996}. This approach exploits the strong discontinuity in a regime where the Lyman-$\alpha$ forest is sufficiently thick ($z>4$), often called the Lyman break. This feature shifts to progressively longer wavelengths with increasing redshift, allowing for the classification of galaxies into specific redshift intervals based on their photometric colors. We divide our sample into three redshift bins based on the position of the Lyman break:
\begin{itemize}

\item $5.5 < z < 8.5$: galaxies where the break is redward of the \textit{HST}/ACS F814W filter but the rest-UV continuum is detected in F115W and in the redder \textit{JWST} filters.
\item $8.5 < z < 12$: galaxies where the break is redward of the \textit{JWST}/NIRCam F115W filter but the rest-UV continuum is detected in F150W and in the redder \textit{JWST} filters.
\item $12 < z < 15$: galaxies where the break is redward of the \textit{JWST}/NIRCam F150W filter but the rest-UV continuum is detected in F277W and in the redder \textit{JWST} filter.
     
\end{itemize}
We then apply the following selection criteria for each high-redshift candidate: (i) A signal-to-noise ratio (S/N) greater than 5$\sigma$ in JWST/NIRCam filters with wavelengths redward of the break. (ii) An S/N less than 2$\sigma$ in the \textit{HST}/ACS and \textit{JWST}/NIRCam filters blueward of the break, as well as in the Hyper Suprime-Cam (HSC) $grizy$ images blueward of the break.
These thresholds ensure the robust detection of the Lyman-$\alpha$ break and minimize contamination from low-redshift interlopers or spurious sources.

To exclude stars, hot pixels, and artifacts, we adopt the flagging scheme described in \citet{Shuntov2025} that identifies non-extragalactic sources based on their size, surface brightness, and morphological parameters. Sources exhibiting features indicative of artifacts or hot pixels—such as irregular shapes or extreme surface brightness—were visually inspected and removed from the final sample when deemed non-physical. 

 \begin{deluxetable*}{ccll}
 \tabletypesize{\scriptsize}
 \tablecaption{Sample Selection Criteria \label{tab:selection_criteria}}
  \tablehead{
  \colhead{Dropout filter} &  \colhead{Redshift range} & \colhead{Criteria} }
 \startdata
 \multirow{5}{*}{F814W} &  \multirow{5}{*}{$z=5.5-8.5$}  & S/N$_{\rm F115W}$,  S/N$_{\rm F150W}$,  S/N$_{\rm F277W}$,  S/N$_{\rm F444W} > 5$,\\
  & & S/N$_{\rm g}$, S/N$_{\rm i}$, S/N$_{\rm r}$, S/N$_{\rm z}$, S/N$_{\rm y} <$ 2, \\
  & & S/N$_{\rm F814W}<2$ OR F814W-F115W $>$0.75, \\
  & & $-$1.0 $<$ F115W - F150W $<$ 0.5, \\
  & & F814W - F115W $>$ 0.5, (F814W - F115W) $>$ (F115W - F150W) + 1.0\\
\hline
 \multirow{5}{*}{F115W} & \multirow{5}{*}{$z=8.5-12$}  &   S/N$_{\rm F150W}$,  S/N$_{\rm F277W}$,  S/N$_{\rm F444W} > 5$,\\
  & & S/N$_{\rm F814W}$,S/N$_{\rm g}$, S/N$_{\rm i}$, S/N$_{\rm r}$, S/N$_{\rm z}$, S/N$_{\rm y} <$ 2, \\
  & & S/N$_{\rm F115W} < 2$ OR F115W - F150W $>$0.75, \\
  & & $-$1.5 $<$ F150W - F277W $<$ 1.0, \\
  & & F115W - F150W $>$ 0.5, (F115W - F150W) $>$ 0.44$\times$(F150W - F277W + 0.8) +0.5\\
\hline
 \multirow{5}{*}{F150W} &  \multirow{5}{*}{$z=12-15$}  &  S/N$_{\rm F277W}$,  S/N$_{\rm F444W} > 5$,\\
  & & S/N$_{\rm F814W}$, S/N$_{\rm F115W}$, S/N$_{\rm F814W}$, S/N$_{\rm g}$, S/N$_{\rm i}$, S/N$_{\rm r}$, S/N$_{\rm z}$, S/N$_{\rm y} <$ 2, \\
  & &  S/N$_{\rm F150W}< 2$ OR F150W - F277 $>$ 0.75, \\
  & & $-$1.0 $<$ F277W - F444W $<$ 0.5, \\
  & & F150W - F277W $>$ 0.5, (F150W - F277W ) $>$ (F277W - F444W) +1.0\\
  \hline
  All & - & $\chi^2$(gal) $<$  1.2 $\chi^2$(star), warn flag =0, $\chi^2$(gal) $<$ 100
\enddata
\tablecomments{Sample selection criteria where S/N is taken as the ratio of flux density to error in flux density for model-based photometric extraction from the COSMOS-Web catalog in \citet{Shuntov2025}.}
\end{deluxetable*}

To refine the selection criteria in color-color space, we used two complementary approaches: first, leveraging the colors of galaxies with robust spectroscopic confirmations at these redshifts; second, using simulated datasets and galaxy models to further validate our selection criteria.

We conducted a systematic search in the DAWN \textit{JWST} Archive (DJA; \footnote{\url{https://dawn-cph.github.io/dja/index.html}}\citealt{Brammer2023,deGraaff2024, Heintz2024}), querying the database for all galaxies having a spectrum with a grade parameter of 3, which corresponds to sources exhibiting robust spectral features and thus having a secure spectroscopic redshift.

\begin{figure*}
    \centering
    \includegraphics[width=\linewidth]{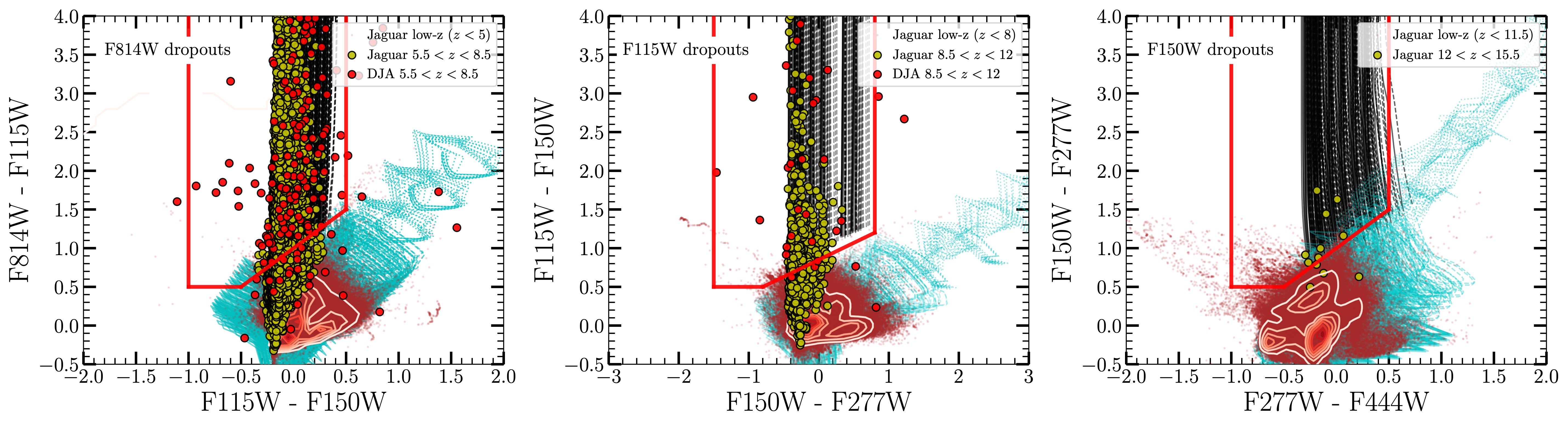}
    \caption{Color-color diagram illustrating the selection regions for high-redshift galaxies. The galaxy populations from the \bagpipes\ models described in Sect.~\ref{sec::color_criteria} are shown as traces, with black lines representing galaxies at the targeted redshifts and cyan lines highlighting the distribution of lower-redshift interlopers (dotted line for the massive galaxies, dashed line for the constant SFH and solid line for the delayed SFH). Additionally, we include mock galaxies from the JAGUAR simulation \citep{Williams2018}, shown in yellow for redshifts corresponding to our different dropout samples and as small brown points overlaid with density contours for lower-redshift galaxies. Furthermore, spectroscopically confirmed galaxies from independent \textit{JWST} observations, compiled in the DAWN \textit{JWST} Archive \citep{Brammer2023,deGraaff2024,Heintz2024}, are marked as red points. These empirical and simulated datasets provide an extra validation of our selection criteria in order to maximize the identification of high-redshift galaxies while minimizing contamination from lower-redshift sources. }
    \label{fig:color_selection}
\end{figure*}

For all identified galaxies, we convolved their spectra with the transmission curves of the filters used in this study, allowing us to place them within our color-color selection diagrams (in red in Fig.~\ref{fig:color_selection}).

We also used \bagpipes\  
\citep{Carnall2018} to simulate galaxy SEDs spanning a wide range of physical properties using the ``Calzetti" attenuation curve \citep{calzetti94, Calzetti2000}. The model parameters included:
\begin{itemize}
\item Stellar mass: $6 \leq \log(M/M_\odot) \leq 10$ in steps of 0.5 dex.
\item Nebular emission: $\log(U)$ varying from $-3$ to $-1$ in steps of 1 dex.
\item Metallicity: [$Z/Z_\odot$] = [0.001, 0.01, 0.05, 0.10, 0.20].
\item Dust attenuation ($A_V$): [0.0, 0.1, 0.2, 0.4, 0.6, 0.8, 1.0, 1.2].
\end{itemize}
We adopted two star formation histories (SFHs): (1) a constant SFH starting 100 Myr before the observed redshift and (2) a delayed SFH with a $\tau$ of 0.4 Gyr. These SED templates were used to populate the color-color diagrams and define regions optimized for high-redshift galaxy selection. 
In addition, to account for the possibility that dusty, massive star-forming galaxies at $z\sim3$–5 may masquerade as high-redshift sources \citep[e.g.,][]{Zavala2023, Arrabal-Haro2023}, we also simulate the SEDs of low-redshift ($0.5<z<5.5$), massive, dusty galaxies. For these, we adopt a delayed star formation history (SFH) with a characteristic timescale $\tau = 0.4$ Gyr, using the following parameters:
\begin{itemize}
\item Stellar mass: $9 \leq \log(M/M_\odot) \leq 11$ in steps of 0.5 dex.
\item Nebular emission: $\log(U)$ varying from $-3$ to $-1$ in steps of 1 dex.
\item Metallicity: [$Z/Z_\odot$] varying from 0.4 to 1 in steps of 0.2.
\item Dust attenuation ($A_V$): from 1 to 5 in steps of 1.
\end{itemize}

Based on these model galaxy spectra, we constructed color–color diagrams to optimize our selection regions, aiming to maximize the number of high-redshift sources while minimizing contamination by outliers. 
Figure~\ref{fig:color_selection} shows the positions of galaxies in different color-color spaces for different redshifts, displaying our separation criterion for splitting galaxies into bins of expected redshifts compared to lower-redshift interlopers.

The choice of axes in each diagram was designed to best capture the dropout feature along the Y-axis and the UV continuum slope (the $\beta$ slope) along the X-axis. Although it is not possible to fully separate high-redshift galaxies from lower-redshift contaminants using strict color boundaries, our selection criteria were defined to closely follow the model-predicted limits.

To further refine and validate these color selections, we overlaid the positions of high-redshift galaxies from semi-empirical models in three redshift bins ($5.5 < z < 8.5$, $8.5 < z < 12$, and $12 < z < 15$) in Figure~\ref{fig:color_selection}. We used JAGUAR semi-empirical models \citep{Williams2018}: a mock \textit{JWST} catalog that includes empirical SEDs and photometry for galaxies up to $z \sim 15$, with a wide range of physical properties. We also checked that the DREaM simulation \citep{Drakos2022}, a semi-empirical simulation that incorporates the physical properties of galaxies and their environments, optimized for understanding \textit{JWST} observational strategies, gives consistent results.

The final color criteria (see Fig.~\ref{fig:color_selection}) were defined as follows:
For the F814W dropouts:
\begin{equation} \label{F814W_dropout}
\begin{split}
     F814W - F115W > 0.5, \rm{AND} \\
     -1.0 < F115W - F150W < 0.5, \rm{AND} \\
     (F814W - F115W) > \\
     (F115W - F150W) + 1.0
\end{split}
\end{equation}

For the F115W dropouts:
\begin{equation} \label{F115W_dropout}
\begin{split}
     F115W - F150W > 0.5, \rm{AND} \\
     -1.5 < F150W - F277W < 1.0, \rm{AND} \\
     (F115W - F150W)>  \\
     0.44\times(F150W - F277W + 0.8) +0.5\
\end{split}
\end{equation}

For the F150W dropouts:
\begin{equation} \label{F150W_dropout}
\begin{split}
     F150W - F277W > 0.5, \rm{AND} \\
     -1.0 < F277W - F444W < 0.5, \rm{AND} \\
     (F150W - F277W) > \\
     (F277W - F444W)  + 1.0
\end{split}
\end{equation}

These selection criteria are summarized in Table~\ref{tab:selection_criteria}.

\begin{figure*}
    \centering
    \includegraphics[width=\linewidth]{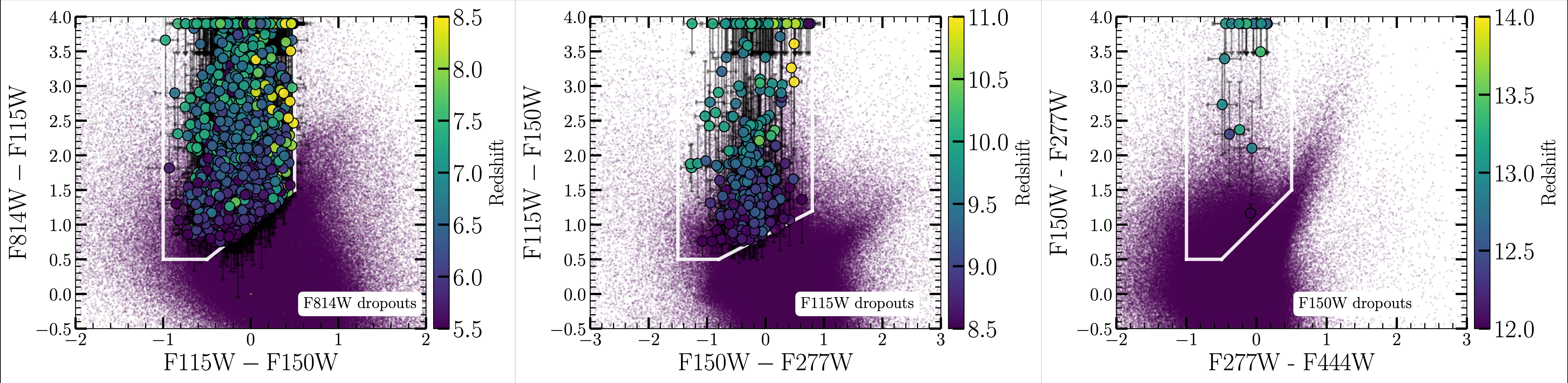}
    \caption{Color-color distribution of galaxies within the selection boundaries defined in Sect.~\ref{sec:selection}, color-coded by redshift for the three dropout samples (F814W dropouts on the left, F115W dropouts in the center, and F150W dropouts on the right). Galaxies that meet our final selection criteria are highlighted as large points, while the full photometric catalog is shown as smaller points.}
    \label{fig:color_selection_CW}
\end{figure*}

\subsection{Spectral Energy Distribution Fitting}

To further reduce contamination and improve the robustness of our high-redshift sample, we employed the photometric redshift code \lephare\  \citep{Arnouts1999, Ilbert2006}. The description of the set of spectral templates, dust attenuation laws, and star formation histories is provided in \citet{Shuntov2025}.

By comparing \lephare\ photometric redshifts with spectroscopic redshifts over the range $0 < z < 8$, they find an overall good agreement, with a normalized median absolute deviation of $\sigma_{\rm MAD} = 0.012$ and an outlier fraction (defined as $\Delta z >$ 0.15(1+z$_{spec}$)) below 2\% for galaxies brighter than $m_{\rm F444W} = 28$. The accuracy decreases at fainter magnitudes, with $\sigma_{\rm MAD}$ increasing from 0.011 at $m_{\rm F444W} < 23$ to 0.030 at $26 < m_{\rm F444W} < 28$, where an increased fraction of catastrophic outliers is observed. These outliers predominantly arise from confusion between the Lyman and Balmer breaks at lower S/N, but no significant redshift bias is detected as a function of magnitude. We describe in Section~\ref{sec::purity} the tests we carried out to quantify the redshift uncertainties for our very high–redshift sources. We refer the reader to \citet{Shuntov2025} for a detailed assessment of the photometric redshift performance in the COSMOS field.

To ensure a robust high-redshift sample, we impose a probabilistic selection criterion based on the redshift probability distribution function (PDF). Specifically, we select galaxies for which the integrated probability of the redshift solution satisfies:

\begin{equation}
\int_{z_{\rm min}}^{z_{\rm max}} P(z) \, dz > 0.6,
\label{eq:pz}
\end{equation}

where we adopt $z_{\rm min} = 5$ for the F814W dropout sample, $z_{\rm min} = 8$ for the F115W dropout sample, and $z_{\rm min} = 11$ for the F150W dropout sample, $z_{\rm max}$ is set to 20 for the three samples. $z_{\rm min}$ corresponds to the minimum redshift that can be detected by taking a margin of $\Delta z$ = 0.5. This criterion ensures that the selected galaxies have a high probability of residing in the targeted redshift ranges, minimizing contamination from lower-redshift interlopers while maintaining a statistically significant sample size.

We note that 60\% of the redshift PDF in the desired redshift range, as given by Eq.~\ref{eq:pz}, may seem low and prone to significant low-redshift source contamination, but we note that the \lephare\ approach to photometric redshift fitting uses a broad range of templates with significant flexibility to increased attenuation and emission line strength relative to stellar continua.  This functionally results in much more weight given to low-redshift solutions (where the range of star formation histories represented by one set of photometry is much more broad than in the EoR) and thus the $P(z)$ distribution, relative to some alternate approaches in the literature \citep[e.g.\ those using {\sc EaZY}][]{brammer2008}, is skewed more conservatively toward low-$z$ solutions.  For example, a \lephare\ $\int P(z)dz>0.6$ corresponds more directly to an {\sc EaZY} $\int P(z)dz>0.9$ (from independent runs).

\begin{figure}
    \centering
    \includegraphics[width=0.9\linewidth]{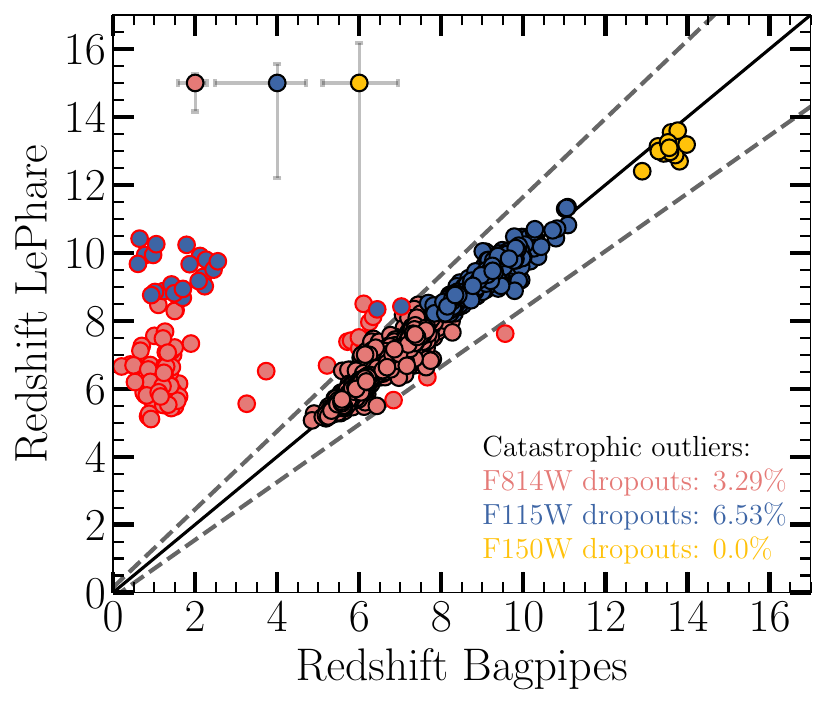}
    \caption{Comparison of photometric redshifts derived from the two SED-fitting codes, \bagpipes\ and \lephare. Each point is color-coded according to the dropout filter used for selection: F814W, F115W, or F150W in pink, blue, and yellow respectively. The black solid line indicates the one-to-one relation. Circles with red edge colors correspond to catastrophic outliers defined in Eq.~\ref{eq::outlier}. The two methods show good agreement for the majority of sources, with more than 95\% of the full sample lying close to the one-to-one relation (black solid line).
}
    \label{fig:comparison_bagpipes_lephare}
\end{figure}

To complement our photometric redshift estimates and ensure the robustness of our derived physical parameters, we computed the redshifts using another SED fitting tool: \bagpipes\ \citep{Carnall2018}. This dual approach allows us to assess systematic uncertainties arising from different assumptions in SED modeling.

The \bagpipes\ analysis follows a delayed exponentially declining star formation history (SFH), where the star formation rate (SFR) evolves as SFR($t$) $\propto$ $t \, e^{-t/\tau}$. To ensure a balanced representation of both young and old stellar populations across cosmic time, we slightly modified the default \bagpipes\ implementation, parametrizing the age of the delayed-$\tau$ SFH as a fraction of the Hubble time at each redshift rather than an absolute age in gigayears. We assume a \citet{Calzetti2000} dust attenuation law and adopt \citet{Bruzual2003} stellar population synthesis models. The total attenuation in the $V$-band is allowed to vary between 0 and 3 magnitudes. Additionally, the ionization parameter is sampled over the range $\log U = [-3, -1]$, the stellar mass spans from $10^{7} M_\odot$ to $10^{11} M_\odot$, and the metallicity varies from 0.001 to 1.0 times the solar value. To account for nebular line emission, we incorporate the updated version (v17.00) of the \textsc{Cloudy} photoionization models \citep{Ferland2017}.
In Fig.~\ref{fig:comparison_bagpipes_lephare}, we compare the photometric redshifts of the two SED-fitting codes. After applying the selection criteria described in Section~\ref{sec:selection}, we find that the results from \bagpipes\ and \lephare\ are largely consistent. To quantify discrepancies, we assess the fraction of catastrophic outliers, defined as those satisfying the condition:
\begin{equation}\label{eq::outlier}
\frac{|z_{\mathrm{bagpipes}} - z_{\mathrm{lephare}}|}{0.5(z_{\mathrm{bagpipes}} + z_{\mathrm{lephare}}) + 1} > 0.15.
\end{equation}
This fraction is found to be 3.29\% (90/2733) for F814W dropouts, 6.53\% (23/352) for F115W dropouts, and 0\% (0/14) for F150W dropouts. Furthermore, we observe that these values exhibit a magnitude dependence: when considering only the brightest galaxies in each sample, the fraction of catastrophic outliers drops to below 2\% for the F814W dropout sample while remaining relatively stable for the other two dropout categories.

\begin{figure*}
    \centering
    \includegraphics[width=1\linewidth]{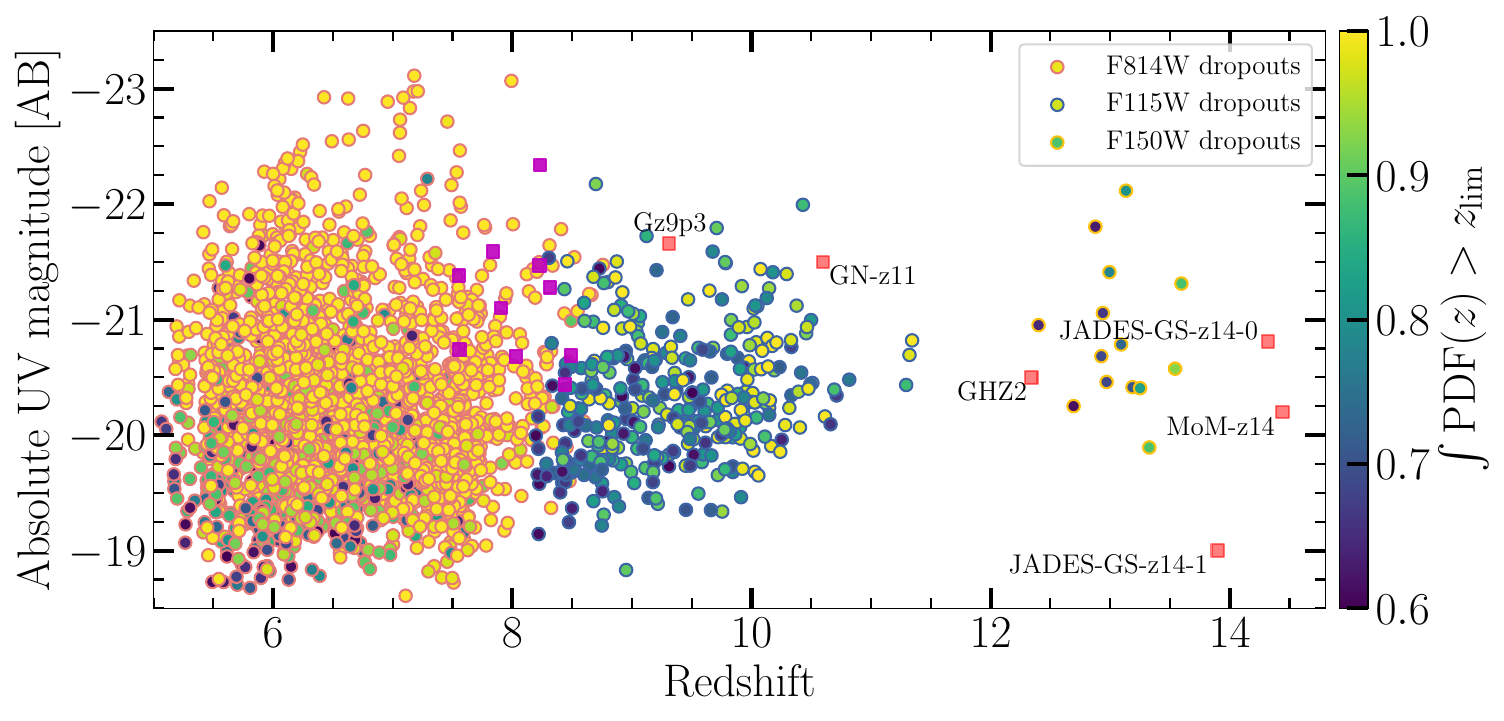}
    \caption{Absolute UV magnitude ($M_{\rm UV}$) as a function of redshift for the three dropout-selected galaxy samples: F814W dropouts (surrounded by a red circle), F115W dropouts (surrounded by a blue circle), and F150W dropouts (surrounded by a yellow circle), color-coded by the probability that the galaxy belongs to the expected redshift range, computed as the integrated probability $\int_{z_{\rm lim}}^{z_{\rm max}} P(z) \, dz$, where $z_{\rm lim} = 5$, $8$, and $11$ for the F814W, F115W, and F150W dropout samples, respectively. For comparison, bright spectroscopically confirmed sources are indicated by red squares, including Gz9p3 \citep{Boyett2024}, GN-z11 \citep{Bunker2023}, GHZ2 \citep{Castellano2024}, JADES-GS-z14-0 \citep{Carniani2024} and MoM-z14 \citep{Naidu2025}, while primary targets from the BoRG-\textit{JWST} Survey \citep{Roberts-Borsani2025} are shown as magenta squares.}
    \label{fig:MUV_redshift}
\end{figure*}

\subsection{Identifying and Removing Stars and Brown Dwarfs}

Cool dwarf stars can contaminate our high-redshift galaxy samples due to their similar colors in near-infrared bands (e.g., \citealt{Bunker2004, Langeroodi2023, Hainline2024, Akins2024}).  Their intrinsic faintness and red colors, resulting from low effective temperatures, mimic the photometric properties of high-redshift galaxies, particularly in \textit{JWST}/NIRCam observations. This is particularly relevant for the selection of $z\sim6-7$ galaxies given the color-space overlap with $\sim$1000\,K M-dwarfs, and less of a concern for higher redshift samples where brown dwarf colors are distinct from LBGs.

\begin{figure*}
\centering
\begin{minipage}{.99\textwidth}
\resizebox{\hsize}{!} { 
\includegraphics[width=0.5\linewidth]{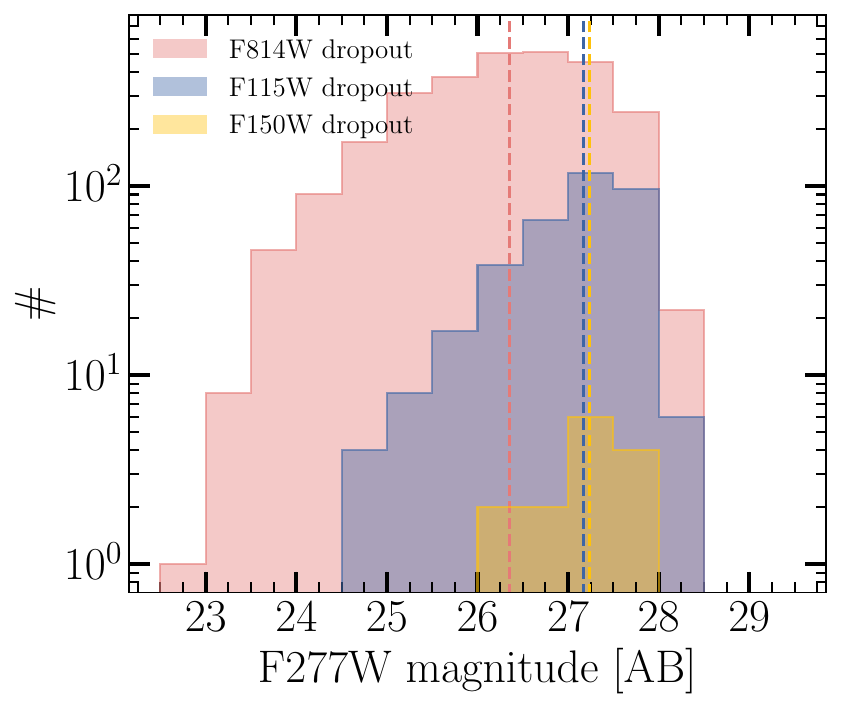} 
\includegraphics[width=0.5\linewidth]{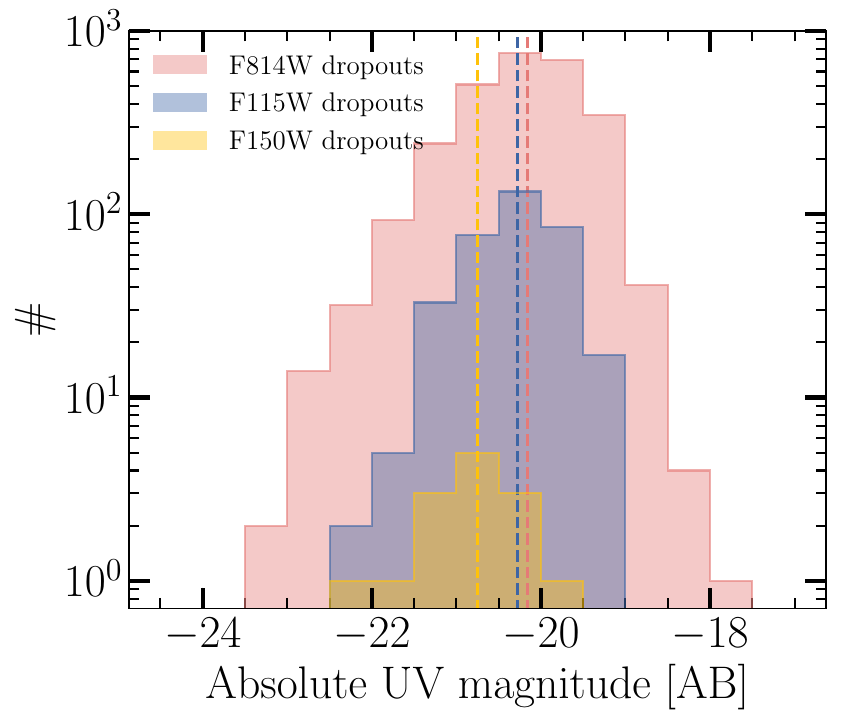} 
}
\end{minipage}
      \caption{Histograms of the F277W apparent magnitude (left) and absolute UV magnitude $M_{\rm UV}$ (right) for our three dropout samples. In each panel, the median of the distribution is marked by a dotted vertical line.}
         \label{Fig::histo_F277}
\end{figure*}

To mitigate this contamination, we fit all color-selected sources with a library of stars and brown dwarf templates. For each source, we evaluated the goodness-of-fit using a $\chi^2$ minimization approach. Sources that exhibited either (\textit{i}) a value of the $\chi_{\rm gal}^2$ at least 20\% higher than the $\chi_{\rm star}^2$ value to fit to brown dwarf templates;  (\textit{ii}) unresolved (i.e. with an effective radius $R_e <$ 3 mas) and with  $\chi_{\rm gal}^2$ $>$  0.5 $\times$ $\chi_{\rm star}^2$, were flagged as likely brown dwarf contaminants (based on a visual examination) and excluded from the final sample. 

This method ensures a robust separation of extragalactic high-redshift candidates from local substellar/stellar objects, minimizing contamination in the ultraviolet luminosity function analysis. Future spectroscopic follow-up of some of these sources will provide further validation of this approach by confirming the stellar or extragalactic nature of ambiguous sources.

\begin{figure*}
    \centering
    \includegraphics[width=\linewidth]{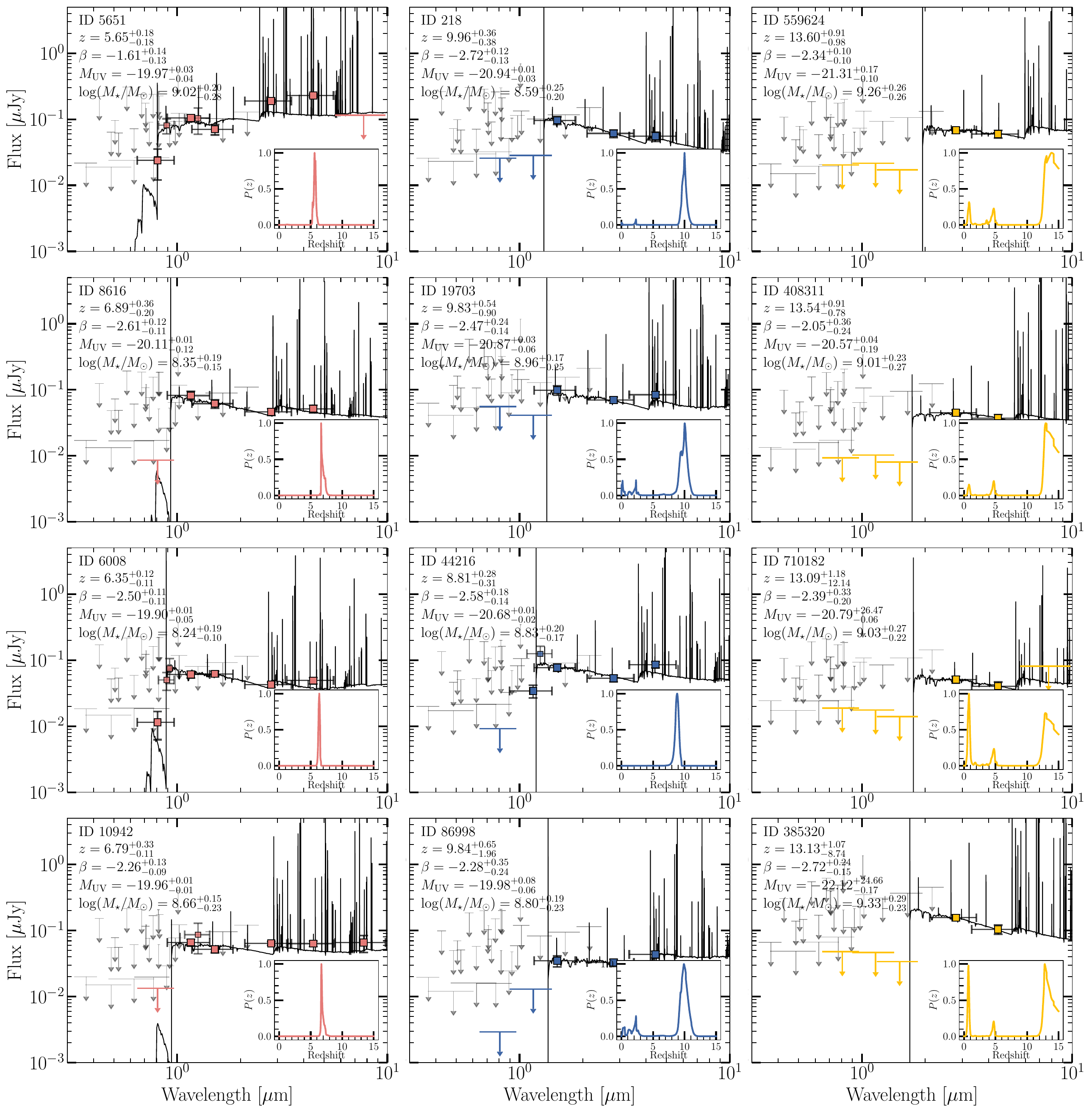}
    \caption{Examples of SED fitting are shown for the three dropout samples (F814W dropouts on the left, F115W dropouts in the center, and F150W dropouts on the right). The best‐fit SED templates were obtained with \lephare\ \citep{Arnouts1999, Ilbert2006}. In each panel, photometric measurements from \textit{HST}/F814W, \textit{JWST}/NIRCam, and \textit{JWST}/MIRI are plotted as larger symbols (or with a color in the case of a non-detection); non‐detections are indicated as 2$\sigma$ upper limits. An inset in each panel shows the redshift probability distribution function, $\mathrm{PDF}(z)\,$. We also indicate the redshift, the UV continuum slope ($\beta$), the absolute UV magnitude ($M_{\rm UV}$), and the stellar mass. The associated uncertainties correspond to the 16th and 84th percentiles of the posterior distributions.}
    \label{fig:test_SEDs}
\end{figure*}

\section{Final Sample}\label{sec:final_sample}

Our final sample consists of $N_{\mathrm{tot}} = 3,099$ sources spanning a redshift range of $z = 5.1$ to $z = 13.6$ (where these values correspond to the \lephare\ median PDF redshifts). These sources were selected based on the criteria described in Section~\ref{sec:selection} and displayed in Fig.~\ref{fig:color_selection_CW}. Without applying a color-based selection and relying solely on the photometric redshifts of \lephare, our sample would have been much more complete, albeit at the expense of purity. So we opted for a more conservative selection. The resulting number of sources is about a factor of two lower than the numbers predicted by \citet{Casey2023}, but this study did not take into account incompleteness effects.

In Figure~\ref{fig:MUV_redshift}, we show $M_{\mathrm{UV}}$ as a function of redshift for all galaxies in the final selection. For comparison, we include bright spectroscopically confirmed sources from recent \textit{JWST} studies \citep[e.g.,][]{Boyett2024, Bunker2023, Carniani2024, Castellano2024, Roberts-Borsani2025, Naidu2025}.

The detection of 74 candidates with $z > 10$ represents a significant addition to the current census of high-redshift galaxies, highlighting the power of \textit{JWST}'s large imaging capabilities over large contiguous areas. In the following subsections, we detail the different properties of these three galaxy samples. 

The left panel of Figure~\ref{Fig::histo_F277} shows the observed F277W magnitude distribution. As expected, galaxies in the F814W sample appear significantly brighter in F277W--by nearly one order of magnitude on average--compared to the higher-redshift dropout samples. The right panel displays the distribution of absolute UV magnitudes. Despite the difference in observed magnitudes (F814W extends to brighter magnitudes, as would be expected given the evolution of UVLF), the $M_{\mathrm{UV}}$ distributions for F814W and F115W samples are broadly similar, with the F150W sample showing a slightly brighter median.

Figure~\ref{fig:test_SEDs} presents representative spectral energy distributions (SEDs) for sources from each dropout sample, color-coded consistently with other figures. Each panel also displays the photometric redshift probability distribution function (PDF), UV continuum slope $\beta$, absolute UV magnitude ($M_{\mathrm{UV}}$) using \lephare, and stellar mass derived using the \cigale\ code \citep{Boquien2019}, taking advantage of its flexible SFHs. Table~\ref{tab:summary_samples} presents the median values of the main properties for these three samples. 

In the subsections below, we provide a brief overview of each dropout sample before discussing their physical properties in more detail in Section~\ref{sec:properties}.

\subsection{F814W dropout sample}

Our F814W dropout sample comprises 2,733 galaxies with photometric redshifts in the range $5.1 < z < 8.8$, representing the largest such sample in our study. Of these, 83\% have a redshift probability with  $\int_{z_{\rm min}}^{z_{\rm max}} P(z) \, dz > 90\%$ indicating a high level of confidence in the redshift solutions. This sample has a mean (median) redshift of 6.6 (6.4). This dataset represents a significant increase relative to pre-\textit{JWST} high-redshift galaxy samples, which included $\sim$2000 sources in this redshift range \citep{Bouwens2021}. The F814W sample also shows considerable diversity, with F277W magnitudes ranging from 22.5 to 28.5 AB, reflecting a broad dynamic range in stellar mass and star formation activity.

\subsection{F115W dropout sample}

The F115W dropout sample includes 352 galaxies in the redshift range $8.2 < z < 11.3$. Here, the median redshift is 9.29, with a mean of 9.33.  The volume of the sample is primarily governed by the depth of the F115W filter and the confidence with which a source is strongly detected at longer wavelengths and not in F115W.  Due to the limited number of available filters for the F115W dropout sample, the robustness of the photometric redshift estimates decreases. Consequently, only about 40\% (62\%) of the galaxies satisfy $\int_{z_{\rm min}}^{z_{\rm max}} P(z)\,dz > 0.90 (0.80)$. Greater depth in the F115W imaging, as will occur for a fraction of the field in the COSMOS-3D program, will likely lead to a larger and more robust sample of F115W dropouts.

 \begin{deluxetable}{ccccc}[h!]
 \tabletypesize{\scriptsize}
 \tablecaption{Median properties for our samples\label{tab:summary_samples}}
  \tablehead{
  \colhead{Dropout filter} & \colhead{N} & \colhead{z$_{\rm {phot, med}}$} & \colhead{M$_{\rm {UV,med}}$}  & \colhead{$\beta_{\rm {UV,med}}$}}
\startdata
F814W &    2733 &  6.41$_{-0.53}^{+1.02}$ & -20.16$_{-0.78}^{+0.62}$ & -2.28$_{-0.29}^{+0.58}$ \\
F115W &     352 &  9.30$_{-0.64}^{+0.72}$ & -20.28$_{-0.55}^{+0.51}$ & -2.52$_{-0.20}^{+0.35}$ \\
F150W &      14 & 13.04$_{-0.16}^{+0.28}$ & -20.73$_{-0.67}^{+0.33}$ & -2.30$_{-0.42}^{+0.64}$ \\
\enddata
\tablecomments{Summary of the median properties of our three samples. The uncertainties on redshift, $M_{\rm UV}$, and $\beta_{\rm UV}$ correspond to the 16th and 84th percentiles.}
\end{deluxetable}

\subsection{F150W dropouts sample}

The F150W dropout sample includes 14 galaxies with redshifts between $11.9 < z < 13.6$. The mean and median redshift are both 13.0. This sample primarily consists of galaxies {\it only} detected in the two LW filters of COSMOS-Web, F277W and F444W.  Two-band detections are intrinsically limiting \citep[as discussed in ][]{Casey2024}, and potentially more prone to contamination that is difficult to constrain.  None of these sources have detections in the UltraVISTA imaging that could help constrain their photometric redshifts. This sample is limited in size and may be affected by contamination biases; it should therefore be interpreted with caution.

\subsection{F277W dropouts sample}

In theory, galaxies at even higher redshift may exist as F277W dropouts in COSMOS-Web. In various surveys, F200W and even F277W dropouts have already been detected and identified as (possible) very high-redshift galaxies \citep[e.g.,][]{Yan2023, Gandolfi2025, Kokorev2025, Castellano2025, Perez-Gonzalez2025}. In COSMOS-Web, these would only be detectable in a single filter, F444W, which prevents reliable identification and any reliable constraints on color or photometric redshift. Thus, we do not analyze or include F277W dropouts in this paper.

\begin{deluxetable*}{cccccccccc}
\tabletypesize{\scriptsize}
\tablecaption{Properties of the galaxies for our final sample (excerpt of 10 galaxies each)\label{tab::final_sample}}
\tablewidth{0pt}
\tablehead{
\colhead{Filter}  & \colhead{ID}  & \colhead{RA}& \colhead{DEC}& \colhead{Redshift}& \colhead{M$_{\rm UV}$} & \colhead{$\beta$}  & \colhead{Radius} & \colhead{Sérsic} & \colhead{M$_\star$} 
\\[-0.2cm]
  \colhead{} & \colhead{} & \colhead{J2000} & \colhead{J2000} & \colhead{} & \colhead{[AB]} & \colhead{} & \colhead{[Arcsec]} & \colhead{} & \colhead{[M$_\odot$]}}
\colnumbers
\startdata
F814W &       93 & 149.777243 & 2.134751 &     6.07$_{-0.15}^{+ 0.17}$  &   -19.56$_{-0.01}^{+ 0.01}$  &    -2.53$_{- 0.10}^{+ 0.14}$ & 0.05 $\pm$ 0.01 & 0.60 $\pm$ 0.43 &     8.49$_{-0.17}^{+ 0.20}$\\
F814W &      131 & 149.736967 & 2.149768 &     6.29$_{-0.12}^{+ 0.11}$  &   -20.57$_{-0.00}^{+ 0.05}$  &    -2.66$_{- 0.09}^{+ 0.10}$ & 0.06 $\pm$ 0.02 & 6.86 $\pm$ 2.27 &     8.64$_{-0.16}^{+ 0.16}$\\
F814W &      663 & 149.816167 & 2.125527 &     7.23$_{-0.08}^{+ 0.15}$  &   -21.81$_{-0.01}^{+ 0.02}$  &    -2.25$_{- 0.08}^{+ 0.08}$ & 0.09 $\pm$ 0.00 & 0.63 $\pm$ 0.07 &     9.69$_{-0.54}^{+ 0.16}$\\
F814W &      886 & 149.750981 & 2.150902 &     6.23$_{-0.11}^{+ 0.08}$  &   -21.14$_{-0.01}^{+ 0.04}$  &    -2.32$_{- 0.10}^{+ 0.11}$ & 0.11 $\pm$ 0.01 & 1.98 $\pm$ 0.34 &     9.53$_{-0.06}^{+ 0.13}$\\
F814W &      912 & 149.830479 & 2.122331 &     6.74$_{-0.14}^{+ 0.25}$  &   -19.79$_{-0.03}^{+ 0.02}$  &    -2.36$_{- 0.13}^{+ 0.22}$ & 0.05 $\pm$ 0.01 & 1.75 $\pm$ 0.99 &     8.73$_{-0.18}^{+ 0.18}$\\
F814W &     1224 & 149.735190 & 2.158941 &     5.66$_{-0.10}^{+ 0.07}$  &   -21.28$_{-0.02}^{+ 0.02}$  &    -1.75$_{- 0.14}^{+ 0.12}$ & 0.30 $\pm$ 0.01 & 0.95 $\pm$ 0.09 &     9.50$_{-0.18}^{+ 0.12}$\\
F814W &     1242 & 149.842992 & 2.119944 &     6.03$_{-0.14}^{+ 0.16}$  &   -20.70$_{-0.01}^{+ 0.01}$  &    -1.98$_{- 0.13}^{+ 0.15}$ & 0.12 $\pm$ 0.01 & 2.75 $\pm$ 0.41 &     9.08$_{-0.17}^{+ 0.18}$\\
F814W &     1373 & 149.790218 & 2.139910 &     5.53$_{-0.12}^{+ 0.08}$  &   -19.89$_{-0.03}^{+ 0.04}$  &    -2.56$_{- 0.13}^{+ 0.11}$ & 0.02 $\pm$ 0.00 & 3.94 $\pm$ 1.58 &     8.73$_{-0.10}^{+ 0.12}$\\
F814W &     1391 & 149.843203 & 2.120775 &     6.07$_{-0.16}^{+ 0.17}$  &   -20.34$_{-0.04}^{+ 0.02}$  &    -1.69$_{- 0.14}^{+ 0.15}$ & 0.08 $\pm$ 0.01 & 2.32 $\pm$ 0.50 &     9.03$_{-0.15}^{+ 0.17}$\\
F814W &     1413 & 149.823691 & 2.127955 &     6.30$_{-0.16}^{+ 0.13}$  &   -20.51$_{-0.01}^{+ 0.06}$  &    -2.48$_{- 0.10}^{+ 0.12}$ & 0.10 $\pm$ 0.02 & 3.80 $\pm$ 1.17 &     8.81$_{-0.18}^{+ 0.11}$\\
\hline
F115W &      218 & 149.805233 & 2.125900 &     9.96$_{-0.38}^{+ 0.36}$  &   -20.94$_{-0.03}^{+ 0.01}$  &    -2.72$_{- 0.13}^{+ 0.12}$ & 0.12 $\pm$ 0.01 & 0.45 $\pm$ 0.20 &     8.59$_{-0.20}^{+ 0.25}$\\
F115W &     2000 & 149.739737 & 2.162444 &    10.02$_{-0.45}^{+ 0.24}$  &   -19.87$_{-0.03}^{+ 0.01}$  &    -2.72$_{- 0.11}^{+ 0.11}$ & 0.06 $\pm$ 0.01 & 0.55 $\pm$ 0.29 &     8.19$_{-0.19}^{+ 0.28}$\\
F115W &     2905 & 149.845232 & 2.129600 &     9.29$_{-7.02}^{+ 0.37}$  &   -20.71$_{-0.06}^{+24.57}$  &    -2.58$_{- 0.23}^{+ 0.24}$ & 0.10 $\pm$ 0.01 & 0.38 $\pm$ 0.14 &     8.47$_{-0.20}^{+ 0.06}$\\
F115W &     9577 & 149.874318 & 2.164015 &     9.93$_{-0.69}^{+ 0.36}$  &   -20.22$_{-0.04}^{+ 0.03}$  &    -2.72$_{- 0.14}^{+ 0.22}$ & 0.10 $\pm$ 0.01 & 0.65 $\pm$ 0.31 &     8.55$_{-0.25}^{+ 0.23}$\\
F115W &    14763 & 149.894364 & 2.196045 &     8.70$_{-1.46}^{+ 0.49}$  &   -20.07$_{-0.03}^{+ 0.05}$  &    -2.56$_{- 0.15}^{+ 0.33}$ & 0.07 $\pm$ 0.01 & 0.47 $\pm$ 0.25 &     8.56$_{-0.23}^{+ 0.16}$\\
F115W &    15571 & 149.835673 & 2.223525 &     8.69$_{-0.16}^{+ 0.14}$  &   -20.64$_{-0.01}^{+ 0.01}$  &    -2.04$_{- 0.11}^{+ 0.09}$ & 0.09 $\pm$ 0.01 & 0.73 $\pm$ 0.31 &     8.85$_{-0.20}^{+ 0.21}$\\
F115W &    15844 & 149.837365 & 2.224930 &     8.64$_{-0.25}^{+ 0.22}$  &   -20.31$_{-0.05}^{+ 0.07}$  &    -1.79$_{- 0.13}^{+ 0.14}$ & 0.01 $\pm$ 0.01 & 4.70 $\pm$ 3.48 &     9.43$_{-0.28}^{+ 0.15}$\\
F115W &    16445 & 149.887843 & 2.210773 &     9.94$_{-0.65}^{+ 0.61}$  &   -20.17$_{-0.17}^{+ 0.06}$  &    -2.58$_{- 1.33}^{+ 0.90}$ & 0.08 $\pm$ 0.01 & 0.51 $\pm$ 0.24 &     8.31$_{-0.11}^{+ 0.19}$\\
F115W &    16523 & 149.796564 & 2.244404 &     9.95$_{-0.48}^{+ 0.47}$  &   -20.38$_{-0.03}^{+ 0.02}$  &    -2.64$_{- 0.14}^{+ 0.25}$ & 0.11 $\pm$ 0.02 & 1.03 $\pm$ 0.44 &     8.74$_{-0.22}^{+ 0.22}$\\
F115W &    17496 & 149.799551 & 2.250041 &    10.33$_{-8.35}^{+ 0.74}$  &   -20.34$_{-0.06}^{+24.43}$  &    -2.43$_{- 0.19}^{+ 0.86}$ & 0.09 $\pm$ 0.01 & 0.44 $\pm$ 0.22 &     8.70$_{-0.20}^{+ 0.17}$\\
\hline
F150W &    50358 & 149.957622 & 2.151699 &    13.33$_{-1.00}^{+ 1.06}$  &   -19.89$_{-0.29}^{+ 0.06}$   &    -1.64$_{- 0.31}^{+ 0.38}$ & 0.00 $\pm$ 0.00 & 0.99 $\pm$ 1.13 &     9.11$_{-0.28}^{+ 0.24}$\\
F150W &   408311 & 149.923702 & 2.659578 &    13.54$_{-0.78}^{+ 0.91}$  &   -20.57$_{-0.19}^{+ 0.04}$   &    -2.05$_{- 0.24}^{+ 0.36}$ & 0.04 $\pm$ 0.01 & 0.91 $\pm$ 1.01 &     9.01$_{-0.27}^{+ 0.23}$\\
F150W &   418915 & 149.908088 & 2.619981 &    12.93$_{-12.13}^{+ 1.28}$ &   -20.68$_{-0.05}^{+26.58}$   &    -2.72$_{- 0.21}^{+ 0.36}$ & 0.06 $\pm$ 0.02 & 1.68 $\pm$ 1.42 &     8.87$_{-0.30}^{+ 0.25}$\\
F150W &   424931 & 150.048117 & 2.459083 &    13.19$_{-8.57}^{+ 1.14}$  &   -20.41$_{-0.06}^{+23.01}$   &    -1.79$_{- 0.36}^{+ 0.33}$ & 0.05 $\pm$ 0.01 & 0.53 $\pm$ 0.38 &     9.19$_{-0.24}^{+ 0.26}$\\
F150W &   525080 & 150.411917 & 2.475959 &    12.97$_{-11.56}^{+ 1.28}$ &   -20.46$_{-0.13}^{+24.74}$   &    -1.37$_{- 0.33}^{+ 0.32}$ & 0.05 $\pm$ 0.01 & 0.74 $\pm$ 0.78 &     9.37$_{-0.29}^{+ 0.24}$\\
F150W &   559624 & 150.484800 & 2.427054 &    13.60$_{-0.98}^{+ 0.91}$  &   -21.31$_{-0.10}^{+ 0.17}$   &    -2.34$_{- 0.26}^{+ 0.34}$ & 0.11 $\pm$ 0.01 & 0.69 $\pm$ 0.37 &     9.26$_{-0.26}^{+ 0.26}$\\
F150W &   588049 & 149.842454 & 2.384471 &    12.99$_{-12.05}^{+ 1.19}$ &   -21.41$_{-0.18}^{+27.34}$   &    -2.72$_{- 0.14}^{+ 0.24}$ & 0.44 $\pm$ nan & 6.96 $\pm$ 2.13 &     9.08$_{-0.24}^{+ 0.29}$\\
F150W &   596146 & 149.906133 & 2.415415 &    13.25$_{-8.38}^{+ 1.11}$  &   -20.41$_{-0.10}^{+22.91}$   &    -2.26$_{- 0.37}^{+ 0.41}$ & 0.04 $\pm$ 0.02 & 1.04 $\pm$ 1.13 &     9.10$_{-0.28}^{+ 0.26}$\\
F150W &   651377 & 150.038831 & 2.242625 &    12.40$_{-11.38}^{+ 1.24}$ &   -20.95$_{-0.15}^{+26.46}$   &    -2.72$_{- 0.24}^{+ 0.32}$ & 0.07 $\pm$ 0.02 & 1.12 $\pm$ 0.89 &     9.10$_{-0.21}^{+ 0.14}$\\
F150W &   710182 & 150.302796 & 2.167558 &    13.09$_{-12.14}^{+ 1.18}$ &   -20.79$_{-0.06}^{+26.47}$   &    -2.39$_{- 0.20}^{+ 0.33}$ & 0.10 $\pm$ 0.02 & 0.79 $\pm$ 0.57 &     9.03$_{-0.22}^{+ 0.27}$\\
\enddata
\tablecomments{(1) Filters where the dropout takes place; (2) ID; (3) and (4) Coordinates of the sources; (5) photometric redshifts from \lephare{} with uncertainties representing the 16th and 84th percentiles (6) UV magnitude; (7) rest-frame UV slopes; (8) and (9) effective radius in arcsec and \sersic\ index;  (9) Stellar mass. Only 10 sources per sample are indicated; the complete set of 3,099 sources comprising our sample is available upon request.}
\end{deluxetable*}

\section{Properties of the Sample}\label{sec:properties}

We report the coordinates and key physical properties derived from SED fitting, including the photometric redshift, UV continuum slope ($\beta$), absolute UV magnitude ($M_{\rm UV}$), stellar mass, effective radius (in arcseconds), and \sersic\ index in Table~\ref{tab::final_sample}. Uncertainties on all parameters correspond to the 16th and 84th percentiles of their posterior probability distributions. What follows is a brief description of the derived properties of the sample.

\subsection{UV $\beta$ slope}

The rest-frame UV spectral slope, $\beta$, provides a key diagnostic of early galaxy populations. It is commonly modeled as a power law, $f_\lambda \propto \lambda^\beta$ \citep{calzetti94, meurer99}.  In practice, we determine $\beta$ for each galaxy by fitting a power law across the rest-frame wavelength range between 1268\,\AA\ and 2580\,\AA, applying the continuum windows prescribed in \citet{calzetti94} to the best-fit spectral energy distribution from \lephare. Since the template library only extends down to $\beta \sim -2.8$, the bluest slopes are effectively truncated, potentially biasing the $\beta$ measurements of intrinsically very blue galaxies.

\begin{figure*}
\centering
\begin{minipage}[t]{1.\textwidth}
\resizebox{\hsize}{!} { 
\includegraphics[width=0.5\linewidth]{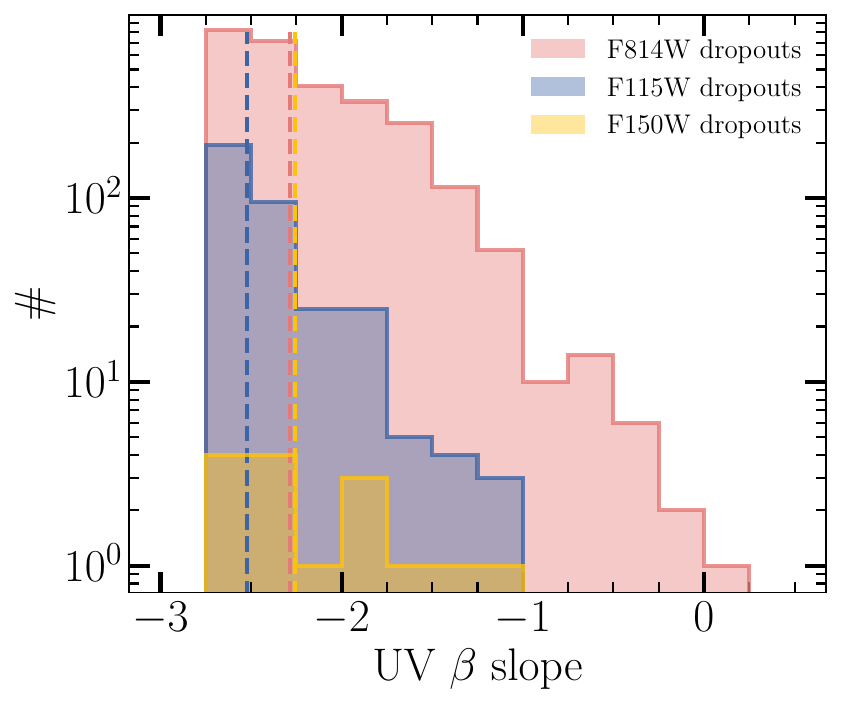} 
\includegraphics[width=0.5\linewidth]{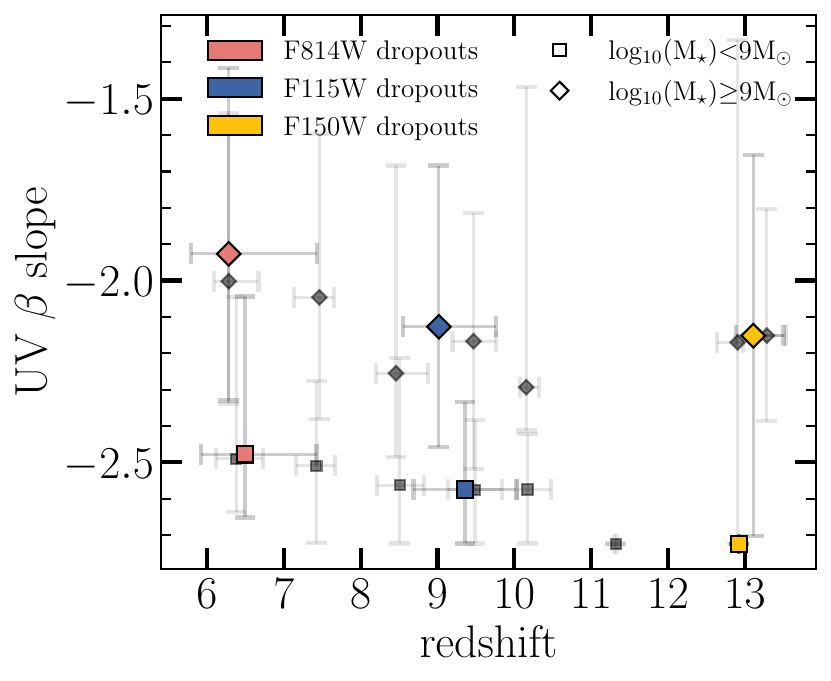} 
}
\end{minipage}
      \caption{Left: Histograms of the ultraviolet spectral slope ($\beta$) for our three dropout samples. Each panel shows the distribution of best-fit $\beta$ values from SED fitting for the F814W, F115W, and F150W dropouts. The median of each sample is indicated by a dashed vertical line. For the F814W and F115W samples, we measure $\beta_{\rm F814W, med}$ = -2.28$_{-0.29}^{+0.58}$ and $\beta_{\rm F115W, med}$ = -2.52$_{-0.20}^{+0.35}$, where uncertainties correspond to the 16th and 84th percentiles. The F150W sample shows a slightly different slope trend ($\beta_{\rm F150W, med}$ = -2.30$_{-0.42}^{+0.64}$), likely due to its larger dispersion.
      Right: Evolution of the UV slope ($\beta$) with redshift for our three dropout samples. To refine this analysis, we binned the full sample in redshift intervals of $\Delta z = 1$. Within each bin, we split galaxies into two stellar-mass bins: $M_\star \ge 10^9\,M_\odot$ (diamonds) and $M_\star < 10^9\,M_\odot$ (squares). More massive galaxies exhibit significantly redder UV slopes than less massive ones.
      }
         \label{fig:histogram_beta_slope}
\end{figure*}

Figure~\ref{fig:histogram_beta_slope} shows the distribution of $\beta$  for our three dropout samples: F814W, F115W, and F150W. For the first two samples, we find a slight evolution for median value of the UV slopes: $\beta_{\rm F814W, med}$ = -2.28$_{-0.29}^{+0.58}$ and $\beta_{\rm F115W, med}$ = -2.52$_{-0.20}^{+0.35}$, with uncertainties representing the 16th and 84th percentiles. These values suggest a gradual evolution of the typical UV color of galaxies for the two first samples. For the full sample, we find a weak anti-correlation between redshift and the UV continuum slope $\beta$, with a Pearson correlation coefficient of $r = -0.3$. Fitting a linear relation, we derive an evolution rate of $d\beta/dz = -0.09 \pm 0.02$, consistent with pre-\textit{JWST} results reported at lower redshifts (up to $z \sim 6$) by \citet{bouwens14}.

However, when restricting the analysis to galaxies at $z > 8$, we no longer observe a significant evolution with redshift, with $d\beta/dz = 0.00 \pm 0.02$ suggesting that UV slopes remain roughly constant beyond $z \sim 8$.

This plateau may signal a transition in the physical properties of galaxies in the early Universe. In particular, the lack of $\beta$ evolution at high redshift could indicate that these galaxies are starting to be ``dust-free" (dust attenuation free  or with regions with lack of dust; \citealt{Faisst2017}), allowing UV photons to escape easily, or evidence for extremely low metallicities \citep{Bouwens2010}.  To further emphasize this point, we subdivide the data into smaller redshift bins of $\Delta z = 1$ (Fig.~\ref{fig:histogram_beta_slope}-right). We find a nearly flat trend, consistent with the results of \citet{Topping2024}, who report $\mathrm{d}\beta/\mathrm{d}z = -0.038 \pm 0.017$ for a sample of bright galaxies (M$_{\rm UV}$=-20), typical of the M$_{\rm UV}$ of our analysis.

For the F150W dropout sample, the derived slope is $\beta_{\rm F150W}$ = $-2.26^{+0.58}_{-0.47}$. While this estimate is more uncertain due to the limited sample size and significant dispersion, it remains consistent with no redshift evolution of the $\beta$ slope at higher redshifts.

\begin{figure*}
\centering
\begin{minipage}[t]{1.\textwidth}
\resizebox{\hsize}{!} { 
\includegraphics[width=0.5\linewidth]{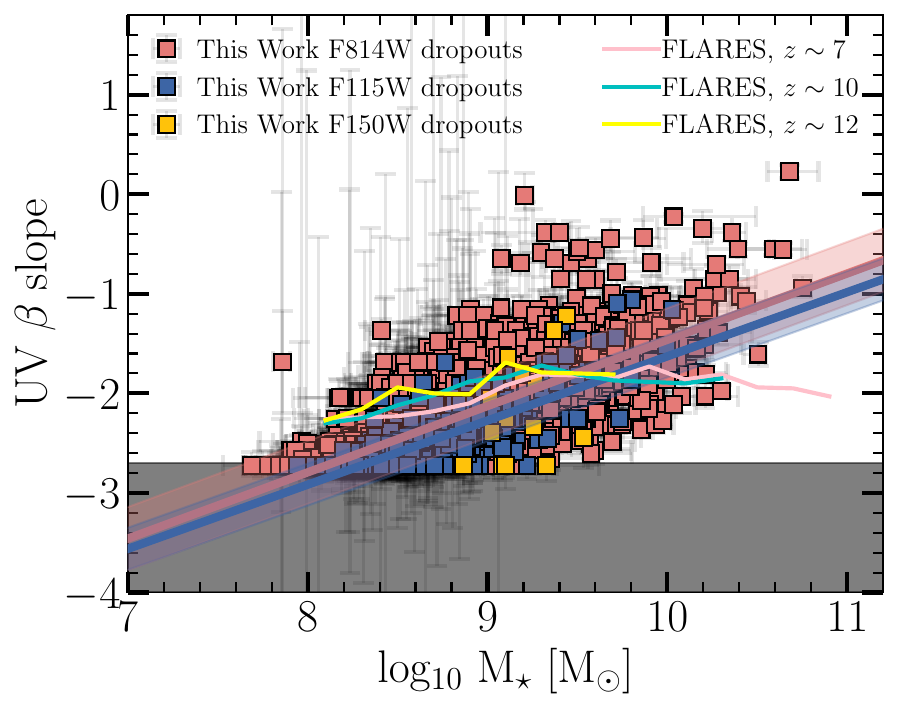} 
\includegraphics[width=0.5\linewidth]{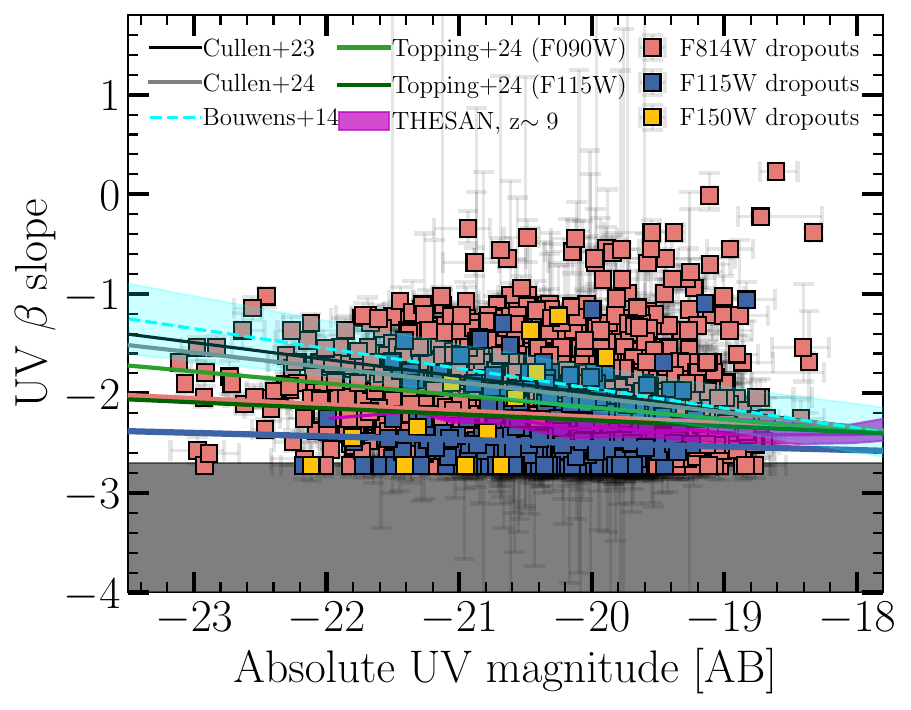} 
}
\end{minipage}
      \caption{Left: Ultraviolet spectral slope ($\beta$) versus stellar mass for our three samples. A clear trend exists between $\beta$ and stellar mass.  A linear fit is shown for the F814W and F115W samples. For the full sample, the fit yields a slope of 0.67 $\pm$ 0.01  indicating that more massive galaxies at these epochs tend to have redder UV continua.  For comparison, we also show predictions from the hydrodynamic simulations of galaxy formation and evolution \textsc{Flares} \citep{Lovell2021, Vijayan2021}. Right panel: UV continuum slope ($\beta$) as a function of absolute UV magnitude ($M_{\rm UV}$). We show, for comparison, the measured relations derived by \citet{bouwens14} at $z \sim 7$ (brown), \citet{Cullen2023, Cullen2024} at $8 < z < 16$ (black), \citet{Topping2024} for the F090W and F115W dropouts (light green and dark green respectively)  as well as the theoretical prediction from the \textsc{THESAN} simulation suite at $z \sim 9$ \citep[magenta;][]{kannan22}.}
         \label{Fig::beta_mass}
\end{figure*}

We detect a clear correlation between UV slope and stellar mass (derived from \cigale, see Sect.~\ref{sec::stellar_mass} and \citealt{Shuntov2025}) in our sample, shown in Fig.~\ref{Fig::beta_mass} (left). Galaxies with higher $M_\star$ tend to have systematically redder UV slopes, as more massive galaxies have typically formed a larger number of stars at earlier epochs. To illustrate this trend, we divide the sample into stellar mass bins. The lowest-mass galaxies ($M_\star < 10^9\,M_\odot$) exhibit significantly bluer slopes, while the highest-mass galaxies ($M_\star \ge 10^9\,M_\odot$) show notably redder spectra. This suggests that more massive galaxies, even at high redshifts, experience greater dust attenuation or/and host stellar populations of a more heterogeneous age distribution, with metallicity potentially contributing to this trend. Similar trends have been reported by \citet{Bouwens2010, Finkelstein2012, Tacchella2022, Roberts-Borsani2024}. However, this trend is slightly stronger than that reported by \citet{Finkelstein2012}, but it is consistent with the results of \citet{Tacchella2022}. We quantify this relation for the F814W and F115W dropout samples as:

\begin{align}
    \beta_{\rm F814W} &= (0.67 \pm 0.01)\, \log_{10} (\text{M}_{\star}/\text{M}_\odot) - 8.12 \pm 0.11 \\
    \beta_{\rm F115W} &= (0.64 \pm 0.03)\, \log_{10} (\text{M}_{\star}/\text{M}_\odot) - 8.07 \pm 0.25 
\end{align}

We observe that for the F814W and F115W dropout samples, which benefit from robust statistics, the evolution of the UV continuum slope $\beta$ with redshift follows a consistent trend across both samples. For the F150W sample, the statistics are insufficient to provide reliable constraints.

In Fig.~\ref{Fig::beta_mass}-right panel, we investigate the relation between the UV continuum slope and the absolute UV magnitude. Across our full sample, we find no strong correlation between $\beta$ and $M_{\rm UV}$, with a Pearson correlation coefficient of $r = -0.1$, indicating only a weak trend. However, we observe that the $\beta$–$M_{\rm UV}$ relation is more pronounced at lower redshift. Specifically, for the F814W sample, corresponding to $z \sim 6$, we measure a slope of $d\beta/dM_{\rm UV} = -0.06 \pm 0.01$, suggesting a mild increase in $\beta$ toward brighter magnitudes. However, this trend is flatter than that reported by \citet{bouwens14} for similar redshifts. In contrast, the F115W sample at $z \sim 9$ exhibits a relatively flat trend, with $d\beta/dM_{\rm UV} = -0.03 \pm 0.03$, indicating no significant correlation between UV slope and luminosity. This may reflect a shift in the dominant physical processes at $z > 8$, where UV-luminous galaxies could be less affected by dust attenuation (becoming progressively dust-free, or having regions in which dust is spatially segregated from the stellar populations) compared to their lower-redshift counterparts.

Taking into account that the $\beta$ slope depends only secondarily on redshift and absolute UV magnitude (see Fig.~\ref{fig:histogram_beta_slope} and \ref{Fig::beta_mass}), we derive a global relation using the full sample:
\begin{equation}
    \beta_{\rm tot} = (0.67 \pm 0.01)\, \log_{10} (\text{M}_{\star}/\text{M}_\odot) -8.18 \pm 0.10
\end{equation}

This slope is slightly steeper than predicted by the \flares\ simulation. In particular, \flares\ tends to produce UV continuum slopes that are redder for low-mass galaxies (log$(M_*/M_\odot) \lesssim 9.5$), while more massive galaxies are predicted to be slightly bluer than observed in our data (see Fig.~\ref{Fig::beta_mass}). We note that our $\beta$ measurements, derived from photometric SEDs, are subject to systematic uncertainties. 
While some studies have suggested that photometric $\beta$ estimates can be systematically redder than their spectroscopic counterparts \citep{Austin2024}, recent analyses \citep[e.g.,][]{Morales2025} indicate that power-law fits with three or more bands, as well as SED-based methods, yield reliable results. Additionally, our SED templates may not fully capture extremely blue populations such as those found in \citet{Cullen2024}. When the nebular continuum is included in population synthesis models, the bluest achievable UV slope is limited to $\beta \approx -2.6$ \citep{Topping2022, Cullen2024}. Our models do not allow $\beta$ values below this limit (see gray shaded region in Fig.~\ref{Fig::beta_mass}). It is important to note that completeness can influence this scenario, since our dataset tends to detect low-mass blue galaxies more readily than low-mass red ones.

For intrinsically UV-bright galaxies ($M_{\rm UV} < -21$), our measured $\beta$ slopes tend to be slightly bluer than those measured from previous studies, such as \citet{Cullen2023, Cullen2024}, and \citet{bouwens14}, as shown in Fig.~\ref{Fig::beta_mass}-right. This deviation may suggest differences in dust content, stellar population age, or metallicity in the most luminous galaxies compared to the average population at similar redshifts. This difference could arise from the fact that, in these studies, the brightest galaxies were primarily identified using ground-based observations over significantly larger survey volumes. As a result, these sources could be more massive on average than those in our sample. Given the correlation between stellar mass and UV continuum slope, such more massive galaxies could tend to exhibit redder $\beta$ values. Interestingly, the observed trend is more consistent with predictions from the \textsc{THESAN} simulations \citep{kannan22a, garaldi22, smith22}, which model the physical conditions of early galaxies ($z\sim9$), including their emission line properties and dust attenuation (Fig.~\ref{Fig::beta_mass}).

\subsection{M$_{\rm UV}$}

To determine each galaxy’s luminosity at 1500\AA, we use the best fit from \lephare. From that spectrum, we measure the flux density through a 100\AA‐wide ``top‐hat" filter centered on 1500\AA, and convert the result into an apparent magnitude, \(m_{1500}\).

We then convert \(m_{1500}\) into an absolute UV magnitude via
\begin{equation}
M_{\rm UV} \;=\; m_{1500} \;-\; 5\,\log_{10}\!\biggl(\frac{D_L}{10\,\mathrm{pc}}\biggr) \;+\; 2.5\,\log_{10}(1+z),
\end{equation}
where $D_L$ is the luminosity distance.

Figure~\ref{Fig::histo_F277}-right shows the distribution of M$_{\rm UV}$ values for our three dropout samples. Although the typical observed F277W magnitudes differ significantly across the three dropout samples—with galaxies in the F814W sample exhibiting substantially brighter fluxes compared to their higher-redshift counterparts—all three samples peak in absolute UV magnitude at $-21 < M_{\rm UV} < -20$ AB. This convergence primarily reflects the incompleteness of our survey at high redshift, which limits our ability to detect intrinsically fainter UV sources at $z \gtrsim 8$.

\subsection{Stellar mass}\label{sec::stellar_mass}

Estimating stellar masses of high-redshift galaxies is affected by substantial biases arising both from observational constraints and model assumptions \citep[e.g.,][]{Pforr2012}. On the observational side, limited rest-frame wavelength coverage shifts into bluer bands at increasing redshift, making SED fits increasingly sensitive to the light of young, massive stars and less sensitive to older, mass-dominant stellar populations (outshining effect; \citealt{Maraston2010}). This could lead to systematic underestimation of stellar masses by $\sim$0.4 dex at $z\sim5-9$ for sources without MIRI counterpart \citep{Papovich2023, Song2023, Wang2024, Akins2024}. On the modeling side, assuming smooth, continuous star formation histories exacerbates this bias, as recent starbursts can dominate the SED and obscure older stellar components. Incorporating the MIRI observations modifies this and further refines the assessment of nebular emission line \citep{Stefanon2022, Papovich2023}. Moreover, the choice of SFHs strongly influences mass estimates; simplistic assumptions such as constant SFHs or exponentially declining models may misrepresent the true assembly histories of high-redshift galaxies, as recent studies suggest bursty and rapidly evolving SFHs at $z>8$ \citep{Whitler2023, Mirocha2023, Cullen2023}. Without such constraints, SED fits may overestimate the contributions of young, UV-bright stellar populations, biasing the inferred stellar masses \citep{Papovich2023, Wang2024}. Dust attenuation adds another layer of complexity, as high-redshift galaxies likely exhibit varying dust contents that can affect mass-to-light ratios as well as the assumed stellar population template \citep{Maraston2006}.

The stellar masses of these galaxies were derived using \cigale\ \citep{Boquien2019} with the parameters described in \cite{Shuntov2025} and \cite{Arango2025}. The median stellar mass for the F814W sample is log$_{10}$M$_{\rm \star, F814W}$/M$_\odot$ = 8.81$_{-0.43}^{+0.52}$, with uncertainties representing the 16th and 84th percentiles. For the F115W sample, this median value slightly decreases to  log$_{10}$M$_{\rm \star, F115W}$/M$_\odot$ = 8.69$_{-0.28}^{+0.39}$. However, for the F150W sample, the median increases to log$_{10}$M$_{\rm \star, F150W}$/M$_\odot$ = 9.30$_{-0.21}^{-0.60}$ which appears to contradict the anticipated trend of decreasing stellar mass with increasing redshift. We will address this discrepancy further in the following section.

Considering the potential biases in estimating stellar masses at high redshift, we compared the derived stellar masses for our three galaxy samples with theoretical predictions derived using extreme value statistics (EVS; \citealt{gumbel58, kotz00}). 

\begin{figure}
    \centering
    \includegraphics[width=1\linewidth]{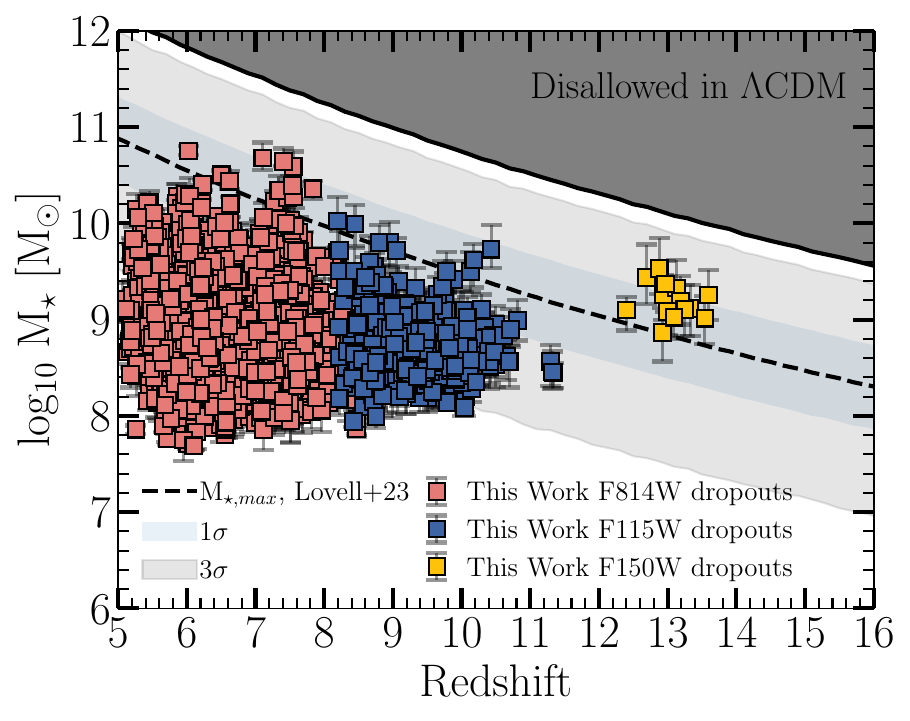}
    \caption{Stellar mass as a function of the redshift for our 3 samples. We superimposed the predicted extreme value statistics PDF of the galaxy stellar mass distribution computed in \cite{Lovell2023} for the size of COSMOS-Web. The shaded regions represent the 1$\sigma$ and 3$\sigma$ confidence intervals for the maximum stellar mass expected in a $\Lambda$CDM cosmology at a given redshift. All the galaxies in the  F814W and F115W dropout samples are below the extreme values given the uncertainties. However, we note that for the F150W dropout sample, the typical stellar masses lie slightly above the values predicted by the EVS (although they remain within $\sim$1$\sigma$ of the expected range). This offset could be attributed to a systematic overestimation of stellar masses in the absence of photometric constraints at rest-frame wavelengths longer than 0.35\,$\mu$m, which are crucial for robustly anchoring the stellar mass estimates (see Sect.\ref{sec::stellar_mass}). The parameter space disallowed by $\Lambda$CDM for the most massive galaxies assuming a baryon-to-stellar conversion rate of 1 is displayed by a solid line.}
    \label{fig:redshift_mass}
\end{figure}

Fig.~\ref{fig:redshift_mass} illustrates the measured stellar masses as a function of redshift alongside predictions based on the EVS framework developed by \citet{Lovell2023}, scaled to the survey's area of 0.54 deg$^2$. The EVS method estimates the probability distribution of extreme (maximum) stellar mass galaxies expected from a volume-limited sample, derived initially from halo mass functions (\citealt{harrison11}) and subsequently extended to stellar mass via assumptions about the stellar-to-halo mass relation. If observed galaxy masses significantly exceed these theoretical predictions, it could indicate tension with either the standard $\Lambda$CDM cosmological paradigm or our astrophysical assumptions regarding galaxy formation at high redshift.

The dashed line in Fig.~\ref{fig:redshift_mass} shows the median of the predicted maximum stellar mass distribution for the survey area, while shaded regions represent 1$\sigma$ and 3$\sigma$ confidence intervals. These predictions assume a baryon fraction of 0.16 \citep{Planck2016} and a log-normal distribution for the stellar-to-halo mass fraction. 

For the F814W and F115W dropout samples, our results show that the most massive galaxies within each redshift bin generally align closely with the median predictions, with no individual galaxy significantly exceeding the 1$\sigma$ deviation (within the error bars). This indicates that our observed galaxy sample remains fully consistent with the theoretical expectations under standard cosmological assumptions. Conversely, our F150W sample includes galaxies with notably high stellar masses, exceeding the limiting stellar mass predicted by the EVS approach (but still more than a factor of five below the parameter space ruled out by $\Lambda$CDM for the most massive galaxies, assuming a baryon-to-stellar conversion rate of 1). This discrepancy is intriguing and might originate from biases in stellar mass estimates, as discussed above, from incorrect redshift identifications, or from potential AGN contributions that bias stellar mass estimates high. All of these galaxies lie at redshifts greater than 12. At such high redshifts, the strongest constraint on their SEDs comes from the F444W filter, which probes rest-frame wavelengths shorter than 350\,nm, that is within the ultraviolet regime. This presents a fundamental limitation for deriving reliable stellar mass estimates, as rest-frame optical or near-infrared coverage is required to constrain the older stellar populations that dominate the mass budget. Physically meaningful stellar mass estimates for these sources will require much deeper mid-infrared imaging, such as that provided by future MIRI observations. These galaxies thus warrant further detailed investigation. However, in all cases, these galaxies remain well within the physically permissible range predicted by the $\Lambda$CDM model.

\subsection{Sizes}

The source sizes and morphological parameters (\sersic\ index, ellipticity, and position angle) were measured directly by \texttt{SourceXtractor++} \citep{Bertin2020, Kummel2020, Kummel2022} during the source extraction process. The detailed configuration parameters, including priors applied during this extraction, are comprehensively described in \citet{Shuntov2025} and \citet{Yang2025}.

\begin{figure}
    \centering
    \includegraphics[width=\linewidth]{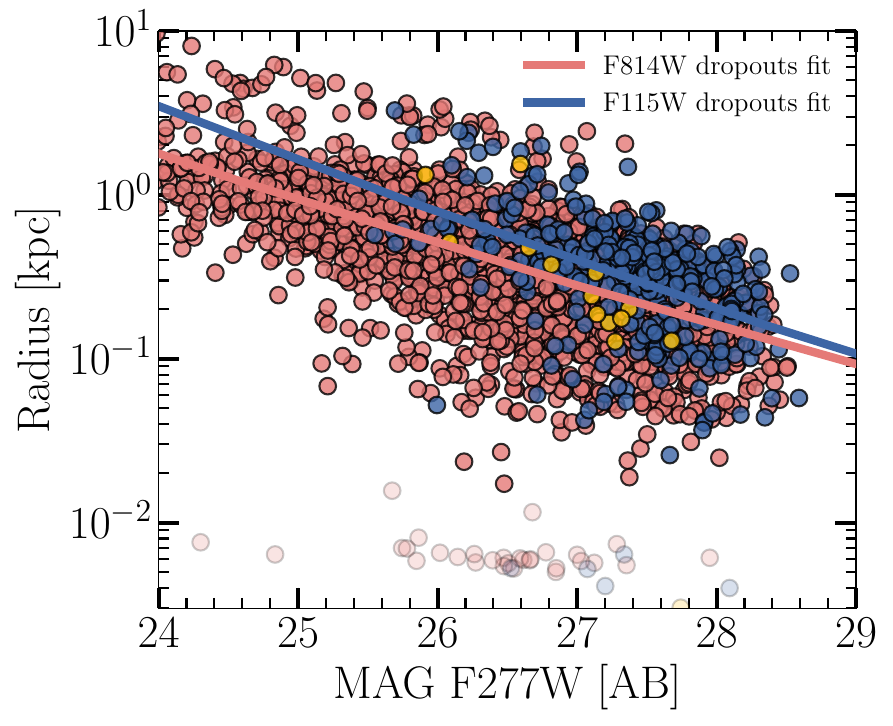}
    \caption{Relation between the F277W magnitude and effective radius (in kpc) for galaxies in our three samples. Galaxies that are unresolved ($R_e < 3$mas) are shown as faded points. }
    \label{fig:size_magnitude}
\end{figure}

In Fig.~\ref{fig:size_magnitude}, we present the effective radius of our sample as a function of the F277W magnitude. For the three samples, we fit this trend after excluding the clear unresolved sources (see Fig.~\ref{fig:size_magnitude}). 
This correlation between magnitude and size can be parameterized by the following relation: 
\begin{equation}
R_e = r_0 \left( \frac{\rm mag_{F277W}}{\rm mag_0} \right) ^ {\alpha} 
\end{equation}
where \(R_e\) is the effective radius in kpc, $\mathrm{mag}_{\mathrm{F277W}}$ is the apparent magnitude in the F277W filter, $r_0$ is a normalization constant corresponding to the effective radius at a reference magnitude ($\mathrm{mag}_0$), and $\alpha$ quantifies the strength of the size–magnitude correlation. The slope becomes progressively steeper with increasing redshift with $\alpha_{F814W} = -15.6\pm0.3$, $\alpha_{F115W} = -18.4\pm1.6$ and $\alpha_{F150W} = -24.9\pm2.8$.

This correlation reflects the well-established trend that brighter galaxies tend to be more extended, consistent with previous \textit{JWST} results \citep[e.g.,][]{Yang2022, Ono2023, Adams2023}. The absence of large, faint galaxies likely reflects detection biases rather than intrinsic properties, as extended sources become increasingly difficult to observe at high redshift. These sources are generally compact, with median sizes ranging between 0.07 and 0.11 arcseconds across the three samples. 
In terms of physical size, this corresponds to 390$_{-228}^{+422}$pc for the F814W sample, 344$_{-206}^{+240}$pc for the F115W sample and 241$_{- 78}^{+338}$pc for the F150W sample, with uncertainties corresponding to the 16th and 84th percentiles for each sample.

A small fraction of the sources are unresolved, having effective radii ($R_e$) smaller than 3 milliarcseconds, representing less than 1.5\% of the F814W and F115W dropout samples, and a single source in the F150W dropout sample. These unresolved sources predominantly exhibit red F277W–F444W colors and are most likely associated with the ``Little Red Dots" population identified in previous studies \citep[e.g.,][]{Labbe2023, Matthee2024}. Due to their unresolved nature, they have been excluded from our size fitting analysis. However, since their origin remains the subject of active debate \citep[e.g.,][]{Furtak2023, Kocevski2023, Labbe2023, Akins2024, Taylor2025}, and a stellar origin cannot be definitively ruled out, we have retained them in the rest of our analysis.

The \sersic\ indices for the full sample are typically close to 1, with a median value of 0.9 and 16th–84th percentiles of 0.5 and 2.9, respectively. This is consistent with expectations for disky, star-forming galaxies \citep[e.g.,][]{Quilley2025}. A small number of sources exhibit larger sizes and appear as outliers relative to the size–mass relation; these cases almost always correspond to notably high \sersic\ indices. This may reflect a degeneracy between galaxy size and \sersic\ index in the fitting process. Both low \sersic\ indices and compact radii could be indicative of high star formation efficiencies, compatible with a feedback-free starburst scenario as suggested by recent studies \citep{Dekel2023}.

\section{UV Luminosity function}\label{sec:UVLF}

\begin{figure}
    \centering
    \includegraphics[width=\linewidth]{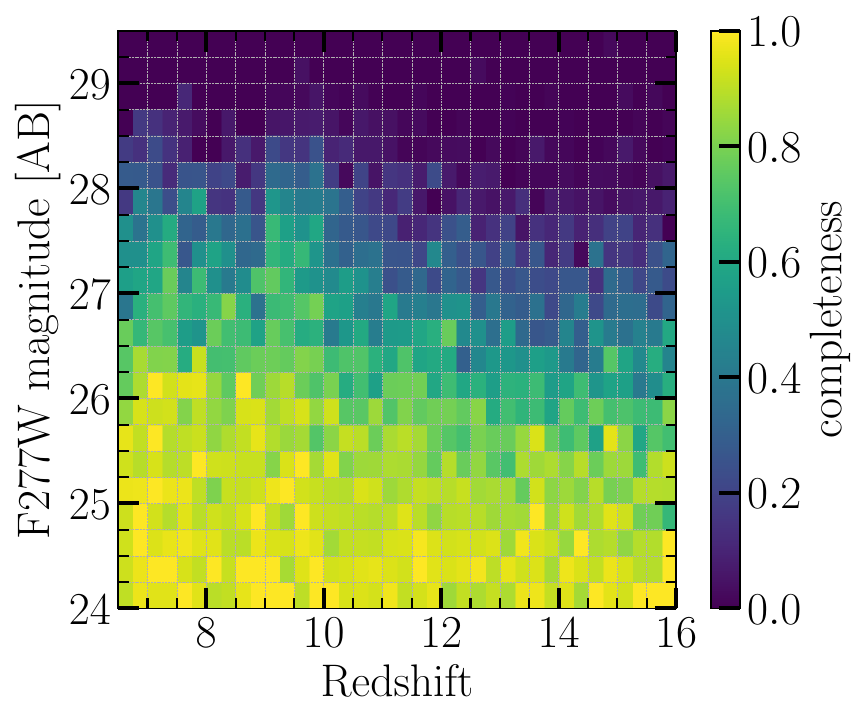}
    \caption{Completeness as a function of input F277W magnitude and input redshift for the simulated sources injected following the methodology described in Sect.~\ref{sec:completeness}. The color scale represents the completeness fraction, defined as the ratio of detected sources to the total number of injected sources in each magnitude-redshift bin.} 
    \label{fig:complteness_F277W}
\end{figure}

\subsection{Completeness correction}\label{sec:completeness}

The accurate determination of the UVLF is critically dependent on understanding both contamination and incompleteness, especially at high redshifts. To address these challenges, we performed rigorous completeness simulations using the DREaM \citep{Drakos2022} simulation. DREaM is a forward-modeling framework that generates realistic mock galaxy catalogs, finely tuned to the characteristics of \textit{JWST} observations, providing detailed predictions for galaxy photometry, morphology, and spatial distributions. These simulations incorporate empirical constraints to ensure a high degree of fidelity in the representation of real galaxy populations.

\begin{figure}
\centering
\begin{minipage}[t]{.5\textwidth}
\resizebox{\hsize}{!} { 
\includegraphics[width=.5\linewidth]{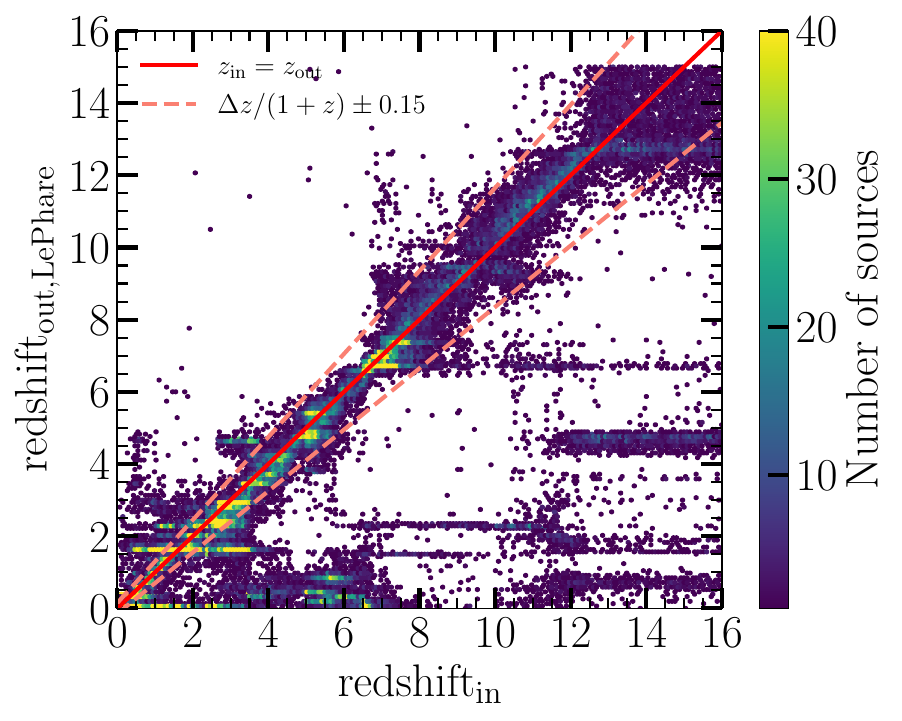} 
}
\end{minipage}
      \caption{Comparison of input photometric redshift $z_{\mathrm{in}}$ and recovered redshift $z_{\mathrm{out}}$ from \lephare~on simulated sources described in Sect.~\ref{sec:completeness} with mag$_{\rm F277W} <$ 27.5 AB.  Hexagonal bins encode the number of sources per bin. The solid red line marks the one–to–one relation $z_{\mathrm{out}} = z_{\mathrm{in}}$, while the dashed black lines show the limits of catastrophic outliers (defined as $|\Delta z|/(1 + z) > 0.15$).
}
         \label{Fig::comparison_simulations}
\end{figure}

In DREaM, individual sources are characterized by five key parameters: their flux in each filter, effective radius, S{\'e}rsic index, ellipticity, and position angle. To model these sources realistically, we employed the modular galaxy image simulation toolkit \texttt{GalSim} \citep{Rowe2015}, coupled with the wavelength-dependent point spread function (PSF) of \textit{JWST} (see \citealt{Shuntov2025} for the details on the creation of the PSFs). This combination allows for accurate simulation of galaxy images as they would appear in NIRCam observations, accounting for instrumental effects such as PSF broadening and detector noise.

However, as DREaM (like any semi-analytical model for galaxy formation) is tuned to reproduce the stellar mass function and UVLF at high-$z$, it provides limited statistics in the EoR. 
To ensure robust completeness statistics even out to $z\sim 16$, we supplement the DREaM catalog with mock sources constructed from \bagpipes\ SEDs over custom distributions of galaxy properties. 
In particular, we use a uniform distribution of redshifts from $z=5$--$16$ and UV luminosities from $M_{\rm UV} = -24$ to $-18$, and log-normal distributions for other physical parameters including metallicity, stellar age and $\tau$, dust attenuation $A_V$, and escape fraction $f_{\rm esc}$. 
We adopt log-normal distributions for the effective radius, S{\'e}rsic index, and axis ratio, tuned to match the distributions of the real COSMOS-Web catalog.  

We injected $N=120,000$ mock sources into a single tile (Tile A8) of the COSMOS-Web survey, spanning 20.2 by 26.2 arcmin. The sources were inserted in two batches of 60,000 each, yielding a simulated surface density comparable to that observed in the real data. We chose tile A8 for its relative uniform depth in the ground-based data (particularly the UltraVISTA deep/ultra-deep tiers) and relative lack of masked area from bright stars.

We ran our detection algorithms on these simulated datasets using the same methodology applied to the COSMOS2025 catalog \citep{Shuntov2025}. This included source extraction, photometric measurements, and application of the color-selection criteria described in Section~\ref{sec:selection}. 

To ensure consistency and mitigate contamination from bright stars and imaging artifacts, we applied the same masking strategy used during the catalog extraction process. Regions affected by bright stars, including their diffraction spikes and extended halos, were masked using star masks based on the segmentation map. Similarly, areas impacted by imaging artifacts, such as residual hot pixels or scattered light, were excluded from the analysis. These masks were carefully constructed to balance the removal of contaminated regions while maximizing the effective survey area for high-redshift galaxy selection. By applying these masks consistently across all analyses, we minimized systematic biases introduced by observational effects and ensured the reliability of our measurements.

To ensure that the completeness estimates were not biased by contamination from real sources, we cross-matched the detected sources with the previously constructed catalog of real sources. Any detected source within 0.2" of a source in the COSMOS-Web catalog \citep{Shuntov2025} was excluded from the analysis. This step ensured that the completeness calculation accurately reflected the detectability of injected sources alone, free from blending or confusion with real objects. These simulations allow us to quantify completeness of our galaxy samples. In Fig.~\ref{fig:complteness_F277W}, we present the completeness as a function of magnitude, and redshift.

\subsection{Purity Simulations}\label{sec::purity}

Similarly, we utilize our simulated datasets to quantify the contamination from low-redshift interlopers within our high-redshift samples. Photometric scatter and degeneracies between spectral energy distributions (SEDs) can mimic the Lyman-break feature, potentially causing redshift misidentifications. In Fig.\ref{Fig::comparison_simulations}, we show the comparison between the injected (true) redshifts and those recovered by \lephare{} for all sources brighter than $m_{\rm F277W}<27.5$ AB. For this comparison, we employ exactly the same detection techniques and photometric extraction procedures as described in \citet{Shuntov2025}. Overall, we find good agreement between the input and recovered redshifts. 

Importantly, our analysis demonstrates that our high-redshift sample selection remains robust, as it is generally more common for genuinely high-redshift sources to be erroneously assigned a low photometric redshift than the converse. In particular, galaxies injected at $z>10$ are occasionally misclassified as sources at $z\sim5$ by \lephare. This misclassification arises from templates featuring very strong emission lines superimposed on a dusty continuum, which can mimic a flat or blue broadband SED with an apparent sharp spectral break resembling the Lyman discontinuity. The fraction of catastrophic outliers identified in this manner is consistent with the photometric redshift probability distributions (PDF$(z)$) of our sources, and this uncertainty is accounted for when deriving the UV luminosity function.

We further benchmark our photometric redshifts against spectroscopic measurements from high-$z$ surveys (COSMOS-3D, CAPERS and transient programs), for $6.5 < z < 10$, as detailed in \citet{Shuntov2025}. That study finds a normalized median absolute deviation of $\sigma_{\rm MAD}=0.038$ and an outlier fraction $\eta=7.9\%$, indicating overall good agreement. However, a median bias of $b=-0.3$ reveals a systematic tendency for \lephare\ to overestimate redshifts. 

\subsection{Construction of the UVLF}\label{sec:construction_UVLF}
To determine the contribution of our high-redshift galaxy sample to the ultraviolet luminosity function, we calculated the co-moving volume probed by the survey, combining its observed area and the redshift distribution of the sources. The UVLF was constructed using the $V_{\rm max}$ method \citep{Schmidt1968}, which estimates the number density of galaxies in a given magnitude range by accounting for the maximum volume within which each galaxy could have been detected given the survey’s sensitivity.

The co-moving number density of galaxies per magnitude bin, $\Phi(M_{\mathrm{UV}})$, was calculated as:
\begin{equation}
\Phi(M_{\mathrm{UV}}) \Delta M_{\mathrm{UV}} = \sum_{i=1}^N \frac{1}{V_{\mathrm{max},i}},
\end{equation}
where $N$ is the number of galaxies in the UV magnitude bin $\Delta M_{\mathrm{UV}}$, and $V_{\mathrm{max},i}$ is the maximum comoving volume over which galaxy $i$ could have been observed. The volume is computed as:
\begin{equation}
V_{\mathrm{max},i} = \int_{\Omega} \int_{z_{\mathrm{min},i}}^{z_{\mathrm{max},i}} \frac{\mathrm{d}V}{\mathrm{d}\Omega \mathrm{d}z} \, \mathrm{d}\Omega \, \mathrm{d}z,
\end{equation}
where $\Omega$ is the area of the survey, and $z_{\mathrm{min},i}$ and $z_{\mathrm{max},i}$ correspond to the 95\% confidence interval of the photometric redshift for galaxy $i$. 

The completeness correction is applied to each galaxy. Uncertainties on the UVLF were estimated using Poissonian statistics, taking into account the uncertainties on $V_{\mathrm{max}}$ for each source:
\begin{equation}
\sigma_\Phi(M_{\mathrm{UV}}) \Delta M_{\mathrm{UV}} = \sqrt{\frac{1}{N} \sum_{i=1}^N \frac{1}{V_{\mathrm{max},i}^2}}.
\end{equation}

%\begin{figure*}
%\centering
%\begin{minipage}[t]{1.\textwidth}
%\resizebox{\hsize}{!} { 
%\includegraphics[width=0.33\linewidth]{img/UVLF_F814W_with_lit_mcmc_v12.pdf} 
%\includegraphics[width=0.33\linewidth]{img/UVLF_F115W_with_lit_v11_mcmc.pdf} 
%\includegraphics[width=0.33\linewidth]{img/UVLF_F150W_with_lit_v11_mcmc.pdf} 
%}

\begin{figure*}
%\centering
\includegraphics[width=1\linewidth]{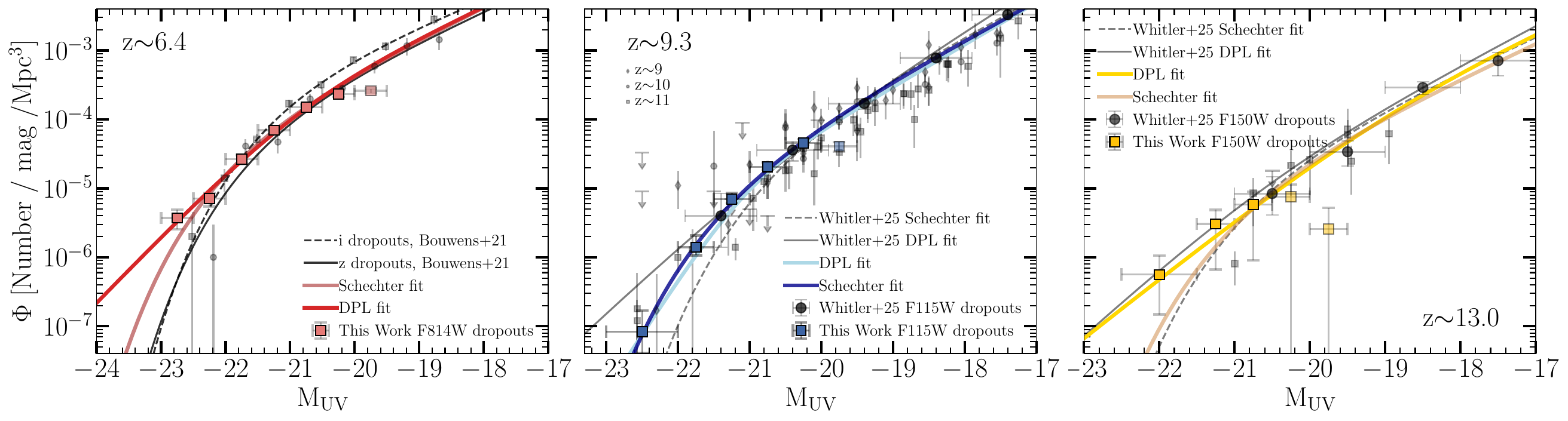} 
\caption{Left: UV luminosity function for the F814W dropout sample. Overplotted are the Schechter fits from \citet{Bouwens2021} for an $i$–dropout sample at $\langle z\rangle=5.9$ and a $z$–dropout sample at $\langle z\rangle=6.8$, derived from multiple Hubble deep fields covering a total of 1136\,arcmin$^2$ around the median redshift of our sample ($z=6.4$). Except for the brightest magnitude bin, our measurements lie between those two Schechter fits. Center: UV luminosity function for the F115W dropout sample. We included previous UVLF estimates in this redshift range from \cite{Leung2023, Perez-Gonzalez2023, Donnan2023, Finkelstein2024, Adams2024, Castellano2023, McLeod2024, Donnan2024, Casey2024}. The sources from the literature at $z\sim9$, $z\sim10$, and $z\sim11$ are represented by diamonds, circles, and squares respectively. In addition, we overplot measurements from JADES \citep{Eisenstein2023}, which probe fainter magnitudes but sample a much smaller comoving volume using a dropout selection  approach closely matching our own \citep{Whitler2025}. A continuity can be seen between the JADES faint-end points and our bright-end measurements.  Right:  UV luminosity function for the F150W-dropout sample. We include existing UVLF estimates at $z\approx12.5$–14 from \citet{Finkelstein2024,McLeod2024,Adams2024, Casey2024}, and overplot the \citet{Whitler2025} fits and data for direct comparison.  For each dropout sample, we mark the median redshift of the selected galaxies and overplot the best‐fit Schechter and double power‐law parameterizations.  In the F115W and F150W samples, we perform a joint fit by combining our measurements with those of \citet{Whitler2025} to more robustly constrain the model parameters. Similarly, for the F814W sample, we jointly fitted our data with the \citep{Bouwens2021} galaxy samples at $z\sim6$ (squares) and $z\sim7$ (circles), considering only galaxies with $M_{\rm UV} > -21.5$.}
    \label{fig:UVLF_tot}
\end{figure*}

To account for the contribution of cosmic variance to the uncertainties in the UVLF, we estimated the fractional cosmic variance ($\sigma_{\rm cv}$) as a function of redshift and UV luminosity.  We have relied on the predictions of analytical model of high-z galaxies from \citet{Trapp2020}. These results are consistent with those obtained from the \textsc{BLUETIDES} cosmological hydrodynamic simulation \citep{Bhowmick2020} or with those from the \textsc{ASTRAEUS} simulation \citep{Ucci2021}. Specifically, we used the cosmic variance values provided for a volume with $\Delta z = 1$ and rescaled them to match the effective comoving volumes probed by our different dropout samples. This rescaling was performed following the methodology outlined in \citet{Bhowmick2020}, which recommends adjusting the variance amplitude according to the survey volume:
\begin{equation}
\sigma_g(\Delta z) / \sigma_g\left(\Delta z_{\text {ref }}\right)=\left(\Delta z / \Delta z_{\text {ref }}\right)^{-0.32}
\end{equation}
\noindent where $\Delta z_{\text {ref }}$ = 1.

As discussed in \citet{Finkelstein2023, Finkelstein2024}, these cosmic variance estimates should be regarded as upper limits. This is because simulations and models done before the observations of the \textit{JWST} tend to underpredict the number densities of galaxies during the epoch of reionization, particularly at the bright end of the UV luminosity function. Consequently, the clustering strength, and hence the cosmic variance, may be overestimated in these models relative to the real Universe. The resulting cosmic variance contribution was then  propagated into the total UVLF uncertainties via:

\begin{equation}
    \sigma_{\rm tot}^2 = \sigma_{\Phi}^2 + \sigma_{\rm cv}^2,
\end{equation}

where $\Phi$ is the galaxy number density in a given magnitude bin. This approach captures both the shot noise and the field-to-field variance due to large-scale structure fluctuations. For the brightest bins ($M_{\rm UV} < -21.5$), we find that cosmic variance reaches typical values of $\sigma_{\rm cv} \sim 10-15\%$. These corrections are essential for robustly interpreting the shape and evolution of the bright end of the UVLF, especially when comparing to theoretical models and simulations. The resulting binned UVLF values and associated uncertainties—including Poisson errors, cosmic variance, and completeness corrections are listed in Table~\ref{tab:UVLF} and plotted in Figure~\ref{fig:UVLF_tot}.

 \begin{deluxetable}{ccc}[h!]
 \tabletypesize{\scriptsize}
 \tablecaption{COSMOS-Web UVLF\label{tab:UVLF}}
  \tablehead{
  \colhead{Dropout filter} & \colhead{M$_\textrm{UV}$ [mag]} & \colhead{$\Phi$ [10$^{-6}$mag$^{-1}$ Mpc$^{3}$]}}
\startdata
\multirow{7}{*}{F814W}  & -22.75 $\pm$ 0.25        &    3.72$_{- 1.16}^{+ 1.16}$\\ 
                        & -22.25 $\pm$ 0.25        &    7.10$_{- 1.61}^{+ 1.55}$\\
                        & -21.75 $\pm$ 0.25        &   26.58$_{- 3.98}^{+ 3.91}$\\
                        & -21.25 $\pm$ 0.25        &   69.76$_{- 7.81}^{+ 7.72}$\\
                        & -20.75 $\pm$ 0.25        &  151.07$_{-13.73}^{+13.25}$\\
                        & -20.25 $\pm$ 0.25        &  232.66$_{-18.53}^{+17.84}$\\
                        & -19.75 $\pm$ 0.25$^\ast$ &  260.12$_{-19.35}^{+18.83}$\\
\hline
\multirow{6}{*}{F115W} & -22.50 $\pm$ 0.50         &   0.08$_{- 0.08}^{+ 0.08}$\\
                        &-21.75 $\pm$ 0.25         &   1.40$_{- 0.66}^{+ 0.66}$\\
                        &-21.25 $\pm$ 0.25         &   6.94$_{- 1.71}^{+ 1.59}$\\
                        &-20.75 $\pm$ 0.25         &  20.62$_{- 3.91}^{+ 3.24}$\\
                        &-20.25 $\pm$ 0.25         &  45.59$_{- 6.56}^{+ 6.13}$\\
                        &-19.75 $\pm$ 0.25$^\ast$  &  40.75$_{- 6.54}^{+ 5.90}$\\
\hline 
\multirow{5}{*}{F150W}  &-22.00 $\pm$ 0.50         &   0.56$_{- 0.41}^{+ 0.50}$\\ 
                        &-21.25 $\pm$ 0.25         &   3.05$_{- 2.38}^{+ 1.83}$\\
                        &-20.75 $\pm$ 0.25         &   5.86$_{- 4.96}^{+ 3.06}$\\
                        &-20.25 $\pm$ 0.25$^\ast$  &   7.55$_{- 8.50}^{+ 3.91}$\\
                        &-19.75 $\pm$ 0.25$^\ast$  &   2.59$_{- 2.61}^{+ 2.61}$\\
\enddata
\tablecomments{Measurements of the UV luminosity function for our three dropout samples.  For each absolute UV magnitude bin, we give the comoving number density, together with uncertainties. Bins with completeness below 20\% are indicated with an asterisk and were excluded from the Schechter and double power–law fits presented in Section~\ref{sec:UVLF}.}
\end{deluxetable}

\begin{deluxetable*}{ccccccc}
\tabletypesize{\scriptsize}
\tablecaption{Parametrization of the UVLF\label{tab:parametrization}}
\tablehead{
  \colhead{Dropout filter} & & \colhead{M$_\textrm{UV}^*$ [mag]} & \colhead{$\Phi^*$ [10$^{-5}$ mag$^{-1}$ Mpc$^{-3}$]} & \colhead{$\alpha$} & \colhead{$\beta$} & \colhead{$\rho_{\rm UV}$ [10$^{25}$erg s$^{-1}$ Hz$^{-1}$ Mpc$^{-3}$}
}
\startdata
\multirow{2}{*}{F814W} & Schechter & -21.66$_{-0.16}^{+0.16}$ &  9.50$_{- 2.36}^{+ 3.09}$   & -2.16$_{-0.05}^{+0.06}$  & -                         & 10.35$_{- 3.37}^{+ 4.30}$ \\ 
                       & DPL       & -20.88$_{-0.64}^{+0.64}$ & 23.39$_{-15.26}^{+37.67}$  & -2.10$_{-0.11}^{+0.11}$  & -3.42$_{-0.55}^{+0.24}$   & 10.53$_{- 10.31}^{+20.97}$ \\
\hline 
\multirow{2}{*}{F115W} & Schechter & -20.97$_{-0.22}^{+0.26}$ & 2.65$_{-1.71}^{+2.28}$     & -2.40$_{-0.10}^{+0.10}$  & -                           &  2.65$_{- 1.72}^{+ 2.08}$ \\ 
                       & DPL       & -21.07$_{-0.20}^{+0.35}$ & 1.61$_{-0.53}^{+2.15}$     & -2.53$_{-0.08}^{+0.11}$  & -5.03$_{-0.97}^{+0.73}$     &  2.39$_{- 2.05}^{+ 2.47}$ \\ 
\hline 
\multirow{2}{*}{F150W} & Schechter & -20.56$_{-0.55}^{+0.40}$ & 2.30$_{-1.73}^{+2.79}$     & -2.25$_{-0.20}^{+0.24}$  & -                          &  0.65$_{- 0.65}^{+ 2.94}$ \\ 
                       & DPL       & -18.96$_{-0.73}^{+0.33}$ & 22.27$_{-15.82}^{+12.67}$  & -2.22$_{-0.54}^{+0.49}$  & -3.18$_{-0.78}^{+0.57}$    &  0.86$_{- 0.62}^{+ 1.22}$ \\ 
\hline \enddata
\tablecomments{Best-fit parameters for the UVLF derived from our three dropout samples. The UVLF is parameterized using both a Schechter function and a Double Power Law (DPL).}
\end{deluxetable*}

\subsection{Parametrization of the UVLF}\label{sec:parametrization}
We construct the UVLF following the methodology described in Section~\ref{sec:construction_UVLF}.  For each dropout sample, galaxies are binned in absolute UV magnitude.  In the F115W and F150W samples, the brightest magnitude bin is chosen to be $\Delta M_{\rm UV}=0.5$ to account for the lower source counts at the extreme bright end, whereas all other bins adopt $\Delta M_{\rm UV}=0.25$.  At the faintest magnitudes, completeness corrections become unreliable beyond $M_{\rm UV}>-20.0$ for the F814W and F115W samples, and $M_{\rm UV}>-20.5$ for F150W; these bins are shown as shaded points in Figure~\ref{fig:UVLF_tot}.   We model the UVLF in each dropout sample using both a Schechter function \citep{Schechter1976}:

\begin{equation}
\begin{aligned}
\Phi\left(M_{\mathrm{UV}}\right)= & \frac{\ln 10}{2.5} \phi^* 10^{-0.4\left(M_{\mathrm{UV}}-M_{\mathrm{UV}}^*\right)(\alpha+1)} \\
& \times \exp \left(-10^{-0.4\left(M_{\mathrm{UV}}-M_{\mathrm{UV}}^*\right)}\right),
\end{aligned}
\end{equation}
\noindent where: $\Phi^*$ is the normalization, $M^*$ is the characteristic magnitude where the function transitions from a power-law to an exponential decline,
$\alpha$ is the faint-end slope, as well as a double power-law (DPL) function:

\begin{equation}
\Phi\left(M_{\mathrm{UV}}\right)= \frac{\phi^*}{10^{0.4(\alpha+1)\left(M_{\mathrm{UV}}-M_{\mathrm{UV}}^*\right)}+10^{0.4(\beta+1)\left(M_{\mathrm{UV}}-M_{\mathrm{UV}}^*\right)}},
\end{equation}

\noindent where $\alpha$ describes the faint-end slope and
$\beta$ defines the bright-end slope. We used the Python package \texttt{emcee} \citep{Foreman2013} to perform Markov Chain Monte Carlo (MCMC) fits of both the Schechter function and the DPL model to our binned UVLF data.

For the F115W and F150W samples, we leverage the \textit{JWST} Advanced Deep Extragalactic Survey (JADES; \citealt{Eisenstein2023}) results of \citet{Whitler2025}, which employ a dropout selection relatively similar to our own.  As shown in Fig.~\ref{fig:UVLF_tot}, the JADES measurements extend the UVLF to fainter magnitudes, while our COSMOS-Web data anchor the bright end; the two datasets join seamlessly to yield a continuous luminosity function over $\sim$5 mag in $M_{\rm UV}$. 
By performing a joint fit to both datasets, we exploit this enlarged dynamic range to place tight constraints on the Schechter and double power–law parameters ($\phi^*$, $M^*$, $\alpha$, and $\beta$) at $z>8$. The parametrization of these fits are given in Table~\ref{tab:parametrization}. 

For the F814W–dropout sample (median $z\approx6.4$) we fit only our COSMOS-Web measurements.  The best–fit Schechter parameters (Table~\ref{tab:parametrization}) show a good agreement with the pre-\textit{JWST} determinations for a sample of over 24,000 galaxies from 1136\,arcmin$^2$ of \textit{HST} legacy fields \citet{Bouwens2021}, their $i$– and $z$–dropout samples with a mean redshift of 5.9 and 6.8 respectively corresponding to the redshift of our sample. Except in the brightest bins ($M_{\rm UV}\lesssim -22$), where our data lie above the \textit{HST}, we are in good agreement with the results from \citet{Bouwens2021}. However, we can note that while this dataset robustly constrains the UVLF, its lack of deep near‐infrared coverage could limit the sensitivity to the highest‐redshift  galaxies in the sample.

For the F115W dropout sample, our bright‐end measurements lie between the double power-law and Schechter extrapolations from \citet{Whitler2025}, whereas for the F150W dropout sample, our results are in excellent agreement with their double power-law extrapolation.  This persistent bright–end excess corroborates early \textit{JWST} findings of an overabundance of luminous galaxies at $z>8$ \citep{Finkelstein2023, Donnan2023, Castellano2022, Harikane2023, Whitler2025}.  In the following sections, we will quantify the significance of this excess, map the redshift evolution of the UVLF parameters, and investigate physical origins—such as enhanced star formation efficiency, reduced dust attenuation, or bursty star‐formation histories, that may drive the unexpectedly large population of UV‐bright galaxies during reionization.

\begin{figure*}
%\centering
\includegraphics[width=1\linewidth]{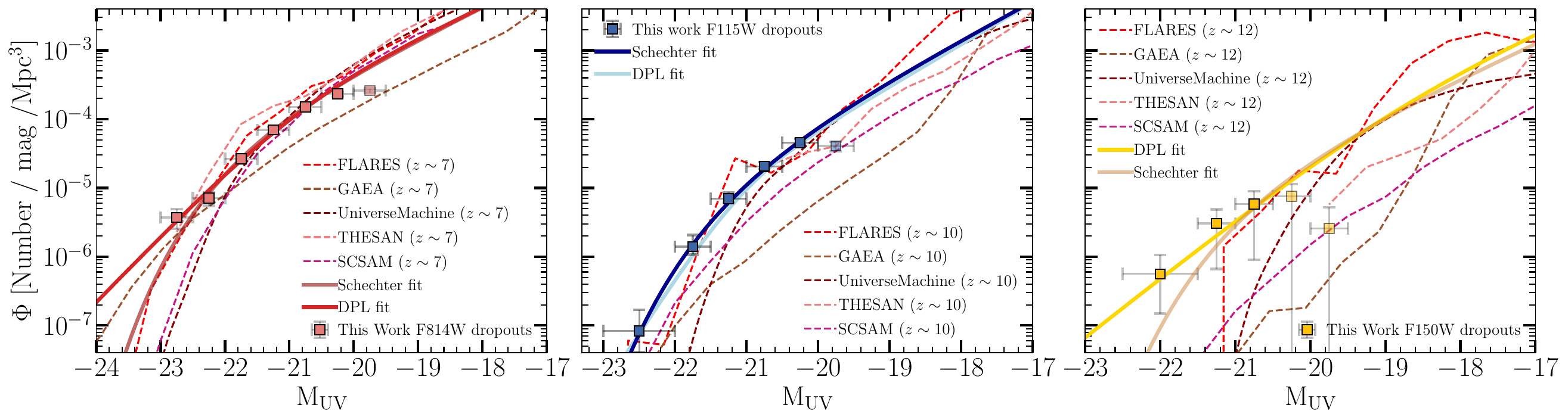} 
    \caption{Comparison of our results with predictions from the hydrodynamic simulations of galaxy formation and evolution \flares\ \citep{Lovell2021, Vijayan2021}, the \textsc{UniverseMachine} \citep{Behroozi2019}, the fiducial GAEA semi-analytic model \citep{De_Lucia2024} and \textsc{THESAN} simulations (\citealt{kannan22}), and the Santa Cruz semi-analytical model (SCSAM) \citep[e.g.,][]{yung19}. For each of these panels, we overplot, the fits used in Fig.~\ref{fig:UVLF_tot} and described in Sect.~\ref{sec:parametrization}.}
    \label{fig:compare_simu}
\end{figure*}

\subsection{Comparison with simulations}

Figure~\ref{fig:compare_simu} compares our derived UV luminosity functions for the three dropout samples (F814W, F115W, and F150W) with theoretical predictions from several models and simulations, including the First Light And Reionisation Epoch Simulations (\flares) zoom–in simulations \citep{Lovell2021, Vijayan2021}  using the \textsc{EAGLE} model \citep{Schaye2015, Crain2009}, the \textsc{UniverseMachine} empirical model of galaxy formation in dark matter halos \citep{Behroozi2019}, the \textsc{THESAN} simulations \citep{kannan22}, a large volume radiation-magnetohydrodynamic simulation, the Santa Cruz semi-analytical model (SCSAM;  \citealt{yung19}) and the GAlaxy Evolution and Assembly (GAEA) semi-analytic model \citep{De_Lucia2024}. Additionally, we overlay the UVLF fits derived solely from our data for the F814W sample, while for the F115W and F150W samples we present fits derived by combining our measurements with those from the deeper JADES survey \citep{Whitler2025}, allowing for stronger constraints across a wider luminosity range.

For the F814W dropout sample ($5.5 < z < 8.5$), our measured number densities are broadly consistent with model predictions at intermediate magnitudes ($-22<M_{\rm UV}<-20$). However, we observe a modest but notable deviation at the brightest magnitudes ($M_{\rm UV}\lesssim-22$), where our observations exceed model predictions. This discrepancy marks the onset of what has been recently described as the “bright-end excess” phenomenon, widely reported in the literature based on pre-\textit{JWST} observations \citep[e.g.,][]{Bowler2020,finkelstein2022b} and early \textit{JWST} observations \citep[e.g.,][]{Finkelstein2023, Donnan2023, Castellano2022, Harikane2023, Whitler2025}. 

At higher redshifts ($z\gtrsim9$) corresponding to our F115W and F150W dropout samples, discrepancies between observations and theoretical predictions become even more pronounced. Model-to-model variations at these redshifts increase significantly, with differences in predicted number densities could exceed an order of magnitude at $z>10$. Despite these larger theoretical uncertainties, our measurements consistently lie above most of the theoretical predictions at the bright end ($M_{\rm UV}\lesssim-21.5$). Among the considered models, the \flares\ simulations generally provide the closest match to our observational data at these highest redshifts. The observed bright-end excess at $z\gtrsim10$ reinforces earlier \textit{JWST} results, highlighting a persistent tension between observations and theoretical frameworks calibrated primarily at lower redshifts. Indeed, this excess in the number of UV-luminous galaxies challenges standard galaxy formation scenarios, pointing towards a need to revise current models of star formation efficiency, stellar feedback mechanisms, and potentially stellar population assumptions in the early Universe \citep{Harikane2023, Mason2023, Dekel2023, Ferrara2024, Somerville2025}.

Thus, our comparison illustrates clearly that while pre-launch galaxy formation models successfully reproduce galaxy abundances at intermediate UV luminosities and moderate redshifts ($z\sim7$), they struggle significantly at higher luminosities and especially at the highest redshifts probed ($z\sim10$–14).  We note that many of the simulation models that our observations are compared against are limited by relatively small volume at the highest redshifts. As a result, rare, bright UV‑luminous galaxies at the bright end of the UV luminosity function may be underrepresented in these simulations due to cosmic variance and limited survey volume. This discrepancy could also underscore our actual limitations in the theoretical understanding of early galaxy formation processes and highlights the necessity for refined modeling approaches. Potential avenues for resolving this tension may involve enhancing the star formation efficiency at early cosmic epochs, relaxing feedback prescriptions, introducing stochastic or bursty star-formation histories, and exploring non-standard stellar IMFs to increase the UV luminosity per unit stellar mass \citep{Trinca2024, Haslbauer2022, Dekel2023}. We will explore this in the following sections.

\subsection{Evolution of the UVLF}

\begin{figure*}
\centering
\setlength{\tabcolsep}{0pt}
\begin{tabular}{ccc}
    \includegraphics[width=0.31\textwidth]{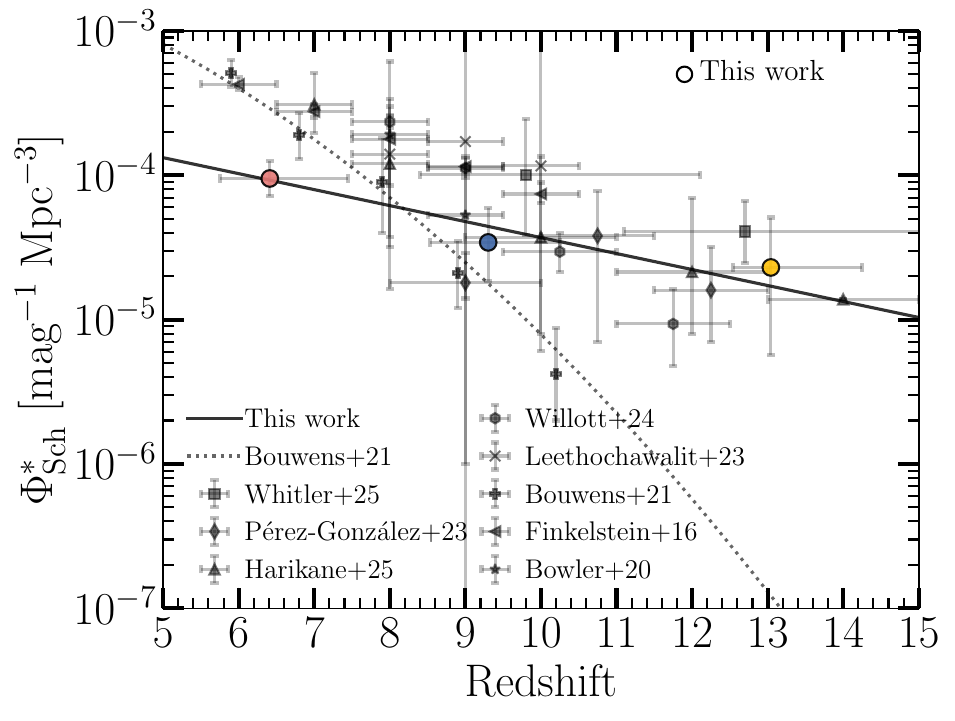} &
    \includegraphics[width=0.31\textwidth]{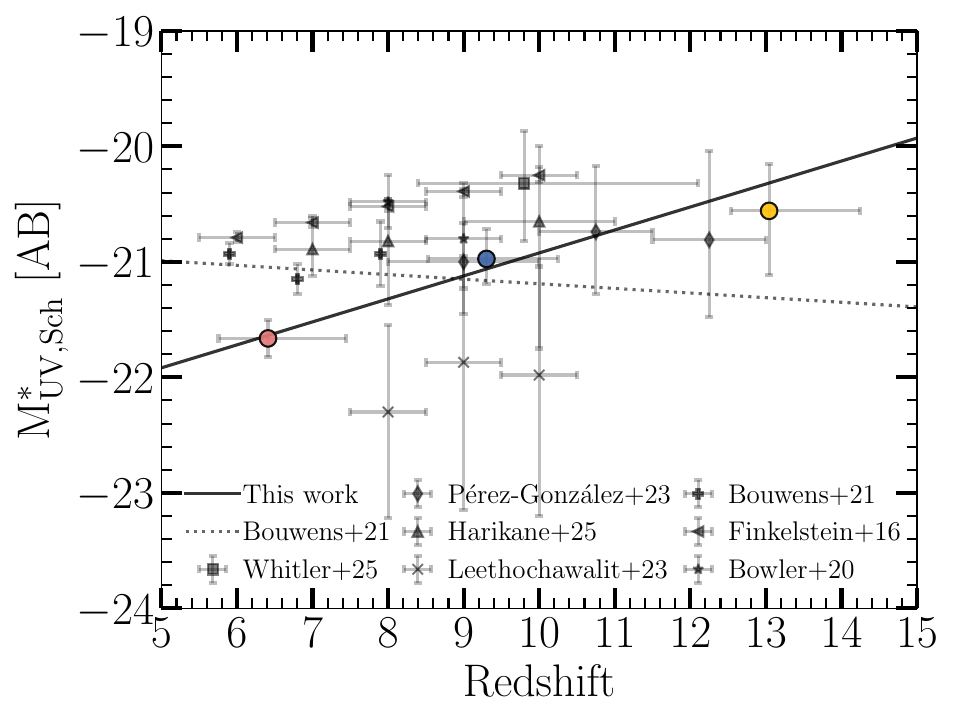} &
    \includegraphics[width=0.35\textwidth]{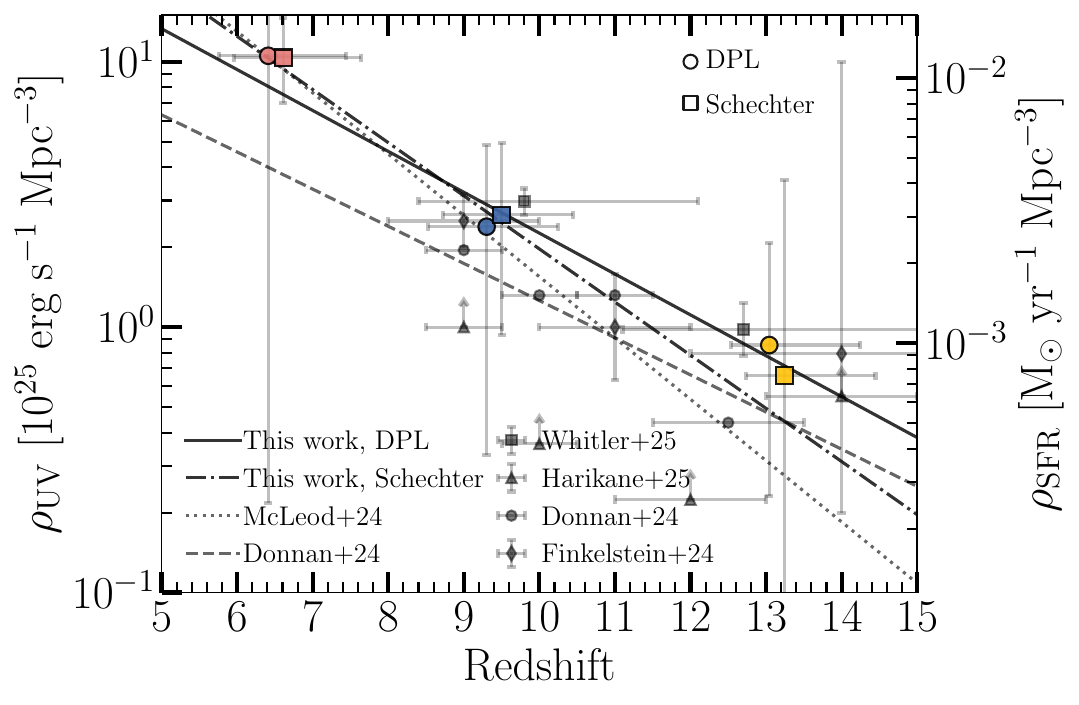} \\ 
    \includegraphics[width=0.31\textwidth]{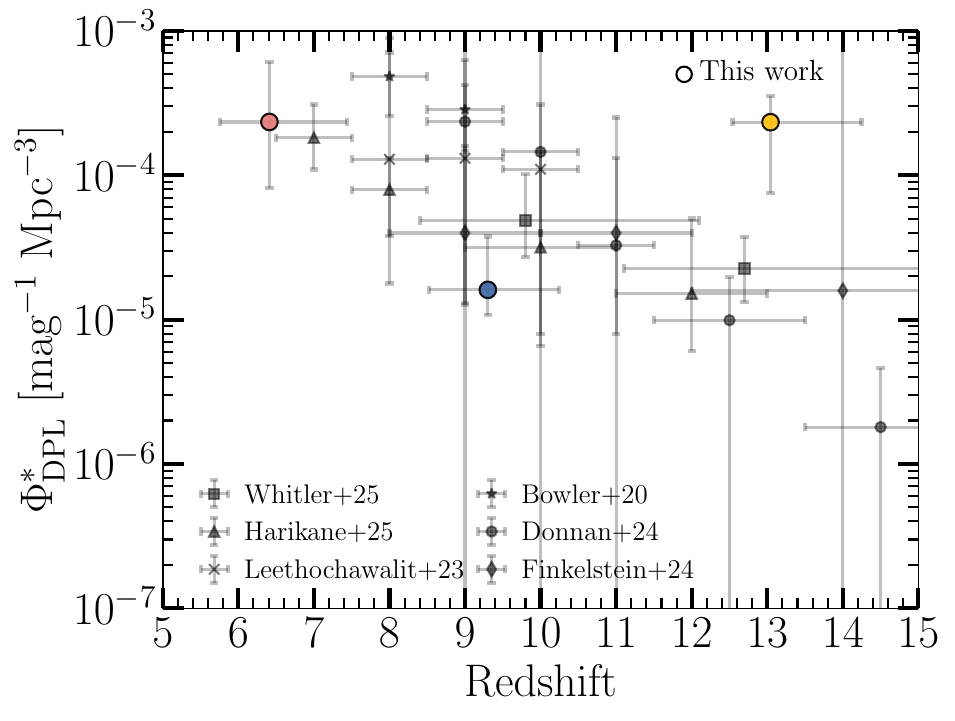} &
    \includegraphics[width=0.31\textwidth]{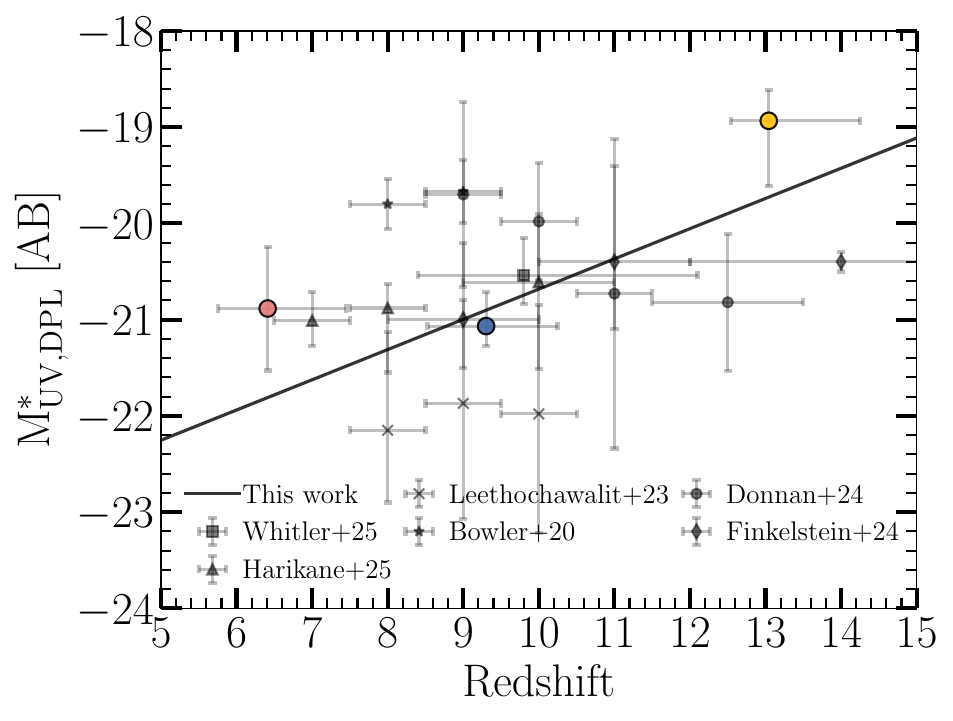} &
    \includegraphics[width=0.31\textwidth]{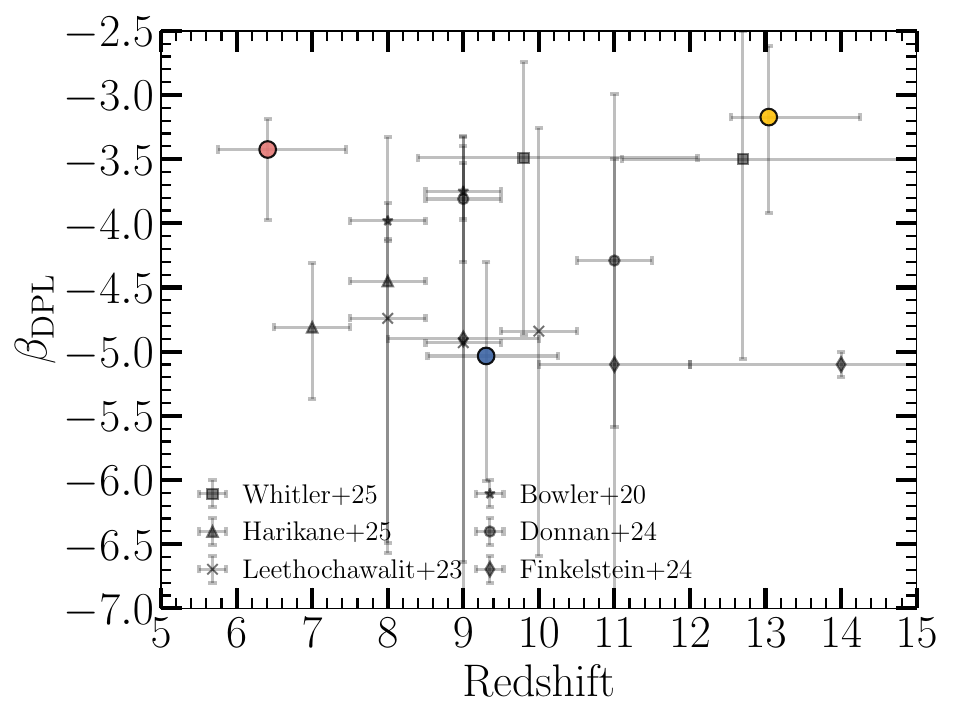} \\

\end{tabular}
      \caption{Evolution of the UVLF parametrization with redshift, including $\phi_{\rm Sch}^*$ (top-left), $\phi_{\rm DPL}^*$ (bottom-left), $M^*_{\rm UV, Sch}$ (top-middle), $M^*_{\rm UV, DPL}$ (bottom-middle), and $\beta_{\rm DPL}$ (bottom-right), as listed in Table~\ref{tab:parametrization}. The ultraviolet luminosity density is also shown (top-right) and compared with recent measurements from the literature \citep{Finkelstein2016, Bowler2020, Bouwens2021, Leethochawalit2023, Perez-Gonzalez2023, Willott2024, Donnan2024, McLeod2024, Finkelstein2024, Whitler2025}. }
         \label{Fig::evolution_UVLF_parameters}
\end{figure*}

In Fig.~\ref{Fig::evolution_UVLF_parameters}, we present the evolution of the Schechter and DPL function parameters ($\phi^*$, $M^*_{\rm UV}$, and $\beta$) as a function of redshift.  The evolution of the normalization parameter $\phi^*$ with redshift has been modeled by \citet{Bouwens2021} using a quadratic function, suggesting an accelerated decline in $\phi^*$ beyond $z \sim 8$. Given the large uncertainties associated with our highest-redshift bin, it is difficult to confirm or refute this accelerated trend. However, a comparison with recent studies \citep{Perez-Gonzalez2023, Harikane2025, Willott2024, Leethochawalit2023, Whitler2025} points to a more gradual, possibly linear evolution of $\phi^*$ with redshift, fitted (only with our data) by  $\log \phi_{\rm Sch}^* = (-0.111\pm0.03)z - (3.32\pm0.21)$, rather than a rapid downturn at the highest redshifts. We emphasize that we deliberately avoided fixing any parameters for the highest redshift bin in the DPL fit, in order to prevent introducing additional bias. As a consequence, the parameter $\phi_{\rm DPL, F150W}^*$ derived for that bin is slightly offset compared to the other redshift bins, and therefore should not be interpreted as reflecting a physical trend.

Regarding the characteristic luminosity $M_{\rm UV}^*$, we observe a mild increase with redshift across our dropout samples with a measured trend of $M_{\rm UV, Sch}^* = (0.20 \pm 0.04) z- (22.91\pm0.33)$. This allows for a slight dimming with increasing redshift. The literature presents a mixed picture on this matter: while \citet{Finkelstein2016} reported a brightening of $M_{\rm UV}^*$ with redshift, consistent with our best-fit trend, other studies such as \citet{Bouwens2021} suggest either a mild dimming, and recent analyses with \textit{JWST} data also remain inconclusive \citep[e.g.,][]{Harikane2025, Leethochawalit2023, Perez-Gonzalez2023} with large uncertainties for the highest redshift bins. When using the DPL parameterization, the result becomes more uncertain, with $dM_{\rm UV, DPL}^*/dz = 0.31 \pm 0.28$.

The faint-end slope $\alpha$ appears to decrease approximately linearly up to $z \sim 9$, consistent with the evolution reported by \citet{Bouwens2021} and \citet{Finkelstein2016}, who derived a relation of $d\alpha/dz = -0.11 \pm 0.01$. This trend mirrors the theoretical expectation from the evolution of the halo mass function, which predicts a similar slope of $d\alpha/dz = -0.12$ \citep{Bouwens2015}. At higher redshifts ($z > 9$), although the uncertainties increase, particularly for the F150W dropout sample, the evolution of $\alpha$ appears to flatten, reaching values of $\alpha \sim -2.25$ to $-2.50$. This apparent plateau is supported by recent analyses, including those of \citet{Perez-Gonzalez2023} and \citet{Whitler2025}, suggesting that the steepening of the faint-end slope may saturate in the earliest epochs of galaxy formation. This supports the idea of a potential transition in the galaxy population at these redshifts, accompanied by a slowdown in the evolution of the UV luminosity function at very high redshift.

\subsection{UV luminosity density}

We present in Fig.~\ref{Fig::evolution_UVLF_parameters}-right the evolution of the UV luminosity density, $\rho_{\mathrm{UV}}$, as a function of redshift, in comparison with recent measurements from the literature. We derive the rest-frame ultraviolet luminosity density ($\rho_{\mathrm{UV}}$) by integrating the best-fitting UVLF down to a limiting magnitude of $M_{\mathrm{UV}} = -17$. This threshold approximately corresponds to the completeness limit of deep \textit{JWST} surveys and allows for meaningful comparisons with previous work \citep[e.g.,][]{Finkelstein2024, Donnan2024, Whitler2025}.  The integration limits span from a bright-end cutoff at $M_{\mathrm{UV}} = -23$ down to a faint-end limit of $M_{\mathrm{UV}} = -17$.

For the F814W sample, we obtain a UV luminosity density of $\rho_{\mathrm{UV,F814W}} =  10.35_{- 3.37}^{+ 4.30}   \times10^{25} (10.53_{- 10.31}^{+20.97} \times10^{25})\ \mathrm{erg\, s}^{-1}\mathrm{Hz}^{-1}\mathrm{Mpc}^{-3}$ using the Schechter (DPL) parametrization.  This value decreases by a factor of four to $\rho_{\mathrm{UV,F115W}} = 2.65_{- 1.72}^{+ 2.08}  \times10^{25} (2.39_{- 2.05}^{+ 2.47} \times10^{25})\ \mathrm{erg\, s}^{-1}\mathrm{Hz}^{-1}\mathrm{Mpc}^{-3}$ for the F115W sample and is again divided by a factor of four to $\rho_{\mathrm{UV,F150W}} = 0.65_{- 0.65}^{+ 2.94} \times10^{25} (0.86_{- 0.62}^{+ 1.22} \times10^{25} )\ \mathrm{erg\, s}^{-1}\mathrm{Hz}^{-1}\mathrm{Mpc}^{-3}$ for the F150W sample. The values obtained using the two parameterizations (Schechter and DPL) are very close and mutually consistent. All results are listed in Table~\ref{tab:parametrization}. We note that our derived values are nonetheless dominated by the faint-end slope. Consequently, the constraints for the F115W and F150W dropout samples are driven by the results of \citet{Whitler2025}, while for the F814W sample by the results of \citet{Bouwens2021}. While our result for the F814W dropout sample remains consistent with pre-\textit{JWST} determinations \citep[e.g.,][]{Mason2015,yung19}, our measurements increasingly diverge from these predictions at higher redshifts ($z>8$), with a discrepancy that grows systematically with redshift. In Fig.~\ref{Fig::evolution_UVLF_parameters}, we show the evolution of the UV luminosity density, $\rho_{\rm UV}$, as a function of redshift, compared to previous literature results. We define the evolution of $\rho_{\mathrm{UV}}$ with redshift as:
\begin{align}
    \log(\rho_{\mathrm{UV, DPL}}) &= (-0.154\pm0.115)z + (26.90 \pm 1.20) \\
    \log(\rho_{\mathrm{UV, Sch}}) &= (-0.200\pm0.115)z + (27.30 \pm 0.82)
\end{align}

To convert $\rho_{\mathrm{UV}}$ into an estimate of the star formation rate density ($\rho_{\mathrm{SFR}}$), we adopt the calibration of \citet{Madau2014}, assuming a constant conversion factor of:
\begin{equation} \kappa_{\mathrm{UV}} = 1.15 \times 10^{-28}\mathrm{M_\odot}\mathrm{yr}^{-1}\mathrm{erg}^{-1}\mathrm{s}^{-1}\mathrm{Hz}^{-1}.
\end{equation}
The resulting $\rho_{\mathrm{SFR}}$ values are presented in Fig.~\ref{Fig::evolution_UVLF_parameters}.

\subsection{Excess of bright galaxies at $z>6$ compared to pre-\textit{JWST} studies}

While the overall goodness-of-fit does not show a strong preference between the Schechter and DPL parameterizations in our data alone, a closer comparison with deeper \textit{JWST} surveys reveals interesting trends. Specifically, in Fig.~\ref{fig:UVLF_tot} (center and right panels), we observe that our measured number densities align more closely with the DPL fits presented by \citet{Whitler2025} than with their corresponding Schechter function fits. By construction, this discrepancy is stronger at the bright end, where the exponential decline imposed by the Schechter form underpredicts the number of luminous galaxies. This result is consistent with previous studies at $z > 6$ \citep{Bowler2020, Harikane2023, Donnan2023}, suggesting that the formation of the most UV-bright galaxies is not as strongly suppressed as the Schechter function implies.  However, the number of galaxies with very high UV luminosities remains intrinsically low—with roughly an order of magnitude fewer galaxies at $M_{\rm UV} = -22$ compared to $M_{\rm UV} = -21$. As a result, the impact of the bright-end excess on the total UV luminosity density ($\rho_{\rm UV}$), and consequently on the ionizing photon budget, is modest (see Sect.~\ref{sec:reionization}).

\section{Discussion}\label{sec:discussion}

In this section, we discuss the broader implications of our findings, comparing them with results from other surveys and theoretical predictions.

\subsection{Star formation efficiency and stochasticity}

\begin{figure*}
\centering
\begin{minipage}[t]{1.\textwidth}
\resizebox{\hsize}{!} { 
\includegraphics[width=0.33\linewidth]{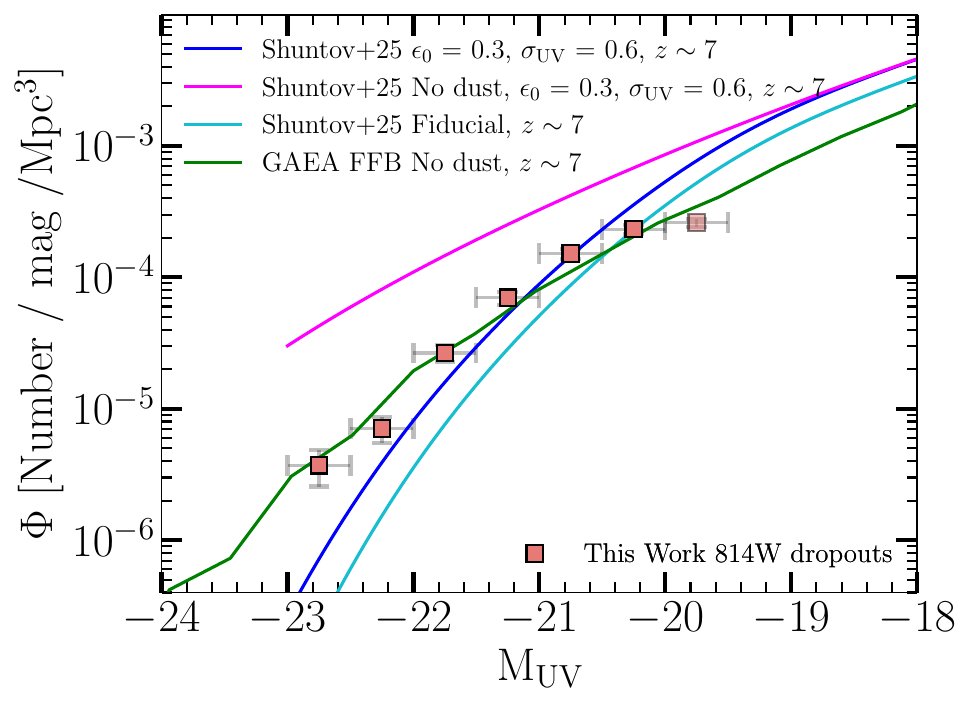}
\includegraphics[width=0.33\linewidth]{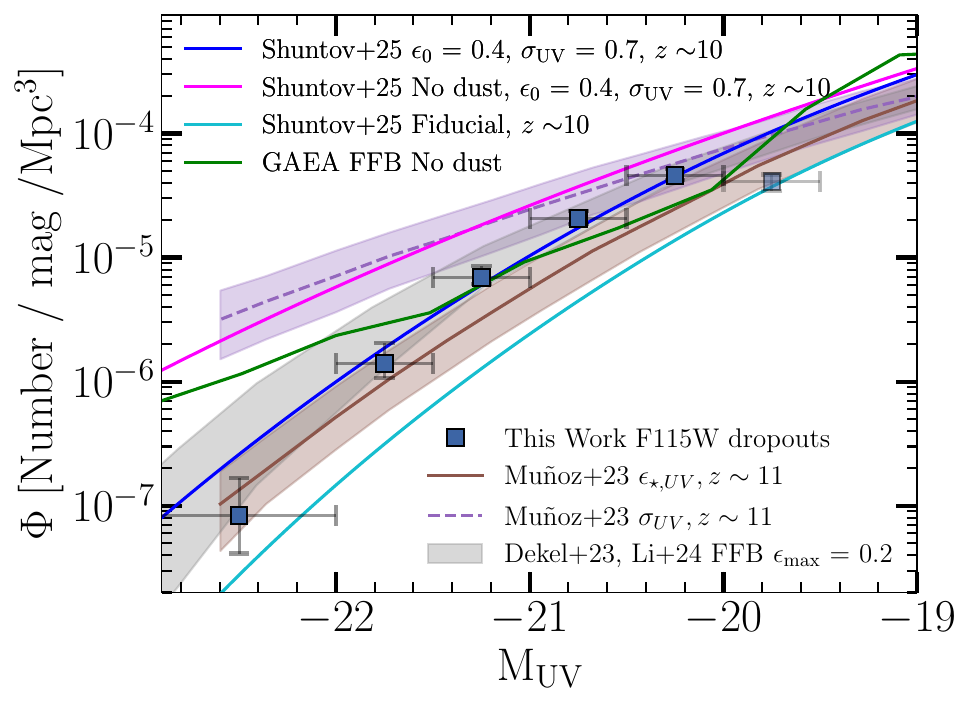}
\includegraphics[width=0.33\linewidth]{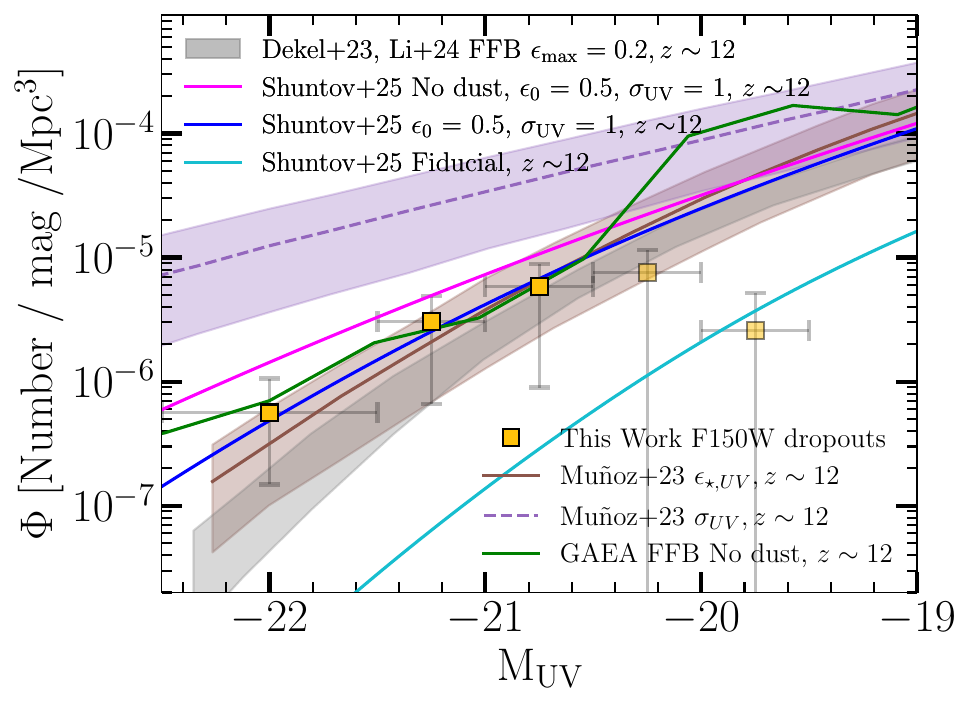}
}

\end{minipage}
      \caption{UV luminosity function measurements for the F814W dropout (left panel), F115W dropout (central panel) and F150W dropout (right panel) samples. In each panel, we compare our results to theoretical predictions exploring different star formation scenarios at the epoch of reionization. Specifically, we include predictions from \citet{Munoz2023} at redshifts closely matching our samples ($z=10.75$, and $z=12.25$). The purple shaded region illustrates the predicted 1$\sigma$ range for a scenario characterized by enhanced stochasticity in star formation ($\sigma_{\rm UV}(z)$), while the brown shaded region shows the expected UV luminosity function for a model with increased star formation efficiency ($\epsilon_{\rm UV}(z)$).
      Despite the slight offset in median redshift for the F115W dropout sample, our observations show a good agreement with models favoring enhanced star formation efficiency, suggesting that the bright galaxy population at these epochs is indeed boosted by highly efficient star formation processes. Additionally, we plot the predictions from the feedback-free starburst (FFB) models proposed by \citet{Dekel2023} and \citet{Li2024}.  In addition, we also add hybrid models from \citet{Shuntov2025b} showing different models of stochasticity (fiducial values with $\sigma_{\rm UV} = 0.53, \epsilon_0 = 0.22$ in cyan, fitted values in blue and models with no dust in magenta). We also display the latest version of the GAlaxy Evolution and Assembly (GAEA; \citealt{De_Lucia2024}) FFB without dust contribution.}
         \label{Fig::Munoz}
\end{figure*}

Several scenarios have been proposed to explain the observed excess of bright galaxies at high redshifts, for example elevated star formation efficiency (SFE) or stochastic star formation processes. \citet{Munoz2023} demonstrated that, at redshifts $z>10$ and within luminosity ranges typical of COSMOS-Web, it is possible to distinguish between an excess of galaxies resulting from high SFE and that caused by stochasticity. Our analysis of the UVLF aligns closely with the predictions of the high-SFE scenario presented by \citet{Munoz2023}, supporting the notion that rapid, efficient conversion of gas into stars likely dominates galaxy growth at these early epochs (see Fig.~\ref{Fig::Munoz}).

Alternative theoretical frameworks, such as the feedback-free starburst (FFB) models proposed by \citet{Dekel2023} and \citet{Li2024}, also successfully reproduce our observed bright-end measurements. These models suggest that at high redshifts, the reduced effectiveness of feedback processes, potentially due to lower metallicity or weaker stellar winds, allows star formation to proceed efficiently and continuously in massive halos, driving galaxies to brighter UV luminosities than previously expected \citep[e.g.,][]{Inayoshi2022c,Harikane2024,Ferrara2024}.

In addition to elevated SFE, stochastic star formation processes have been proposed as a contributing factor to the observed abundance of bright galaxies at high redshifts. \citet{Mason2023, Pallottini2023, Shen2023, Kravtsov2024, Gelli2024} investigated the impact of stochastic variability in SFR of high-redshift galaxies, finding that such variability can lead to significant luminosity fluctuations. Their study quantified the amplitude and timescales of this variability, identifying key physical processes driving these fluctuations. They concluded that while stochastic star formation can boost the luminosity of galaxies, it alone cannot account for the observed overabundance of bright galaxies at $z \gtrsim 10$.

To distinguish empirically between scenarios of enhanced SFE and stochastic star formation, \citet{Munoz2023} proposed that measuring galaxy clustering and associated halo biases is more efficient that comparing with the UVLF. Galaxies residing in more massive dark matter halos exhibit stronger spatial clustering, characterized by higher galaxy bias, compared to their lower-mass counterparts. Consequently, quantifying the clustering of high-redshift UV-bright galaxies provides a direct observational test to discern whether their rarity stems primarily from the scarcity of sufficiently massive halos or from episodic and inefficient star formation activity within more common, lower-mass halos.

\begin{figure*}
\centering
\begin{minipage}[t]{1.\textwidth}
\resizebox{\hsize}{!} { 
\includegraphics[width=0.33\linewidth]{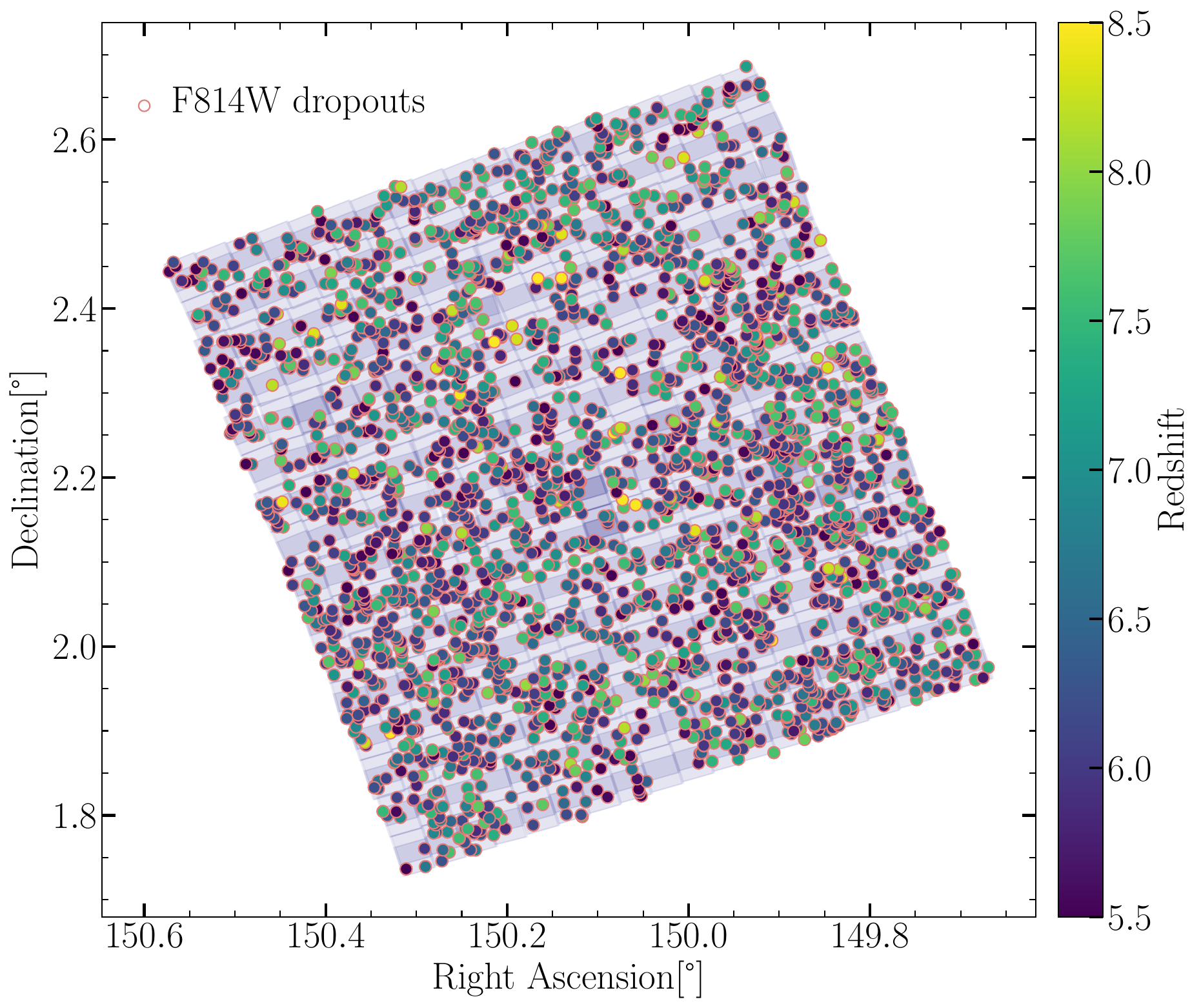} 
\includegraphics[width=0.33\linewidth]{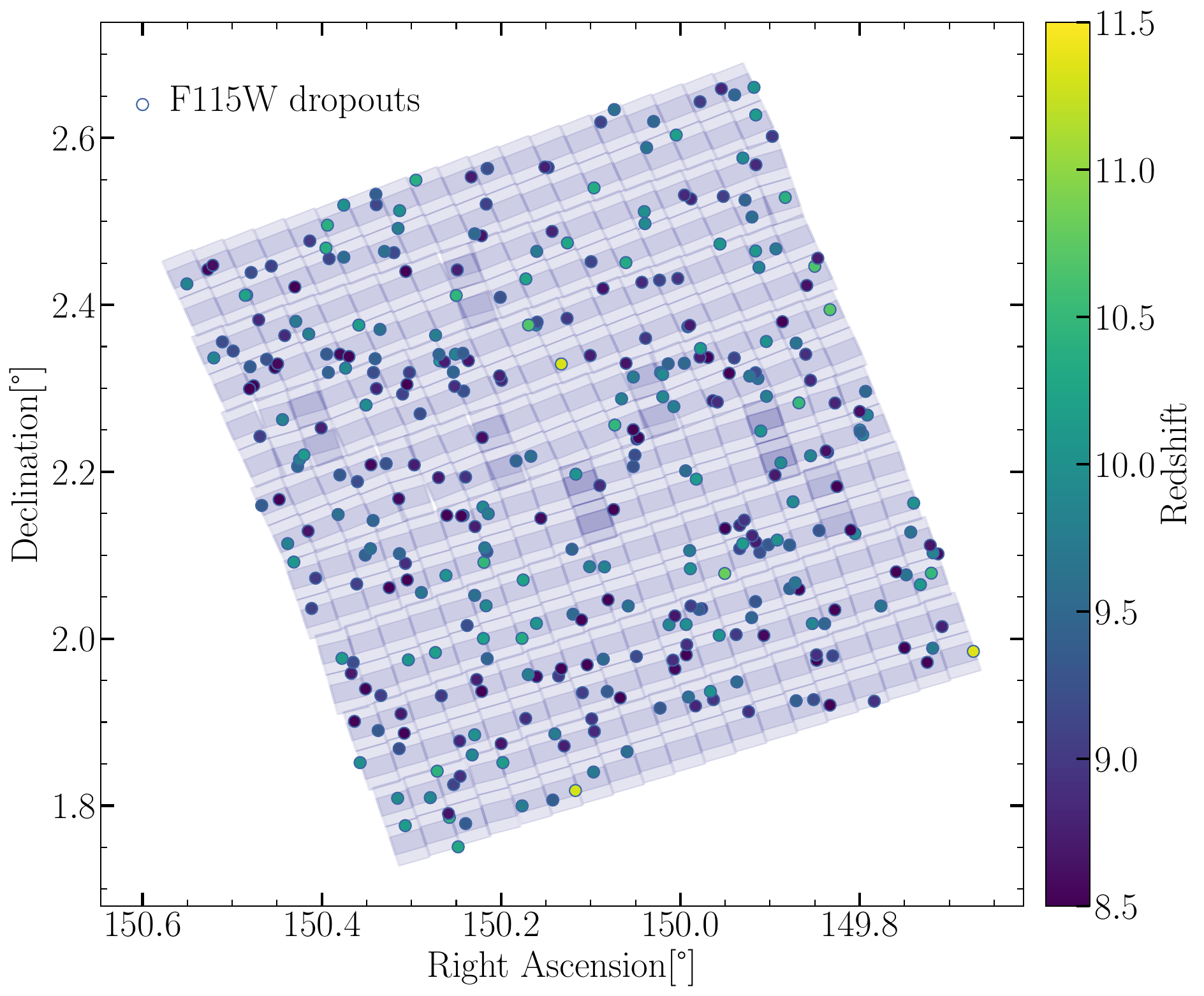} 
\includegraphics[width=0.33\linewidth]{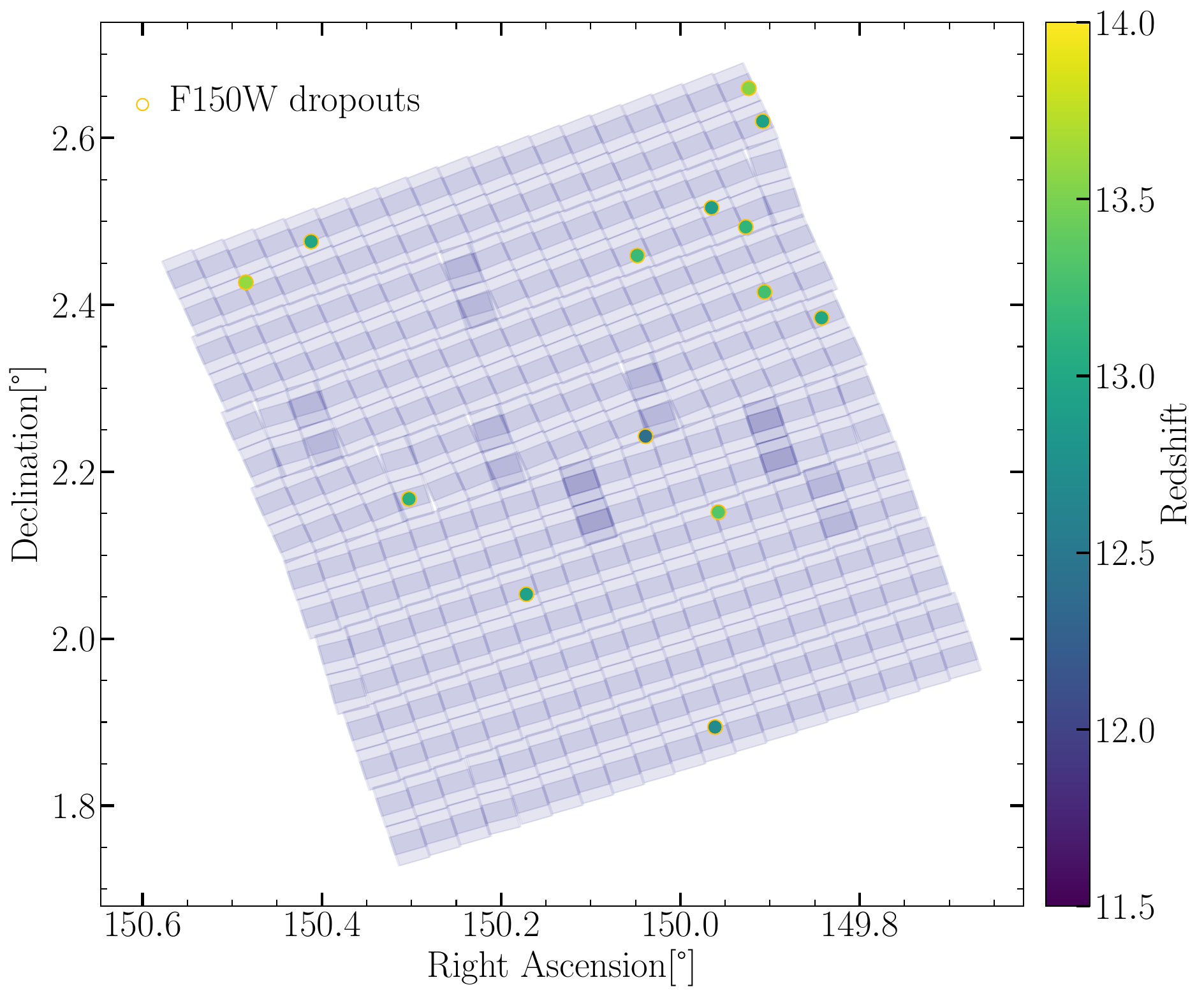} 
}
\end{minipage}
      \caption{Spatial distribution of our dropout-selected galaxy samples (F814W-left, F115W-middle, F150W-right) color-coded by redshift over the coverage map in the F277W \textit{JWST}/NIRCam filter of COSMOS-Web.
}
    \label{fig:coordinates_galaxies}
\end{figure*}

To perform this test with our sample, we computed the angular auto-correlation function using the spatial distribution of our galaxy candidates using the methodology developed by \citet{Paquereau2025}. Despite the more restrictive selection imposed by our color and photometric redshift criteria compared to the full COSMOS-Web galaxy sample, the resulting angular auto-correlation function remains consistent with the measurement presented in \citet{Paquereau2025}. A more detailed clustering analysis, incorporating data from COSMOS-3D \citep{Kakiichi2024}, will be presented in a forthcoming paper (Chase et al., in prep.). The spatial distributions of the galaxies in our F814W, F115W, and F150W dropout samples, used for these clustering measurements, are shown in Fig.~\ref{fig:coordinates_galaxies}. This further reinforces the interpretation that SFE is elevated in galaxies at these early epochs.

A higher-than-anticipated SFE in early galaxies suggests that the processes governing star formation in the early Universe may differ significantly from those in later epochs. This could indicate that early galaxies experienced rapid star formation episodes, leading to the swift build-up of stellar mass observed in the early Universe. Such a scenario aligns with the FFB model proposed by \citet{Dekel2023} and \citet{Li2024}, which posits that the high densities and low metallicities at cosmic dawn facilitated rapid star formation before feedback mechanisms could suppress it. However, it is important to note that recent studies, such as \citet{Donnan2025}, have found no evidence for increased SFE at early times, suggesting that further investigation is needed to fully understand the star formation processes in the early Universe.

To further distinguish between the excess of galaxies due to stochasticity and the increase in SFE, we compared our results in Fig.~\ref{Fig::Munoz} with the hybrid model presented in \citet{Shuntov2025b}. These models are constructed by embedding a physically motivated star–halo connection within a halo occupation distribution framework, explicitly calibrated using both the UVLF and galaxy clustering measurements from \textit{JWST} surveys. The instantaneous SFE is parameterized as a function of halo mass, along with a log-normal scatter in UV luminosity ($\sigma_{\rm UV}$). The fiducial model features an SFE peaking around $\epsilon_0 = 23\%$ at $M_{\rm h} \sim 3 \times 10^{11} M_\odot$ and a scatter $\sigma_{\rm UV} \approx 0.53$, which has been constrained using $z\lesssim8$ UVLF and clustering measurements. We used the fiducial model fixed at $z=8$ and tuned both $\epsilon_0$ and $\sigma_{\rm UV}$ manually for each of our samples. For the F814W-selected galaxies, adjusting the SFE normalization to $\epsilon_0$ = 0.30 and increasing the scatter to $\sigma_{\rm UV}$ = 0.50 brings the model into close agreement with the observed UVLF. Similarly, for F115W dropouts, adopting $\epsilon_0$ = 0.40 and $\sigma_{\rm UV}$ = 0.70 matches our data, and for F150W dropouts, $\epsilon_0$ = 0.50 with $\sigma_{\rm UV}$ = 1.0 yields excellent consistency. In all three cases, the unmodified fiducial models systematically underpredict galaxy abundances by more than an order of magnitude across the full luminosity range. 

We also compared our results with the UV LF predictions from the GAEA semi-analytic model \citep{De_Lucia2024}. The models are computed as follows (see also Cantarella et al. in prep. for more details): for each galaxy, the stellar UV luminosity is estimated using the synthetic stellar population model from \cite{Bruzual2003}. Dust attenuation follows the prescriptions included in \cite{De_Lucia2007}, which are based on the original model proposed by \cite{Charlot2000}. To add the AGN contribution to the UV luminosity,  we first compute the AGN bolometric luminosity from the gas accretion onto the supermassive BH, assuming a radiative efficiency of 15\%. To convert the bolometric AGN luminosity into the B-band luminosity, a bolometric correction as described in \cite{Marconi2004} has been applied. After that, to estimate the UV luminosity, we assume that the luminosity between the B- and UV bands scales as a power law: L$(\lambda) \propto \lambda^{\alpha}$ with \quad $\alpha = 0.44, \lambda_B = 4400\,\text{\AA} $ and $ \lambda_{\text{UV}} = 1600\,\text{\AA}$.
%L(λ)~λ^(α) with α=0.44, λ_B=4400 A and λ_UV = 1600 A. 
For each galaxy, we summed the UV luminosity coming from stars and the AGN in order to have the total UV luminosity per galaxy, then we computed the total UV LF, with and without considering dust attenuation. We did not apply any obscuration correction for the AGN luminosity.  In Fig.\ref{Fig::Munoz}, we compare our measurements to a variant of the model implementing the FFB prescription, introduced by \citet{Dekel2023}. The FFB scenario occurs in halos at $z > 10$ when the gas free-fall time becomes shorter than $\sim$1 Myr, corresponding to densities above a critical threshold of $\tilde{n} = 3\times10^3 \mathrm{cm}^{-3}$. The GAEA models incorporating the FFB scenario, particularly in the dust-free case, provide a better match to our observed UVLF than the fiducial simulation (shown in Fig.\ref{fig:compare_simu}; see Cantarella et al., in prep., for further details). These models (combined with \citealt{Munoz2023}) suggest that early galaxies are primarily forming stars with high efficiency and could also be experiencing burstier star formation than previously assumed. We discuss the consequences of dust presence on these parameters in Section~\ref{sec:dust}.

\subsection{Dust attenuation and its impact on the UV luminosity function during the EoR}\label{sec:dust}

The large abundance of UV-bright galaxies at $z>8$ discovered by \textit{JWST} has sparked intense interest in the role of dust attenuation during the EoR. Observationally, these early galaxies are characterized by very blue UV continuum slopes (often $\beta \lesssim -2.3$) and high luminosities ($M_{\rm UV}\lesssim -21$) despite their significant stellar masses, properties which suggest little-to-no dust obscuration for some of them \citep[e.g.,][]{Cullen2024, Cullen2025}. This stands in tension with theoretical expectations: rapid enrichment from massive stars implies that even at $z>10$, galaxies with $M_\ast \sim 10^8$--$10^9\,M_\odot$ should have produced appreciable dust from supernova ejecta \citep{Nozawa2007,Ferrara2022, Witstok2023}. In the standard picture, such dust would absorb UV starlight and suppress the observed luminosity of the most massive early galaxies, steepening the bright-end of the UVLF. Pre-\textit{JWST} models incorporating dust indeed predicted a sharp decline in the number of bright galaxies towards high redshift, partly due to increasing dust attenuation in more massive halos \citep[e.g.,][]{Dayal2022}. The \textit{JWST} results, however, show a much flatter, extended bright-end UVLF at $z\gtrsim 10$ than anticipated \citep[e.g.,][]{Harikane2023, Finkelstein2023}, hinting that the dust content or its effect on UV light in these early systems could be dramatically reduced. One prominent explanation is that early galaxies undergo an “attenuation-free” phase in which dust has little impact on their UV output \citep{Ferrara2023,Ziparo2023,Fiore2023, Ferrara2025}. In this scenario, galaxies could do still produce dust in normal amounts (consistent with dust-to-stellar mass ratios $\xi_d\sim10^{-3}$–$10^{-2}$ observed up to $z\sim7$; e.g. \citealt{Dayal2022}), but radiative feedback redistributes (or even expels) this dust so that the line-of-sight extinction is low. Radiation pressure from intense star formation can expel dust grains out of the dense star-forming regions, or even drive galaxy-scale dusty outflows, thereby clearing the nucleus of absorbing material \citep{Ziparo2023, Fiore2023}. \citet{Ferrara2023} showed that by assuming essentially zero UV attenuation in galaxies at $z\gtrsim10$, one can nearly perfectly reconcile the observed UVLF at $z=10$--14 with the underlying halo mass function: the loss of dust makes galaxies intrinsically brighter, compensating for the paucity of massive halos at early times. In fact, a conspiracy between decreasing dust attenuation and decreasing host halo abundance, which can explain the small evolution of the bright-end UVLF from $z\approx 7$ to $z\sim 14$ \citep{Ferrara2023}.  The existence of these luminous, metal-enriched yet blue systems (nicknamed “blue monsters”; \citealt{Ziparo2023}) hints that classical dust production must be accompanied by extremely efficient dust removal or suppression in the earliest galaxies. By contrast, dust-rich models in which early galaxies retain their dust in the interstellar medium predict a markedly different UVLF shape. If high-$z$ massive galaxies were as dusty (in terms of obscured fraction) as $z\sim6$–7 galaxies of similar mass, their observed UV luminosities would be heavily diminished. The bright-end of the UVLF would then drop off exponentially (similar to a Schechter function cutoff) because any galaxies above a certain luminosity threshold would have their UV output throttled by extinction. In addition, dust tends to redden the UV spectra of galaxies: a moderately dust-reddened galaxy at $z>10$ would have a UV slope $\beta>-2$, significantly redder than the $\beta\approx -2.5$ observed for the brightest sources in JADES and other \textit{JWST} surveys \citep{Cullen2024, Topping2022}. Thus, the differential impact of dust is most pronounced at the bright end: faint galaxies (low-mass halos) at high redshift likely have little dust in any case, so their UV luminosities are similar in both models, but for the rare massive systems, the dust-rich scenario would hide a large fraction of their star formation, whereas the dust-cleared scenario keeps them visible in UV. This difference provides a clear observational discriminant.

We show the effect of turning off dust attenuation in the \citet{Shuntov2025b} model with $\epsilon_0$ and $\sigma_{\rm UV}$ tuned to match the UVLF in Fig.~\ref{Fig::Munoz}. The resulting UVLF with no dust attenuation is significantly higher than the measurements and showcases the degeneracy with the SFE and stochasticity, if we turn off dust attenuation then lower values of SFE and stochasticity will be required to match the UVLF. For the F814W and F115W samples (galaxies at $z<11$), our measured bright-end of the UV luminosity function lies systematically below predictions from dust-free models, suggesting that these galaxies already experience non-negligible dust attenuation for the \citet{Shuntov2025b} models. 

When dust attenuation is turned off in the GAEA FFB models, the predictions become consistent with our measurements for the F814W sample, as well as for the fainter galaxies (with $M_{\rm UV} > -21$) in the F115W sample. In contrast, for the F150W dropout sample ($z>12$), our measurements are getting increasingly closer to the expectations of dust-free scenarios (both for \citealt{Shuntov2025b} and \citealt{De_Lucia2024} models), consistent with theoretical predictions at these early epochs \citep[e.g.,][]{Ferrara2023, Ziparo2023}. This interpretation is further supported by non-detections of dust continuum emission at very high redshift ($z>9$; \citealt{Bakx2020}), reinforcing the idea that the earliest galaxies may indeed be nearly transparent in the UV \citep{Fiore2023, Ferrara2025}. The presence or absence of dust in early galaxies has far-reaching implications for cosmic reionization and for how we interpret UV-bright populations. If the most luminous galaxies at $z>10$ truly lack significant dust attenuation, they would contribute more to the production of ionizing photons. This could make of these “blue monsters” important agents of reionization despite their low number densities.

\subsection{Reionization}\label{sec:reionization}

The UVLF provides a direct means to assess the contribution of galaxies to the ionizing photon budget necessary for cosmic reionization. Our results indicate that galaxies selected as F115W dropouts, spanning redshifts $8.5 < z < 11.5$, exhibit a significant number density of bright UV sources. These values are pushing the UV magnitude range higher than previous studies and for the $-21.5<M_{\rm UV}<-19.5$ range, are consistent with recent \textit{JWST} constraints on the UVLF at $z \approx 9-12$ \citep[e.g.,][]{Harikane2023, Donnan2023, Perez-Gonzalez2023, Whitler2025}. Moreover, for the most UV-bright galaxies, we find a number density that diverges from pre-\textit{JWST} measurements, with the discrepancy increasing toward higher redshifts. To estimate $f_{\rm esc}$, we followed  the empirical calibrations of \cite{Chisholm2022}. Bluer UV slopes indicate younger, less dust-obscured stellar populations, which correlate with higher $\xi_{\rm ion}$ and significantly larger $f_{\rm esc}$. For galaxies at $z>8$, where the typical $\beta$ shows minimal further evolution with redshift, we accordingly infer a relatively high median escape fraction of $f_{\rm esc}\approx15\%$, compared to $\sim9\%$ for the full sample. These the values are consistent with the results of recent studies \citep[e.g., ][]{Mascia2024, Atek2024}.

While the abundant faint galaxies (which lie below our detection limits) are likely responsible for the bulk of reionizing photons \citep{Atek2024}, the UV-bright systems observed in COSMOS-Web ($M_{\rm UV}<-21.5$) could still play a role. These rare, luminous galaxies residing in the highest-density peaks of the early Universe \citep[e.g.,][]{Paquereau2025}, could generate intense local ionizing flux and thus can initiate sizeable H II regions around themselves. In essence, such UV-bright galaxies could serve as early seeds of reionization bubbles, jump-starting the clearing of neutral hydrogen in their vicinity even as the more numerous faint galaxies gradually complete the reionization of the intergalactic medium.

\section{Conclusions}
We have presented new measurements of the rest-frame UV luminosity function at median redshifts of $z\sim6.4$, $9.3$, and $12.8$, derived from dropout-selected samples identified in three photometric bands (\textit{HST}/ACS F814W, \textit{JWST}/NIRCam F115W, \textit{JWST}/NIRCam F150W) within the 0.54 deg$^2$ COSMOS-Web field. Leveraging the wide survey area, which substantially exceeds that of typical pencil-beam surveys, we constructed a statistically census of 3099 galaxies in the Epoch of Reionization. 

We measured key physical properties, notably the UV continuum slope ($\beta$) and absolute UV magnitude ($M_{\rm UV}$). Despite their bright luminosities, the galaxies in our sample are generally blue, with a median UV slope $\beta \approx -2.34$. We characterized the evolution of the UV slope with stellar mass, finding a strong correlation of ${\rm d}\beta/{\rm d}M_* = 0.67 \pm 0.01$. In contrast, the dependence of the UV slope on UV magnitude and redshift becomes negligible beyond $z\sim8$ (F115W dropout sample), reaching a typical value of $\beta\approx -2.5$ for galaxies at $z>8$. This suggests minimal evolution of the UV slope at the highest redshifts probed by our survey. By uniting the large-area COSMOS-Web survey with the deeper but narrower \textit{JWST} fields from prior studies \citep{Whitler2025} using a relatively similar approach,  our analysis captures the full shape of the UVLF across a wide luminosity baseline – from the faint galaxy regime (M$_{\rm UV}$ $\sim$ -17) to the bright galaxies (M$_{\rm UV}$ $\sim$ -22.5).  We provide a detailed parameterization of the UVLF over this broad luminosity regime.

The resulting UVLF reveals a significant overabundance of bright galaxies compared to earlier observations and theoretical expectations. This bright-end excess becomes noticeable at $z\sim9$ and increasingly pronounced by $z\sim12$, indicating the rapid build-up of stellar mass (and consequently UV luminosity) within the first few hundred million years after the Big Bang. When contextualized with theoretical predictions, this excess challenges current models of early galaxy formation. Several scenarios could alleviate this discrepancy, including strongly bursty star formation histories, a redshift-dependent increase in star formation efficiency, modifications of the stellar initial mass function, or substantially dust-free star-formation environments. By exploring various models, we found that an enhancement in SFE at early epochs, a moderate stochastic scatter, and a reduced dust attenuation appears to be most consistent with our observations. Dust-free models tend to overpredict galaxy counts, although this discrepancy diminishes at our highest-redshift bins, becoming compatible within uncertainties.

%\nocite{*}

\begin{acknowledgments}

We acknowledge that the location where a part of this work took place, the University of Texas at Austin, sits on the Indigenous lands of Turtle Island, the ancestral name for what now is called North America. Moreover, we would like to acknowledge the Alabama-Coushatta, Caddo, Carrizo/Comecrudo, Coahuiltecan, Comanche, Kickapoo, Lipan Apache, Tonkawa and Ysleta Del Sur Pueblo, and all the American Indian and Indigenous Peoples and communities who have been or have become a part of these lands and territories in Texas. This project has received funding from the European Union’s Horizon 2020 research and innovation program under the Marie Sklodowska-Curie grant agreement No 101148925. The French part of the COSMOS team is partly supported by the Centre National d'Etudes Spatiales (CNES).  This work was made possible by utilizing the CANDIDE cluster at the Institut d’Astrophysique de Paris, which was funded through grants from the PNCG, CNES, DIM-ACAV, and the Cosmic Dawn Center and maintained by S. Rouberol. OI acknowledges the funding of the French Agence Nationale de la Recherche for the project
iMAGE (grant ANR-22-CE31-0007). JRW acknowledges support for this work was provided by The Brinson Foundation through a Brinson Prize Fellowship grant. SC acknowledges the use of INAF-OATs computational resources within the framework of the CHIPP project \citep{Taffoni_etal_2020}. GL and RM are supported by UK STFC (grant ST/X001075/1). DBS gratefully acknowledges support from NSF grant 2407752.

\end{acknowledgments}

\bibliography{cw_UVLF}{}

\begin{thebibliography}{}
\expandafter\ifx\csname natexlab\endcsname\relax\def\natexlab#1{#1}\fi
\providecommand{\url}[1]{\href{#1}{#1}}
\providecommand{\dodoi}[1]{doi:~\href{http://doi.org/#1}{\nolinkurl{#1}}}
\providecommand{\doeprint}[1]{\href{http://ascl.net/#1}{\nolinkurl{http://ascl.net/#1}}}
\providecommand{\doarXiv}[1]{\href{https://arxiv.org/abs/#1}{\nolinkurl{https://arxiv.org/abs/#1}}}

\bibitem[{{Adams} {et~al.}(2023){Adams}, {Conselice}, {Ferreira}, {Austin}, {Trussler}, {Juod{\v{z}}balis}, {Wilkins}, {Caruana}, {Dayal}, {Verma}, \& {Vijayan}}]{Adams2023}
{Adams}, N.~J., {Conselice}, C.~J., {Ferreira}, L., {et~al.} 2023, \mnras, 518, 4755, \dodoi{10.1093/mnras/stac3347}

\bibitem[{{Adams} {et~al.}(2024){Adams}, {Conselice}, {Austin}, {Harvey}, {Ferreira}, {Trussler}, {Juod{\v{z}}balis}, {Li}, {Windhorst}, {Cohen}, {Jansen}, {Summers}, {Tompkins}, {Driver}, {Robotham}, {D'Silva}, {Yan}, {Coe}, {Frye}, {Grogin}, {Koekemoer}, {Marshall}, {Pirzkal}, {Ryan}, {Maksym}, {Rutkowski}, {Willmer}, {Hammel}, {Nonino}, {Bhatawdekar}, {Wilkins}, {Bradley}, {Broadhurst}, {Cheng}, {Dole}, {Hathi}, \& {Zitrin}}]{Adams2024}
{Adams}, N.~J., {Conselice}, C.~J., {Austin}, D., {et~al.} 2024, \apj, 965, 169, \dodoi{10.3847/1538-4357/ad2a7b}

\bibitem[{{Aihara} {et~al.}(2022){Aihara}, {AlSayyad}, {Ando}, {Armstrong}, {Bosch}, {Egami}, {Furusawa}, {Furusawa}, {Harasawa}, {Harikane}, {Hsieh}, {Ikeda}, {Ito}, {Iwata}, {Kodama}, {Koike}, {Kokubo}, {Komiyama}, {Li}, {Liang}, {Lin}, {Lupton}, {Lust}, {MacArthur}, {Mawatari}, {Mineo}, {Miyatake}, {Miyazaki}, {More}, {Morishima}, {Murayama}, {Nakajima}, {Nakata}, {Nishizawa}, {Oguri}, {Okabe}, {Okura}, {Ono}, {Osato}, {Ouchi}, {Pan}, {Plazas Malag{\'o}n}, {Price}, {Reed}, {Rykoff}, {Shibuya}, {Simunovic}, {Strauss}, {Sugimori}, {Suto}, {Suzuki}, {Takada}, {Takagi}, {Takata}, {Takita}, {Tanaka}, {Tang}, {Taranu}, {Terai}, {Toba}, {Turner}, {Uchiyama}, {Vijarnwannaluk}, {Waters}, {Yamada}, {Yamamoto}, \& {Yamashita}}]{Aihara2022}
{Aihara}, H., {AlSayyad}, Y., {Ando}, M., {et~al.} 2022, \pasj, 74, 247, \dodoi{10.1093/pasj/psab122}

\bibitem[{{Akins} {et~al.}(2024){Akins}, {Casey}, {Lambrides}, {Allen}, {Andika}, {Brinch}, {Champagne}, {Cooper}, {Ding}, {Drakos}, {Faisst}, {Finkelstein}, {Franco}, {Fujimoto}, {Gentile}, {Gillman}, {Gozaliasl}, {Harish}, {Hayward}, {Hirschmann}, {Ilbert}, {Kartaltepe}, {Kocevski}, {Koekemoer}, {Kokorev}, {Liu}, {Long}, {McCracken}, {McKinney}, {Onoue}, {Paquereau}, {Renzini}, {Rhodes}, {Robertson}, {Shuntov}, {Silverman}, {Tanaka}, {Toft}, {Trakhtenbrot}, {Valentino}, \& {Zavala}}]{Akins2024}
{Akins}, H.~B., {Casey}, C.~M., {Lambrides}, E., {et~al.} 2024, arXiv e-prints, arXiv:2406.10341, \dodoi{10.48550/arXiv.2406.10341}

\bibitem[{{{\'A}lvarez-M{\'a}rquez} {et~al.}(2019){{\'A}lvarez-M{\'a}rquez}, {Colina}, {Marques-Chaves}, {Ceverino}, {Alonso-Herrero}, {Caputi}, {Garc{\'\i}a-Mar{\'\i}n}, {Labiano}, {Le F{\`e}vre}, {Norgaard-Nielsen}, {{\"O}stlin}, {P{\'e}rez-Gonz{\'a}lez}, {Pye}, {Tikkanen}, {van der Werf}, {Walter}, \& {Wright}}]{Alvarez-Marquez2019}
{{\'A}lvarez-M{\'a}rquez}, J., {Colina}, L., {Marques-Chaves}, R., {et~al.} 2019, \aap, 629, A9, \dodoi{10.1051/0004-6361/201935594}

\bibitem[{{Andalman} {et~al.}(2025){Andalman}, {Teyssier}, \& {Dekel}}]{Andalman2025}
{Andalman}, Z.~L., {Teyssier}, R., \& {Dekel}, A. 2025, \mnras, 540, 3350, \dodoi{10.1093/mnras/staf930}

\bibitem[{{Arango-Toro} {et~al.}(2025){Arango-Toro}, {Ilbert}, {Ciesla}, {Shuntov}, {Aufort}, {Mercier}, {Laigle}, {Franco}, {Bethermin}, {Le Borgne}, {Dubois}, {McCracken}, {Paquereau}, {Huertas-Company}, {Kartaltepe}, {Casey}, {Akins}, {Allen}, {Andika}, {Brinch}, {Drakos}, {Faisst}, {Gozaliasl}, {Harish}, {Kaminsky}, {Koekemoer}, {Kokorev}, {Liu}, {Magdis}, {Martin}, {Moutard}, {Rhodes}, {Rich}, {Robertson}, {Sanders}, {Sheth}, {Talia}, {Toft}, {Tresse}, {Valentino}, {Vijayan}, \& {Weaver}}]{Arango2025}
{Arango-Toro}, R.~C., {Ilbert}, O., {Ciesla}, L., {et~al.} 2025, \aap, 696, A159, \dodoi{10.1051/0004-6361/202452519}

\bibitem[{{Arnouts} {et~al.}(1999){Arnouts}, {Cristiani}, {Moscardini}, {Matarrese}, {Lucchin}, {Fontana}, \& {Giallongo}}]{Arnouts1999}
{Arnouts}, S., {Cristiani}, S., {Moscardini}, L., {et~al.} 1999, \mnras, 310, 540, \dodoi{10.1046/j.1365-8711.1999.02978.x}

\bibitem[{{Arrabal Haro} {et~al.}(2023){Arrabal Haro}, {Dickinson}, {Finkelstein}, {Kartaltepe}, {Donnan}, {Burgarella}, {Carnall}, {Cullen}, {Dunlop}, {Fern{\'a}ndez}, {Fujimoto}, {Jung}, {Krips}, {Larson}, {Papovich}, {P{\'e}rez-Gonz{\'a}lez}, {Amor{\'\i}n}, {Bagley}, {Buat}, {Casey}, {Chworowsky}, {Cohen}, {Ferguson}, {Giavalisco}, {Huertas-Company}, {Hutchison}, {Kocevski}, {Koekemoer}, {Lucas}, {McLeod}, {McLure}, {Pirzkal}, {Seill{\'e}}, {Trump}, {Weiner}, {Wilkins}, \& {Zavala}}]{Arrabal-Haro2023}
{Arrabal Haro}, P., {Dickinson}, M., {Finkelstein}, S.~L., {et~al.} 2023, \nat, 622, 707, \dodoi{10.1038/s41586-023-06521-7}

\bibitem[{{Atek} {et~al.}(2024){Atek}, {Labb{\'e}}, {Furtak}, {Chemerynska}, {Fujimoto}, {Setton}, {Miller}, {Oesch}, {Bezanson}, {Price}, {Dayal}, {Zitrin}, {Kokorev}, {Weaver}, {Brammer}, {Dokkum}, {Williams}, {Cutler}, {Feldmann}, {Fudamoto}, {Greene}, {Leja}, {Maseda}, {Muzzin}, {Pan}, {Papovich}, {Nelson}, {Nanayakkara}, {Stark}, {Stefanon}, {Suess}, {Wang}, \& {Whitaker}}]{Atek2024}
{Atek}, H., {Labb{\'e}}, I., {Furtak}, L.~J., {et~al.} 2024, \nat, 626, 975, \dodoi{10.1038/s41586-024-07043-6}

\bibitem[{{Austin} {et~al.}(2024){Austin}, {Conselice}, {Adams}, {Harvey}, {Duan}, {Trussler}, {Li}, {Juodzbalis}, {Ormerod}, {Ferreira}, {Westcott}, {Harris}, {Wilkins}, {Bhatawdekar}, {Caruana}, {Coe}, {Cohen}, {Driver}, {D'Silva}, {Frye}, {Furtak}, {Grogin}, {Hathi}, {Holwerda}, {Jansen}, {Koekemoer}, {Marshall}, {Nonino}, {Ortiz}, {Pirzkal}, {Robotham}, {Ryan}, {Summers}, {Willmer}, {Windhorst}, {Yan}, \& {Zackrisson}}]{Austin2024}
{Austin}, D., {Conselice}, C.~J., {Adams}, N.~J., {et~al.} 2024, arXiv e-prints, arXiv:2404.10751, \dodoi{10.48550/arXiv.2404.10751}

\bibitem[{{Bagley} {et~al.}(2023){Bagley}, {Finkelstein}, {Koekemoer}, {Ferguson}, {Arrabal Haro}, {Dickinson}, {Kartaltepe}, {Papovich}, {P{\'e}rez-Gonz{\'a}lez}, {Pirzkal}, {Somerville}, {Willmer}, {Yang}, {Yung}, {Fontana}, {Grazian}, {Grogin}, {Hirschmann}, {Kewley}, {Kirkpatrick}, {Kocevski}, {Lotz}, {Medrano}, {Morales}, {Pentericci}, {Ravindranath}, {Trump}, {Wilkins}, {Calabr{\`o}}, {Cooper}, {Costantin}, {de la Vega}, {Hilbert}, {Hutchison}, {Larson}, {Lucas}, {McGrath}, {Ryan}, {Wang}, \& {Wuyts}}]{Bagley2023}
{Bagley}, M.~B., {Finkelstein}, S.~L., {Koekemoer}, A.~M., {et~al.} 2023, \apjl, 946, L12, \dodoi{10.3847/2041-8213/acbb08}

\bibitem[{{Bakx} {et~al.}(2020){Bakx}, {Tamura}, {Hashimoto}, {Inoue}, {Lee}, {Mawatari}, {Ota}, {Umehata}, {Zackrisson}, {Hatsukade}, {Kohno}, {Matsuda}, {Matsuo}, {Okamoto}, {Shibuya}, {Shimizu}, {Taniguchi}, \& {Yoshida}}]{Bakx2020}
{Bakx}, T. J.~L.~C., {Tamura}, Y., {Hashimoto}, T., {et~al.} 2020, \mnras, 493, 4294, \dodoi{10.1093/mnras/staa509}

\bibitem[{{Barbary} {et~al.}(2016){Barbary}, {Boone}, {McCully}, {Craig}, {Deil}, \& {Rose}}]{Barbary2016}
{Barbary}, K., {Boone}, K., {McCully}, C., {et~al.} 2016, {kbarbary/sep: v1.0.0}, v1.0.0,  Zenodo, \dodoi{10.5281/zenodo.159035}

\bibitem[{{Becker} {et~al.}(2001){Becker}, {Fan}, {White}, {Strauss}, {Narayanan}, {Lupton}, {Gunn}, {Annis}, {Bahcall}, {Brinkmann}, {Connolly}, {Csabai}, {Czarapata}, {Doi}, {Heckman}, {Hennessy}, {Ivezi{\'c}}, {Knapp}, {Lamb}, {McKay}, {Munn}, {Nash}, {Nichol}, {Pier}, {Richards}, {Schneider}, {Stoughton}, {Szalay}, {Thakar}, \& {York}}]{Becker2001}
{Becker}, R.~H., {Fan}, X., {White}, R.~L., {et~al.} 2001, \aj, 122, 2850, \dodoi{10.1086/324231}

\bibitem[{{Behroozi} {et~al.}(2019){Behroozi}, {Wechsler}, {Hearin}, \& {Conroy}}]{Behroozi2019}
{Behroozi}, P., {Wechsler}, R.~H., {Hearin}, A.~P., \& {Conroy}, C. 2019, \mnras, 488, 3143, \dodoi{10.1093/mnras/stz1182}

\bibitem[{{Beichman} {et~al.}(2012){Beichman}, {Rieke}, {Eisenstein}, {Greene}, {Krist}, {McCarthy}, {Meyer}, \& {Stansberry}}]{beichman12}
{Beichman}, C.~A., {Rieke}, M., {Eisenstein}, D., {et~al.} 2012, in Society of Photo-Optical Instrumentation Engineers (SPIE) Conference Series, Vol. 8442, Space Telescopes and Instrumentation 2012: Optical, Infrared, and Millimeter Wave, ed. M.~C. {Clampin}, G.~G. {Fazio}, H.~A. {MacEwen}, \& J.~{Oschmann}, Jacobus~M., 84422N, \dodoi{10.1117/12.925447}

\bibitem[{{Bertin} \& {Arnouts}(1996)}]{Bertin1996}
{Bertin}, E., \& {Arnouts}, S. 1996, \aaps, 117, 393, \dodoi{10.1051/aas:1996164}

\bibitem[{{Bertin} {et~al.}(2020){Bertin}, {Schefer}, {Apostolakos}, {{\'A}lvarez-Ayll{\'o}n}, {Dubath}, \& {K{\"u}mmel}}]{Bertin2020}
{Bertin}, E., {Schefer}, M., {Apostolakos}, N., {et~al.} 2020, in Astronomical Society of the Pacific Conference Series, Vol. 527, Astronomical Data Analysis Software and Systems XXIX, ed. R.~{Pizzo}, E.~R. {Deul}, J.~D. {Mol}, J.~{de Plaa}, \& H.~{Verkouter}, 461

\bibitem[{{Bhowmick} {et~al.}(2020){Bhowmick}, {Somerville}, {Di Matteo}, {Wilkins}, {Feng}, \& {Tenneti}}]{Bhowmick2020}
{Bhowmick}, A.~K., {Somerville}, R.~S., {Di Matteo}, T., {et~al.} 2020, \mnras, 496, 754, \dodoi{10.1093/mnras/staa1605}

\bibitem[{{Boquien} {et~al.}(2019){Boquien}, {Burgarella}, {Roehlly}, {Buat}, {Ciesla}, {Corre}, {Inoue}, \& {Salas}}]{Boquien2019}
{Boquien}, M., {Burgarella}, D., {Roehlly}, Y., {et~al.} 2019, \aap, 622, A103, \dodoi{10.1051/0004-6361/201834156}

\bibitem[{{Bouwens} {et~al.}(2010){Bouwens}, {Illingworth}, {Oesch}, {Trenti}, {Stiavelli}, {Carollo}, {Franx}, {van Dokkum}, {Labb{\'e}}, \& {Magee}}]{Bouwens2010}
{Bouwens}, R.~J., {Illingworth}, G.~D., {Oesch}, P.~A., {et~al.} 2010, \apjl, 708, L69, \dodoi{10.1088/2041-8205/708/2/L69}

\bibitem[{{Bouwens} {et~al.}(2012){Bouwens}, {Illingworth}, {Oesch}, {Trenti}, {Labb{\'e}}, {Franx}, {Stiavelli}, {Carollo}, {van Dokkum}, \& {Magee}}]{Bouwens2012}
---. 2012, \apjl, 752, L5, \dodoi{10.1088/2041-8205/752/1/L5}

\bibitem[{{Bouwens} {et~al.}(2014){Bouwens}, {Illingworth}, {Oesch}, {Labb{\'e}}, {van Dokkum}, {Trenti}, {Franx}, {Smit}, {Gonzalez}, \& {Magee}}]{bouwens14}
---. 2014, \apj, 793, 115, \dodoi{10.1088/0004-637X/793/2/115}

\bibitem[{{Bouwens} {et~al.}(2015){Bouwens}, {Illingworth}, {Oesch}, {Trenti}, {Labb{\'e}}, {Bradley}, {Carollo}, {van Dokkum}, {Gonzalez}, {Holwerda}, {Franx}, {Spitler}, {Smit}, \& {Magee}}]{Bouwens2015}
---. 2015, \apj, 803, 34, \dodoi{10.1088/0004-637X/803/1/34}

\bibitem[{{Bouwens} {et~al.}(2021){Bouwens}, {Oesch}, {Stefanon}, {Illingworth}, {Labb{\'e}}, {Reddy}, {Atek}, {Montes}, {Naidu}, {Nanayakkara}, {Nelson}, \& {Wilkins}}]{Bouwens2021}
{Bouwens}, R.~J., {Oesch}, P.~A., {Stefanon}, M., {et~al.} 2021, \aj, 162, 47, \dodoi{10.3847/1538-3881/abf83e}

\bibitem[{{Bowler} {et~al.}(2020){Bowler}, {Jarvis}, {Dunlop}, {McLure}, {McLeod}, {Adams}, {Milvang-Jensen}, \& {McCracken}}]{Bowler2020}
{Bowler}, R.~A.~A., {Jarvis}, M.~J., {Dunlop}, J.~S., {et~al.} 2020, \mnras, 493, 2059, \dodoi{10.1093/mnras/staa313}

\bibitem[{{Boyett} {et~al.}(2024){Boyett}, {Trenti}, {Leethochawalit}, {Calabr{\'o}}, {Metha}, {Roberts-Borsani}, {Dalmasso}, {Yang}, {Santini}, {Treu}, {Jones}, {Henry}, {Mason}, {Morishita}, {Nanayakkara}, {Roy}, {Wang}, {Fontana}, {Merlin}, {Castellano}, {Paris}, {Brada{\v{c}}}, {Malkan}, {Marchesini}, {Mascia}, {Glazebrook}, {Pentericci}, {Vanzella}, \& {Vulcani}}]{Boyett2024}
{Boyett}, K., {Trenti}, M., {Leethochawalit}, N., {et~al.} 2024, Nature Astronomy, 8, 657, \dodoi{10.1038/s41550-024-02218-7}

\bibitem[{{Boylan-Kolchin}(2023)}]{Boylan-Kolchin2023}
{Boylan-Kolchin}, M. 2023, Nature Astronomy, 7, 731, \dodoi{10.1038/s41550-023-01937-7}

\bibitem[{Brammer(2023)}]{Brammer2023}
Brammer, G. 2023, msaexp: NIRSpec analyis tools, 0.6.17,  Zenodo, \dodoi{10.5281/zenodo.8319596}

\bibitem[{{Brammer} {et~al.}(2008){Brammer}, {van Dokkum}, \& {Coppi}}]{brammer2008}
{Brammer}, G.~B., {van Dokkum}, P.~G., \& {Coppi}, P. 2008, \apj, 686, 1503, \dodoi{10.1086/591786}

\bibitem[{{Bruzual} \& {Charlot}(2003)}]{Bruzual2003}
{Bruzual}, G., \& {Charlot}, S. 2003, \mnras, 344, 1000, \dodoi{10.1046/j.1365-8711.2003.06897.x}

\bibitem[{{Bunker} {et~al.}(2004){Bunker}, {Stanway}, {Ellis}, \& {McMahon}}]{Bunker2004}
{Bunker}, A.~J., {Stanway}, E.~R., {Ellis}, R.~S., \& {McMahon}, R.~G. 2004, \mnras, 355, 374, \dodoi{10.1111/j.1365-2966.2004.08326.x}

\bibitem[{{Bunker} {et~al.}(2023){Bunker}, {Saxena}, {Cameron}, {Willott}, {Curtis-Lake}, {Jakobsen}, {Carniani}, {Smit}, {Maiolino}, {Witstok}, {Curti}, {D'Eugenio}, {Jones}, {Ferruit}, {Arribas}, {Charlot}, {Chevallard}, {Giardino}, {de Graaff}, {Looser}, {L{\"u}tzgendorf}, {Maseda}, {Rawle}, {Rix}, {Del Pino}, {Alberts}, {Egami}, {Eisenstein}, {Endsley}, {Hainline}, {Hausen}, {Johnson}, {Rieke}, {Rieke}, {Robertson}, {Shivaei}, {Stark}, {Sun}, {Tacchella}, {Tang}, {Williams}, {Willmer}, {Baker}, {Baum}, {Bhatawdekar}, {Bowler}, {Boyett}, {Chen}, {Circosta}, {Helton}, {Ji}, {Kumari}, {Lyu}, {Nelson}, {Parlanti}, {Perna}, {Sandles}, {Scholtz}, {Suess}, {Topping}, {{\"U}bler}, {Wallace}, \& {Whitler}}]{Bunker2023}
{Bunker}, A.~J., {Saxena}, A., {Cameron}, A.~J., {et~al.} 2023, \aap, 677, A88, \dodoi{10.1051/0004-6361/202346159}

\bibitem[{{Bushouse} {et~al.}(2023){Bushouse}, {Eisenhamer}, {Dencheva}, {Davies}, {Greenfield}, {Morrison}, {Hodge}, {Simon}, {Grumm}, {Droettboom}, {Slavich}, {Sosey}, {Pauly}, {Miller}, {Jedrzejewski}, {Hack}, {Davis}, {Crawford}, {Law}, {Gordon}, {Regan}, {Cara}, {MacDonald}, {Bradley}, {Shanahan}, {Jamieson}, {Teodoro}, \& {Williams}}]{Bushouse2023}
{Bushouse}, H., {Eisenhamer}, J., {Dencheva}, N., {et~al.} 2023, {JWST Calibration Pipeline}, 1.9.4,  Zenodo, \dodoi{10.5281/zenodo.7577320}

\bibitem[{{Calzetti} {et~al.}(2000){Calzetti}, {Armus}, {Bohlin}, {Kinney}, {Koornneef}, \& {Storchi-Bergmann}}]{Calzetti2000}
{Calzetti}, D., {Armus}, L., {Bohlin}, R.~C., {et~al.} 2000, \apj, 533, 682, \dodoi{10.1086/308692}

\bibitem[{{Calzetti} {et~al.}(1994){Calzetti}, {Kinney}, \& {Storchi-Bergmann}}]{calzetti94}
{Calzetti}, D., {Kinney}, A.~L., \& {Storchi-Bergmann}, T. 1994, \apj, 429, 582, \dodoi{10.1086/174346}

\bibitem[{{Capak} {et~al.}(2007){Capak}, {Aussel}, {Ajiki}, {McCracken}, {Mobasher}, {Scoville}, {Shopbell}, {Taniguchi}, {Thompson}, {Tribiano}, {Sasaki}, {Blain}, {Brusa}, {Carilli}, {Comastri}, {Carollo}, {Cassata}, {Colbert}, {Ellis}, {Elvis}, {Giavalisco}, {Green}, {Guzzo}, {Hasinger}, {Ilbert}, {Impey}, {Jahnke}, {Kartaltepe}, {Kneib}, {Koda}, {Koekemoer}, {Komiyama}, {Leauthaud}, {Le Fevre}, {Lilly}, {Liu}, {Massey}, {Miyazaki}, {Murayama}, {Nagao}, {Peacock}, {Pickles}, {Porciani}, {Renzini}, {Rhodes}, {Rich}, {Salvato}, {Sanders}, {Scarlata}, {Schiminovich}, {Schinnerer}, {Scodeggio}, {Sheth}, {Shioya}, {Tasca}, {Taylor}, {Yan}, \& {Zamorani}}]{capak07a}
{Capak}, P., {Aussel}, H., {Ajiki}, M., {et~al.} 2007, \apjs, 172, 99, \dodoi{10.1086/519081}

\bibitem[{{Carnall} {et~al.}(2018){Carnall}, {McLure}, {Dunlop}, \& {Dav{\'e}}}]{Carnall2018}
{Carnall}, A.~C., {McLure}, R.~J., {Dunlop}, J.~S., \& {Dav{\'e}}, R. 2018, \mnras, 480, 4379, \dodoi{10.1093/mnras/sty2169}

\bibitem[{{Carniani} {et~al.}(2024){Carniani}, {Hainline}, {D'Eugenio}, {Eisenstein}, {Jakobsen}, {Witstok}, {Johnson}, {Chevallard}, {Maiolino}, {Helton}, {Willott}, {Robertson}, {Alberts}, {Arribas}, {Baker}, {Bhatawdekar}, {Boyett}, {Bunker}, {Cameron}, {Cargile}, {Charlot}, {Curti}, {Curtis-Lake}, {Egami}, {Giardino}, {Isaak}, {Ji}, {Jones}, {Kumari}, {Maseda}, {Parlanti}, {P{\'e}rez-Gonz{\'a}lez}, {Rawle}, {Rieke}, {Rieke}, {Del Pino}, {Saxena}, {Scholtz}, {Smit}, {Sun}, {Tacchella}, {{\"U}bler}, {Venturi}, {Williams}, \& {Willmer}}]{Carniani2024}
{Carniani}, S., {Hainline}, K., {D'Eugenio}, F., {et~al.} 2024, \nat, 633, 318, \dodoi{10.1038/s41586-024-07860-9}

\bibitem[{{Casey} {et~al.}(2023){Casey}, {Kartaltepe}, {Drakos}, {Franco}, {Harish}, {Paquereau}, {Ilbert}, {Rose}, {Cox}, {Nightingale}, {Robertson}, {Silverman}, {Koekemoer}, {Massey}, {McCracken}, {Rhodes}, {Akins}, {Allen}, {Amvrosiadis}, {Arango-Toro}, {Bagley}, {Bongiorno}, {Capak}, {Champagne}, {Chartab}, {Ch{\'a}vez Ortiz}, {Chworowsky}, {Cooke}, {Cooper}, {Darvish}, {Ding}, {Faisst}, {Finkelstein}, {Fujimoto}, {Gentile}, {Gillman}, {Gould}, {Gozaliasl}, {Hayward}, {He}, {Hemmati}, {Hirschmann}, {Jahnke}, {Jin}, {Khostovan}, {Kokorev}, {Lambrides}, {Laigle}, {Larson}, {Leung}, {Liu}, {Liaudat}, {Long}, {Magdis}, {Mahler}, {Mainieri}, {Manning}, {Maraston}, {Martin}, {McCleary}, {McKinney}, {McPartland}, {Mobasher}, {Pattnaik}, {Renzini}, {Rich}, {Sanders}, {Sattari}, {Scognamiglio}, {Scoville}, {Sheth}, {Shuntov}, {Sparre}, {Suzuki}, {Talia}, {Toft}, {Trakhtenbrot}, {Urry}, {Valentino}, {Vanderhoof}, {Vardoulaki}, {Weaver}, {Whitaker}, {Wilkins}, {Yang}, \& {Zavala}}]{Casey2023}
{Casey}, C.~M., {Kartaltepe}, J.~S., {Drakos}, N.~E., {et~al.} 2023, \apj, 954, 31, \dodoi{10.3847/1538-4357/acc2bc}

\bibitem[{{Casey} {et~al.}(2024){Casey}, {Akins}, {Shuntov}, {Ilbert}, {Paquereau}, {Franco}, {Hayward}, {Finkelstein}, {Boylan-Kolchin}, {Robertson}, {Allen}, {Brinch}, {Cooper}, {Ding}, {Drakos}, {Faisst}, {Fujimoto}, {Gillman}, {Harish}, {Hirschmann}, {Jin}, {Kartaltepe}, {Koekemoer}, {Kokorev}, {Liu}, {Long}, {Magdis}, {Maraston}, {Martin}, {McCracken}, {McKinney}, {Mobasher}, {Rhodes}, {Rich}, {Sanders}, {Silverman}, {Toft}, {Vijayan}, {Weaver}, {Wilkins}, {Yang}, \& {Zavala}}]{Casey2024}
{Casey}, C.~M., {Akins}, H.~B., {Shuntov}, M., {et~al.} 2024, \apj, 965, 98, \dodoi{10.3847/1538-4357/ad2075}

\bibitem[{{Castellano} {et~al.}(2022){Castellano}, {Fontana}, {Treu}, {Santini}, {Merlin}, {Leethochawalit}, {Trenti}, {Vanzella}, {Mestric}, {Bonchi}, {Belfiori}, {Nonino}, {Paris}, {Polenta}, {Roberts-Borsani}, {Boyett}, {Brada{\v{c}}}, {Calabr{\`o}}, {Glazebrook}, {Grillo}, {Mascia}, {Mason}, {Mercurio}, {Morishita}, {Nanayakkara}, {Pentericci}, {Rosati}, {Vulcani}, {Wang}, \& {Yang}}]{Castellano2022}
{Castellano}, M., {Fontana}, A., {Treu}, T., {et~al.} 2022, \apjl, 938, L15, \dodoi{10.3847/2041-8213/ac94d0}

\bibitem[{{Castellano} {et~al.}(2023){Castellano}, {Fontana}, {Treu}, {Merlin}, {Santini}, {Bergamini}, {Grillo}, {Rosati}, {Acebron}, {Leethochawalit}, {Paris}, {Bonchi}, {Belfiori}, {Calabr{\`o}}, {Correnti}, {Nonino}, {Polenta}, {Trenti}, {Boyett}, {Brammer}, {Broadhurst}, {Caminha}, {Chen}, {Filippenko}, {Fortuni}, {Glazebrook}, {Mascia}, {Mason}, {Menci}, {Meneghetti}, {Mercurio}, {Metha}, {Morishita}, {Nanayakkara}, {Pentericci}, {Roberts-Borsani}, {Roy}, {Vanzella}, {Vulcani}, {Yang}, \& {Wang}}]{Castellano2023}
---. 2023, \apjl, 948, L14, \dodoi{10.3847/2041-8213/accea5}

\bibitem[{{Castellano} {et~al.}(2024){Castellano}, {Napolitano}, {Fontana}, {Roberts-Borsani}, {Treu}, {Vanzella}, {Zavala}, {Arrabal Haro}, {Calabr{\`o}}, {Llerena}, {Mascia}, {Merlin}, {Paris}, {Pentericci}, {Santini}, {Bakx}, {Bergamini}, {Cupani}, {Dickinson}, {Filippenko}, {Glazebrook}, {Grillo}, {Kelly}, {Malkan}, {Mason}, {Morishita}, {Nanayakkara}, {Rosati}, {Sani}, {Wang}, \& {Yoon}}]{Castellano2024}
{Castellano}, M., {Napolitano}, L., {Fontana}, A., {et~al.} 2024, \apj, 972, 143, \dodoi{10.3847/1538-4357/ad5f88}

\bibitem[{{Castellano} {et~al.}(2025){Castellano}, {Fontana}, {Merlin}, {Santini}, {Napolitano}, {Menci}, {Calabr{\`o}}, {Paris}, {Pentericci}, {Zavala}, {Dickinson}, {Finkelstein}, {Treu}, {Amorin}, {Arrabal Haro}, {Bergamini}, {Bisigello}, {Daddi}, {Dayal}, {Dekel}, {Ferrara}, {Fortuni}, {Gandolfi}, {Giavalisco}, {Grillo}, {Guida}, {Hathi}, {Holwerda}, {Koekemoer}, {Kokorev}, {Li}, {Llerena}, {Lucas}, {Mascia}, {Metha}, {Morishita}, {Nanayakkara}, {Pacucci}, {P{\'e}rez-Gonz{\'a}lez}, {Roberts-Borsani}, {Rodighiero}, {Rosati}, {Salazar}, {Schneider}, {Somerville}, {Taylor}, {Trenti}, {Trinca}, {Wang}, {Watson}, {Yang}, \& {Yung}}]{Castellano2025}
{Castellano}, M., {Fontana}, A., {Merlin}, E., {et~al.} 2025, arXiv e-prints, arXiv:2504.05893, \dodoi{10.48550/arXiv.2504.05893}

\bibitem[{{Chabrier}(2003)}]{Chabrier2003}
{Chabrier}, G. 2003, \pasp, 115, 763, \dodoi{10.1086/376392}

\bibitem[{{Chardin} {et~al.}(2015){Chardin}, {Haehnelt}, {Aubert}, \& {Puchwein}}]{Chardin2015}
{Chardin}, J., {Haehnelt}, M.~G., {Aubert}, D., \& {Puchwein}, E. 2015, \mnras, 453, 2943, \dodoi{10.1093/mnras/stv1786}

\bibitem[{{Charlot} \& {Fall}(2000)}]{Charlot2000}
{Charlot}, S., \& {Fall}, S.~M. 2000, \apj, 539, 718, \dodoi{10.1086/309250}

\bibitem[{{Chisholm} {et~al.}(2022){Chisholm}, {Saldana-Lopez}, {Flury}, {Schaerer}, {Jaskot}, {Amor{\'\i}n}, {Atek}, {Finkelstein}, {Fleming}, {Ferguson}, {Fern{\'a}ndez}, {Giavalisco}, {Hayes}, {Heckman}, {Henry}, {Ji}, {Marques-Chaves}, {Mauerhofer}, {McCandliss}, {Oey}, {{\"O}stlin}, {Rutkowski}, {Scarlata}, {Thuan}, {Trebitsch}, {Wang}, {Worseck}, \& {Xu}}]{Chisholm2022}
{Chisholm}, J., {Saldana-Lopez}, A., {Flury}, S., {et~al.} 2022, \mnras, 517, 5104, \dodoi{10.1093/mnras/stac2874}

\bibitem[{{Ciesla} {et~al.}(2024){Ciesla}, {Elbaz}, {Ilbert}, {Buat}, {Magnelli}, {Narayanan}, {Daddi}, {G{\'o}mez-Guijarro}, \& {Arango-Toro}}]{Ciesla2024}
{Ciesla}, L., {Elbaz}, D., {Ilbert}, O., {et~al.} 2024, \aap, 686, A128, \dodoi{10.1051/0004-6361/202348091}

\bibitem[{{Cole} {et~al.}(2025){Cole}, {Papovich}, {Finkelstein}, {Bagley}, {Dickinson}, {Iyer}, {Yung}, {Ciesla}, {Amor{\'\i}n}, {Arrabal Haro}, {Bhatawdekar}, {Calabr{\`o}}, {Cleri}, {de la Vega}, {Dekel}, {Endsley}, {Gawiser}, {Giavalisco}, {Hathi}, {Hirschmann}, {Holwerda}, {Kartaltepe}, {Koekemoer}, {Lucas}, {Mascia}, {Mobasher}, {P{\'e}rez-Gonz{\'a}lez}, {Rodighiero}, {Ronayne}, {Tacchella}, {Weiner}, \& {Wilkins}}]{Cole2025}
{Cole}, J.~W., {Papovich}, C., {Finkelstein}, S.~L., {et~al.} 2025, \apj, 979, 193, \dodoi{10.3847/1538-4357/ad9a6a}

\bibitem[{{Conselice} {et~al.}(2025){Conselice}, {Adams}, {Harvey}, {Austin}, {Ferreira}, {Ormerod}, {Duan}, {Trussler}, {Li}, {Juod{\v{z}}balis}, {Westcott}, {Harris}, {Seeyave}, {Bluck}, {Windhorst}, {Bhatawdekar}, {Coe}, {Cohen}, {Cheng}, {Driver}, {Frye}, {Furtak}, {Grogin}, {Hathi}, {Holwerda}, {Jansen}, {Koekemoer}, {Marshall}, {Nonino}, {Robotham}, {Summers}, {Wilkins}, {Willmer}, {Yan}, \& {Zitrin}}]{Conselice2025}
{Conselice}, C.~J., {Adams}, N., {Harvey}, T., {et~al.} 2025, \apj, 983, 30, \dodoi{10.3847/1538-4357/ada608}

\bibitem[{{Crain} {et~al.}(2009){Crain}, {Theuns}, {Dalla Vecchia}, {Eke}, {Frenk}, {Jenkins}, {Kay}, {Peacock}, {Pearce}, {Schaye}, {Springel}, {Thomas}, {White}, \& {Wiersma}}]{Crain2009}
{Crain}, R.~A., {Theuns}, T., {Dalla Vecchia}, C., {et~al.} 2009, \mnras, 399, 1773, \dodoi{10.1111/j.1365-2966.2009.15402.x}

\bibitem[{{Cullen} {et~al.}(2023){Cullen}, {McLure}, {McLeod}, {Dunlop}, {Donnan}, {Carnall}, {Bowler}, {Begley}, {Hamadouche}, \& {Stanton}}]{Cullen2023}
{Cullen}, F., {McLure}, R.~J., {McLeod}, D.~J., {et~al.} 2023, \mnras, 520, 14, \dodoi{10.1093/mnras/stad073}

\bibitem[{{Cullen} {et~al.}(2024){Cullen}, {McLeod}, {McLure}, {Dunlop}, {Donnan}, {Carnall}, {Keating}, {Magee}, {Arellano-Cordova}, {Bowler}, {Begley}, {Flury}, {Hamadouche}, \& {Stanton}}]{Cullen2024}
{Cullen}, F., {McLeod}, D.~J., {McLure}, R.~J., {et~al.} 2024, \mnras, 531, 997, \dodoi{10.1093/mnras/stae1211}

\bibitem[{{Cullen} {et~al.}(2025){Cullen}, {Carnall}, {Scholte}, {McLeod}, {McLure}, {Arellano-C{\'o}rdova}, {Stanton}, {Donnan}, {Dunlop}, {Shapley}, {Barrufet}, {Begley}, {Bondestam}, {Cirasuolo}, {Leung}, {Pollock}, \& {Stevenson}}]{Cullen2025}
{Cullen}, F., {Carnall}, A.~C., {Scholte}, D., {et~al.} 2025, \mnras, \dodoi{10.1093/mnras/staf838}

\bibitem[{{Dayal} \& {Ferrara}(2018)}]{Dayal2018}
{Dayal}, P., \& {Ferrara}, A. 2018, \physrep, 780, 1, \dodoi{10.1016/j.physrep.2018.10.002}

\bibitem[{{Dayal} {et~al.}(2020){Dayal}, {Volonteri}, {Choudhury}, {Schneider}, {Trebitsch}, {Gnedin}, {Atek}, {Hirschmann}, \& {Reines}}]{Dayal2020}
{Dayal}, P., {Volonteri}, M., {Choudhury}, T.~R., {et~al.} 2020, \mnras, 495, 3065, \dodoi{10.1093/mnras/staa1138}

\bibitem[{{Dayal} {et~al.}(2022){Dayal}, {Ferrara}, {Sommovigo}, {Bouwens}, {Oesch}, {Smit}, {Gonzalez}, {Schouws}, {Stefanon}, {Kobayashi}, {Bremer}, {Algera}, {Aravena}, {Bowler}, {da Cunha}, {Fudamoto}, {Graziani}, {Hodge}, {Inami}, {De Looze}, {Pallottini}, {Riechers}, {Schneider}, {Stark}, \& {Endsley}}]{Dayal2022}
{Dayal}, P., {Ferrara}, A., {Sommovigo}, L., {et~al.} 2022, \mnras, 512, 989, \dodoi{10.1093/mnras/stac537}

\bibitem[{{de Graaff} {et~al.}(2024){de Graaff}, {Brammer}, {Weibel}, {Lewis}, {Maseda}, {Oesch}, {Bezanson}, {Boogaard}, {Cleri}, {Cooper}, {Gottumukkala}, {Greene}, {Hirschmann}, {Hviding}, {Katz}, {Labb{\'e}}, {Leja}, {Matthee}, {McConachie}, {Miller}, {Naidu}, {Price}, {Rix}, {Setton}, {Suess}, {Wang}, {Whitaker}, \& {Williams}}]{deGraaff2024}
{de Graaff}, A., {Brammer}, G., {Weibel}, A., {et~al.} 2024, arXiv e-prints, arXiv:2409.05948, \dodoi{10.48550/arXiv.2409.05948}

\bibitem[{{De Lucia} \& {Blaizot}(2007)}]{De_Lucia2007}
{De Lucia}, G., \& {Blaizot}, J. 2007, \mnras, 375, 2, \dodoi{10.1111/j.1365-2966.2006.11287.x}

\bibitem[{{De Lucia} {et~al.}(2024){De Lucia}, {Fontanot}, {Xie}, \& {Hirschmann}}]{De_Lucia2024}
{De Lucia}, G., {Fontanot}, F., {Xie}, L., \& {Hirschmann}, M. 2024, \aap, 687, A68, \dodoi{10.1051/0004-6361/202349045}

\bibitem[{{Dekel} {et~al.}(2023){Dekel}, {Sarkar}, {Birnboim}, {Mandelker}, \& {Li}}]{Dekel2023}
{Dekel}, A., {Sarkar}, K.~C., {Birnboim}, Y., {Mandelker}, N., \& {Li}, Z. 2023, \mnras, 523, 3201, \dodoi{10.1093/mnras/stad1557}

\bibitem[{{Donnan} {et~al.}(2025){Donnan}, {Dunlop}, {McLure}, {McLeod}, \& {Cullen}}]{Donnan2025}
{Donnan}, C.~T., {Dunlop}, J.~S., {McLure}, R.~J., {McLeod}, D.~J., \& {Cullen}, F. 2025, arXiv e-prints, arXiv:2501.03217, \dodoi{10.48550/arXiv.2501.03217}

\bibitem[{{Donnan} {et~al.}(2023){Donnan}, {McLeod}, {Dunlop}, {McLure}, {Carnall}, {Begley}, {Cullen}, {Hamadouche}, {Bowler}, {Magee}, {McCracken}, {Milvang-Jensen}, {Moneti}, \& {Targett}}]{Donnan2023}
{Donnan}, C.~T., {McLeod}, D.~J., {Dunlop}, J.~S., {et~al.} 2023, \mnras, 518, 6011, \dodoi{10.1093/mnras/stac3472}

\bibitem[{{Donnan} {et~al.}(2024){Donnan}, {McLure}, {Dunlop}, {McLeod}, {Magee}, {Arellano-C{\'o}rdova}, {Barrufet}, {Begley}, {Bowler}, {Carnall}, {Cullen}, {Ellis}, {Fontana}, {Illingworth}, {Grogin}, {Hamadouche}, {Koekemoer}, {Liu}, {Mason}, {Santini}, \& {Stanton}}]{Donnan2024}
{Donnan}, C.~T., {McLure}, R.~J., {Dunlop}, J.~S., {et~al.} 2024, \mnras, 533, 3222, \dodoi{10.1093/mnras/stae2037}

\bibitem[{{Drakos} {et~al.}(2022){Drakos}, {Villasenor}, {Robertson}, {Hausen}, {Dickinson}, {Ferguson}, {Furlanetto}, {Greene}, {Madau}, {Shapley}, {Stark}, \& {Wechsler}}]{Drakos2022}
{Drakos}, N.~E., {Villasenor}, B., {Robertson}, B.~E., {et~al.} 2022, \apj, 926, 194, \dodoi{10.3847/1538-4357/ac46fb}

\bibitem[{{Eisenstein} {et~al.}(2023){Eisenstein}, {Willott}, {Alberts}, {Arribas}, {Bonaventura}, {Bunker}, {Cameron}, {Carniani}, {Charlot}, {Curtis-Lake}, {D'Eugenio}, {Endsley}, {Ferruit}, {Giardino}, {Hainline}, {Hausen}, {Jakobsen}, {Johnson}, {Maiolino}, {Rieke}, {Rieke}, {Rix}, {Robertson}, {Stark}, {Tacchella}, {Williams}, {Willmer}, {Baker}, {Baum}, {Bhatawdekar}, {Boyett}, {Chen}, {Chevallard}, {Circosta}, {Curti}, {Danhaive}, {DeCoursey}, {de Graaff}, {Dressler}, {Egami}, {Helton}, {Hviding}, {Ji}, {Jones}, {Kumari}, {L{\"u}tzgendorf}, {Laseter}, {Looser}, {Lyu}, {Maseda}, {Nelson}, {Parlanti}, {Perna}, {Pusk{\'a}s}, {Rawle}, {Rodr{\'\i}guez Del Pino}, {Sandles}, {Saxena}, {Scholtz}, {Sharpe}, {Shivaei}, {Silcock}, {Simmonds}, {Skarbinski}, {Smit}, {Stone}, {Suess}, {Sun}, {Tang}, {Topping}, {{\"U}bler}, {Villanueva}, {Wallace}, {Whitler}, {Witstok}, \& {Woodrum}}]{Eisenstein2023}
{Eisenstein}, D.~J., {Willott}, C., {Alberts}, S., {et~al.} 2023, arXiv e-prints, arXiv:2306.02465, \dodoi{10.48550/arXiv.2306.02465}

\bibitem[{{Faisst} {et~al.}(2017){Faisst}, {Capak}, {Yan}, {Pavesi}, {Riechers}, {Bari{\v{s}}i{\'c}}, {Cooke}, {Kartaltepe}, \& {Masters}}]{Faisst2017}
{Faisst}, A.~L., {Capak}, P.~L., {Yan}, L., {et~al.} 2017, \apj, 847, 21, \dodoi{10.3847/1538-4357/aa886c}

\bibitem[{{Ferland} {et~al.}(2017){Ferland}, {Chatzikos}, {Guzm{\'a}n}, {Lykins}, {van Hoof}, {Williams}, {Abel}, {Badnell}, {Keenan}, {Porter}, \& {Stancil}}]{Ferland2017}
{Ferland}, G.~J., {Chatzikos}, M., {Guzm{\'a}n}, F., {et~al.} 2017, \rmxaa, 53, 385, \dodoi{10.48550/arXiv.1705.10877}

\bibitem[{{Ferrara}(2024)}]{Ferrara2024}
{Ferrara}, A. 2024, \aap, 684, A207, \dodoi{10.1051/0004-6361/202348321}

\bibitem[{{Ferrara} {et~al.}(2023){Ferrara}, {Pallottini}, \& {Dayal}}]{Ferrara2023}
{Ferrara}, A., {Pallottini}, A., \& {Dayal}, P. 2023, \mnras, 522, 3986, \dodoi{10.1093/mnras/stad1095}

\bibitem[{{Ferrara} {et~al.}(2025){Ferrara}, {Pallottini}, \& {Sommovigo}}]{Ferrara2025}
{Ferrara}, A., {Pallottini}, A., \& {Sommovigo}, L. 2025, \aap, 694, A286, \dodoi{10.1051/0004-6361/202452707}

\bibitem[{{Ferrara} {et~al.}(2022){Ferrara}, {Sommovigo}, {Dayal}, {Pallottini}, {Bouwens}, {Gonzalez}, {Inami}, {Smit}, {Bowler}, {Endsley}, {Oesch}, {Schouws}, {Stark}, {Stefanon}, {Aravena}, {da Cunha}, {De Looze}, {Fudamoto}, {Graziani}, {Hodge}, {Riechers}, {Schneider}, {Algera}, {Barrufet}, {Hygate}, {Labb{\'e}}, {Li}, {Nanayakkara}, {Topping}, \& {van der Werf}}]{Ferrara2022}
{Ferrara}, A., {Sommovigo}, L., {Dayal}, P., {et~al.} 2022, \mnras, 512, 58, \dodoi{10.1093/mnras/stac460}

\bibitem[{{Finkelstein}(2016)}]{Finkelstein2016}
{Finkelstein}, S.~L. 2016, \pasa, 33, e037, \dodoi{10.1017/pasa.2016.26}

\bibitem[{{Finkelstein} {et~al.}(2012){Finkelstein}, {Papovich}, {Ryan}, {Pawlik}, {Dickinson}, {Ferguson}, {Finlator}, {Koekemoer}, {Giavalisco}, {Cooray}, {Dunlop}, {Faber}, {Grogin}, {Kocevski}, \& {Newman}}]{Finkelstein2012}
{Finkelstein}, S.~L., {Papovich}, C., {Ryan}, R.~E., {et~al.} 2012, \apj, 758, 93, \dodoi{10.1088/0004-637X/758/2/93}

\bibitem[{{Finkelstein} {et~al.}(2019){Finkelstein}, {D'Aloisio}, {Paardekooper}, {Ryan}, {Behroozi}, {Finlator}, {Livermore}, {Upton Sanderbeck}, {Dalla Vecchia}, \& {Khochfar}}]{finkelstein2019}
{Finkelstein}, S.~L., {D'Aloisio}, A., {Paardekooper}, J.-P., {et~al.} 2019, \apj, 879, 36, \dodoi{10.3847/1538-4357/ab1ea8}

\bibitem[{{Finkelstein} {et~al.}(2022{\natexlab{a}}){Finkelstein}, {Bagley}, {Arrabal Haro}, {Dickinson}, {Ferguson}, {Kartaltepe}, {Papovich}, {Burgarella}, {Kocevski}, {Huertas-Company}, {Iyer}, {Koekemoer}, {Larson}, {P{\'e}rez-Gonz{\'a}lez}, {Rose}, {Tacchella}, {Wilkins}, {Chworowsky}, {Medrano}, {Morales}, {Somerville}, {Yung}, {Fontana}, {Giavalisco}, {Grazian}, {Grogin}, {Kewley}, {Kirkpatrick}, {Kurczynski}, {Lotz}, {Pentericci}, {Pirzkal}, {Ravindranath}, {Ryan}, {Trump}, {Yang}, {Almaini}, {Amor{\'\i}n}, {Annunziatella}, {Backhaus}, {Barro}, {Behroozi}, {Bell}, {Bhatawdekar}, {Bisigello}, {Bromm}, {Buat}, {Buitrago}, {Calabr{\`o}}, {Casey}, {Castellano}, {Ch{\'a}vez Ortiz}, {Ciesla}, {Cleri}, {Cohen}, {Cole}, {Cooke}, {Cooper}, {Cooray}, {Costantin}, {Cox}, {Croton}, {Daddi}, {Dav{\'e}}, {de La Vega}, {Dekel}, {Elbaz}, {Estrada-Carpenter}, {Faber}, {Fern{\'a}ndez}, {Finkelstein}, {Freundlich}, {Fujimoto}, {Garc{\'\i}a-Argum{\'a}nez}, {Gardner}, {Gawiser}, {G{\'o}mez-Guijarro}, {Guo}, {Hamblin},
  {Hamilton}, {Hathi}, {Holwerda}, {Hirschmann}, {Hutchison}, {Jaskot}, {Jha}, {Jogee}, {Juneau}, {Jung}, {Kassin}, {Le Bail}, {Leung}, {Lucas}, {Magnelli}, {Mantha}, {Matharu}, {McGrath}, {McIntosh}, {Merlin}, {Mobasher}, {Newman}, {Nicholls}, {Pandya}, {Rafelski}, {Ronayne}, {Santini}, {Seill{\'e}}, {Shah}, {Shen}, {Simons}, {Snyder}, {Stanway}, {Straughn}, {Teplitz}, {Vanderhoof}, {Vega-Ferrero}, {Wang}, {Weiner}, {Willmer}, {Wuyts}, {Zavala}, \& {Ceers Team}}]{finkelstein2022}
{Finkelstein}, S.~L., {Bagley}, M.~B., {Arrabal Haro}, P., {et~al.} 2022{\natexlab{a}}, \apjl, 940, L55, \dodoi{10.3847/2041-8213/ac966e}

\bibitem[{{Finkelstein} {et~al.}(2022{\natexlab{b}}){Finkelstein}, {Bagley}, {Song}, {Larson}, {Papovich}, {Dickinson}, {Finkelstein}, {Koekemoer}, {Pirzkal}, {Somerville}, {Yung}, {Behroozi}, {Ferguson}, {Giavalisco}, {Grogin}, {Hathi}, {Hutchison}, {Jung}, {Kocevski}, {Kawinwanichakij}, {Rojas-Ruiz}, {Ryan}, {Snyder}, \& {Tacchella}}]{finkelstein2022b}
{Finkelstein}, S.~L., {Bagley}, M., {Song}, M., {et~al.} 2022{\natexlab{b}}, \apj, 928, 52, \dodoi{10.3847/1538-4357/ac3aed}

\bibitem[{{Finkelstein} {et~al.}(2023){Finkelstein}, {Bagley}, {Ferguson}, {Wilkins}, {Kartaltepe}, {Papovich}, {Yung}, {Arrabal Haro}, {Behroozi}, {Dickinson}, {Kocevski}, {Koekemoer}, {Larson}, {Le Bail}, {Morales}, {P{\'e}rez-Gonz{\'a}lez}, {Burgarella}, {Dav{\'e}}, {Hirschmann}, {Somerville}, {Wuyts}, {Bromm}, {Casey}, {Fontana}, {Fujimoto}, {Gardner}, {Giavalisco}, {Grazian}, {Grogin}, {Hathi}, {Hutchison}, {Jha}, {Jogee}, {Kewley}, {Kirkpatrick}, {Long}, {Lotz}, {Pentericci}, {Pierel}, {Pirzkal}, {Ravindranath}, {Ryan}, {Trump}, {Yang}, {Bhatawdekar}, {Bisigello}, {Buat}, {Calabr{\`o}}, {Castellano}, {Cleri}, {Cooper}, {Croton}, {Daddi}, {Dekel}, {Elbaz}, {Franco}, {Gawiser}, {Holwerda}, {Huertas-Company}, {Jaskot}, {Leung}, {Lucas}, {Mobasher}, {Pandya}, {Tacchella}, {Weiner}, \& {Zavala}}]{Finkelstein2023}
{Finkelstein}, S.~L., {Bagley}, M.~B., {Ferguson}, H.~C., {et~al.} 2023, \apjl, 946, L13, \dodoi{10.3847/2041-8213/acade4}

\bibitem[{{Finkelstein} {et~al.}(2024){Finkelstein}, {Leung}, {Bagley}, {Dickinson}, {Ferguson}, {Papovich}, {Akins}, {Arrabal Haro}, {Dav{\'e}}, {Dekel}, {Kartaltepe}, {Kocevski}, {Koekemoer}, {Pirzkal}, {Somerville}, {Yung}, {Amor{\'\i}n}, {Backhaus}, {Behroozi}, {Bisigello}, {Bromm}, {Casey}, {Ch{\'a}vez Ortiz}, {Cheng}, {Chworowsky}, {Cleri}, {Cooper}, {Davis}, {de la Vega}, {Elbaz}, {Franco}, {Fontana}, {Fujimoto}, {Giavalisco}, {Grogin}, {Holwerda}, {Huertas-Company}, {Hirschmann}, {Iyer}, {Jogee}, {Jung}, {Larson}, {Lucas}, {Mobasher}, {Morales}, {Morley}, {Mukherjee}, {P{\'e}rez-Gonz{\'a}lez}, {Ravindranath}, {Rodighiero}, {Rowland}, {Tacchella}, {Taylor}, {Trump}, \& {Wilkins}}]{Finkelstein2024}
{Finkelstein}, S.~L., {Leung}, G. C.~K., {Bagley}, M.~B., {et~al.} 2024, \apjl, 969, L2, \dodoi{10.3847/2041-8213/ad4495}

\bibitem[{{Fiore} {et~al.}(2023){Fiore}, {Ferrara}, {Bischetti}, {Feruglio}, \& {Travascio}}]{Fiore2023}
{Fiore}, F., {Ferrara}, A., {Bischetti}, M., {Feruglio}, C., \& {Travascio}, A. 2023, \apjl, 943, L27, \dodoi{10.3847/2041-8213/acb5f2}

\bibitem[{{Foreman-Mackey} {et~al.}(2013){Foreman-Mackey}, {Hogg}, {Lang}, \& {Goodman}}]{Foreman2013}
{Foreman-Mackey}, D., {Hogg}, D.~W., {Lang}, D., \& {Goodman}, J. 2013, \pasp, 125, 306, \dodoi{10.1086/670067}

\bibitem[{{Franco} {et~al.}(2024){Franco}, {Akins}, {Casey}, {Finkelstein}, {Shuntov}, {Chworowsky}, {Faisst}, {Fujimoto}, {Ilbert}, {Koekemoer}, {Liu}, {Lovell}, {Maraston}, {McCracken}, {McKinney}, {Robertson}, {Bagley}, {Champagne}, {Cooper}, {Ding}, {Drakos}, {Enia}, {Gillman}, {Gozaliasl}, {Harish}, {Hayward}, {Hirschmann}, {Jin}, {Kartaltepe}, {Kokorev}, {Laigle}, {Long}, {Magdis}, {Mahler}, {Martin}, {Massey}, {Mobasher}, {Paquereau}, {Renzini}, {Rhodes}, {Rich}, {Sheth}, {Silverman}, {Sparre}, {Talia}, {Trakhtenbrot}, {Valentino}, {Vijayan}, {Wilkins}, {Yang}, \& {Zavala}}]{Franco2024}
{Franco}, M., {Akins}, H.~B., {Casey}, C.~M., {et~al.} 2024, \apj, 973, 23, \dodoi{10.3847/1538-4357/ad5e6a}

\bibitem[{{Franco} {et~al.}(2025){Franco}, {Casey}, {Koekemoer}, {Liu}, {Bagley}, {McCracken}, {Kartaltepe}, {Akins}, {Ilbert}, {Shuntov}, {Harish}, {Robertson}, {Arango-Toro}, {Battisti}, {Chartab}, {Drakos}, {Faisst}, {Flayhart}, {Gozaliasl}, {Hirschmann}, {Massey}, {Rhodes}, {Sattari}, {Scognamiglio}, {Weaver}, {Yang}, {Zavala}, {Berman}, {Gentile}, {Gillman}, {Long}, {Magdis}, {McCleary}, {McKinney}, {Mobasher}, {Paquereau}, {Rest}, {Sanders}, {Toft}, \& {Yu}}]{Franco2025}
{Franco}, M., {Casey}, C.~M., {Koekemoer}, A.~M., {et~al.} 2025, arXiv e-prints, arXiv:2506.03256.
\newblock \doarXiv{2506.03256}

\bibitem[{{Furtak} {et~al.}(2023){Furtak}, {Zitrin}, {Plat}, {Fujimoto}, {Wang}, {Nelson}, {Labb{\'e}}, {Bezanson}, {Brammer}, {van Dokkum}, {Endsley}, {Glazebrook}, {Greene}, {Leja}, {Price}, {Smit}, {Stark}, {Weaver}, {Whitaker}, {Atek}, {Chevallard}, {Curtis-Lake}, {Dayal}, {Feltre}, {Franx}, {Fudamoto}, {Marchesini}, {Mowla}, {Pan}, {Suess}, {Vidal-Garc{\'\i}a}, \& {Williams}}]{Furtak2023}
{Furtak}, L.~J., {Zitrin}, A., {Plat}, A., {et~al.} 2023, \apj, 952, 142, \dodoi{10.3847/1538-4357/acdc9d}

\bibitem[{{Gaia Collaboration} {et~al.}(2023){Gaia Collaboration}, {Vallenari}, {Brown}, {Prusti}, {de Bruijne}, {Arenou}, {Babusiaux}, {Biermann}, {Creevey}, {Ducourant}, {Evans}, {Eyer}, {Guerra}, {Hutton}, {Jordi}, {Klioner}, {Lammers}, {Lindegren}, {Luri}, {Mignard}, {Panem}, {Pourbaix}, {Randich}, {Sartoretti}, {Soubiran}, {Tanga}, {Walton}, {Bailer-Jones}, {Bastian}, {Drimmel}, {Jansen}, {Katz}, {Lattanzi}, {van Leeuwen}, {Bakker}, {Cacciari}, {Casta{\~n}eda}, {De Angeli}, {Fabricius}, {Fouesneau}, {Fr{\'e}mat}, {Galluccio}, {Guerrier}, {Heiter}, {Masana}, {Messineo}, {Mowlavi}, {Nicolas}, {Nienartowicz}, {Pailler}, {Panuzzo}, {Riclet}, {Roux}, {Seabroke}, {Sordo}, {Th{\'e}venin}, {Gracia-Abril}, {Portell}, {Teyssier}, {Altmann}, {Andrae}, {Audard}, {Bellas-Velidis}, {Benson}, {Berthier}, {Blomme}, {Burgess}, {Busonero}, {Busso}, {C{\'a}novas}, {Carry}, {Cellino}, {Cheek}, {Clementini}, {Damerdji}, {Davidson}, {de Teodoro}, {Nu{\~n}ez Campos}, {Delchambre}, {Dell'Oro}, {Esquej},
  {Fern{\'a}ndez-Hern{\'a}ndez}, {Fraile}, {Garabato}, {Garc{\'\i}a-Lario}, {Gosset}, {Haigron}, {Halbwachs}, {Hambly}, {Harrison}, {Hern{\'a}ndez}, {Hestroffer}, {Hodgkin}, {Holl}, {Jan{\ss}en}, {Jevardat de Fombelle}, {Jordan}, {Krone-Martins}, {Lanzafame}, {L{\"o}ffler}, {Marchal}, {Marrese}, {Moitinho}, {Muinonen}, {Osborne}, {Pancino}, {Pauwels}, {Recio-Blanco}, {Reyl{\'e}}, {Riello}, {Rimoldini}, {Roegiers}, {Rybizki}, {Sarro}, {Siopis}, {Smith}, {Sozzetti}, {Utrilla}, {van Leeuwen}, {Abbas}, {{\'A}brah{\'a}m}, {Abreu Aramburu}, {Aerts}, {Aguado}, {Ajaj}, {Aldea-Montero}, {Altavilla}, {{\'A}lvarez}, {Alves}, {Anders}, {Anderson}, {Anglada Varela}, {Antoja}, {Baines}, {Baker}, {Balaguer-N{\'u}{\~n}ez}, {Balbinot}, {Balog}, {Barache}, {Barbato}, {Barros}, {Barstow}, {Bartolom{\'e}}, {Bassilana}, {Bauchet}, {Becciani}, {Bellazzini}, {Berihuete}, {Bernet}, {Bertone}, {Bianchi}, {Binnenfeld}, {Blanco-Cuaresma}, {Blazere}, {Boch}, {Bombrun}, {Bossini}, {Bouquillon}, {Bragaglia}, {Bramante}, {Breedt},
  {Bressan}, {Brouillet}, {Brugaletta}, {Bucciarelli}, {Burlacu}, {Butkevich}, {Buzzi}, {Caffau}, {Cancelliere}, {Cantat-Gaudin}, {Carballo}, {Carlucci}, {Carnerero}, {Carrasco}, {Casamiquela}, {Castellani}, {Castro-Ginard}, {Chaoul}, {Charlot}, {Chemin}, {Chiaramida}, {Chiavassa}, {Chornay}, {Comoretto}, {Contursi}, {Cooper}, {Cornez}, {Cowell}, {Crifo}, {Cropper}, {Crosta}, {Crowley}, {Dafonte}, {Dapergolas}, {David}, {David}, {de Laverny}, {De Luise}, \& {De March}}]{Gaia_Collaboration2023}
{Gaia Collaboration}, {Vallenari}, A., {Brown}, A.~G.~A., {et~al.} 2023, \aap, 674, A1, \dodoi{10.1051/0004-6361/202243940}

\bibitem[{{Gandolfi} {et~al.}(2025){Gandolfi}, {Rodighiero}, {Bisigello}, {Grazian}, {Finkelstein}, {Dickinson}, {Castellano}, {Merlin}, {Calabr{\`o}}, {Papovich}, {Bianchetti}, {Ba{\~n}ados}, {Benotto}, {Buitrago}, {Daddi}, {Girardi}, {Giulietti}, {Hirschmann}, {Holwerda}, {Arrabal Haro}, {Lapi}, {Lucas}, {Lyu}, {Massardi}, {Pacucci}, {P{\'e}rez-Gonz{\'a}lez}, {Ronconi}, {Tarrasse}, {Wilkins}, {Vulcani}, {Yung}, {Zavala}, {Backhaus}, {Bagley}, {Buat}, {Burgarella}, {Kartaltepe}, {Khusanova}, {Kirkpatrick}, {Kocevski}, {Koekemoer}, {Lambrides}, {Pirzkal}, \& {Yang}}]{Gandolfi2025}
{Gandolfi}, G., {Rodighiero}, G., {Bisigello}, L., {et~al.} 2025, arXiv e-prints, arXiv:2502.02637, \dodoi{10.48550/arXiv.2502.02637}

\bibitem[{{Garaldi} {et~al.}(2022){Garaldi}, {Kannan}, {Smith}, {Springel}, {Pakmor}, {Vogelsberger}, \& {Hernquist}}]{garaldi22}
{Garaldi}, E., {Kannan}, R., {Smith}, A., {et~al.} 2022, \mnras, 512, 4909, \dodoi{10.1093/mnras/stac257}

\bibitem[{{Gelli} {et~al.}(2024){Gelli}, {Mason}, \& {Hayward}}]{Gelli2024}
{Gelli}, V., {Mason}, C., \& {Hayward}, C.~C. 2024, \apj, 975, 192, \dodoi{10.3847/1538-4357/ad7b36}

\bibitem[{{Grazian} {et~al.}(2018){Grazian}, {Giallongo}, {Boutsia}, {Cristiani}, {Vanzella}, {Scarlata}, {Santini}, {Pentericci}, {Merlin}, {Menci}, {Fontanot}, {Fontana}, {Fiore}, {Civano}, {Castellano}, {Brusa}, {Bonchi}, {Carini}, {Cusano}, {Faccini}, {Garilli}, {Marchetti}, {Rossi}, \& {Speziali}}]{Grazian2018}
{Grazian}, A., {Giallongo}, E., {Boutsia}, K., {et~al.} 2018, \aap, 613, A44, \dodoi{10.1051/0004-6361/201732385}

\bibitem[{{Grazian} {et~al.}(2024){Grazian}, {Giallongo}, {Boutsia}, {Cristiani}, {Fontanot}, {Bischetti}, {Bisigello}, {Bongiorno}, {Calderone}, {Chiti Tegli}, {Cupani}, {De Lucia}, {D'Odorico}, {Feruglio}, {Fiore}, {Gandolfi}, {Girardi}, {Guarneri}, {Hirschmann}, {Porru}, {Rodighiero}, {Saccheo}, {Simioni}, {Trost}, \& {Viitanen}}]{Grazian2024}
---. 2024, \apj, 974, 84, \dodoi{10.3847/1538-4357/ad6980}

\bibitem[{Gumbel(1958)}]{gumbel58}
Gumbel, E.~J. 1958, Statistics of Extremes (New York Chichester, West Sussex: Columbia University Press), \dodoi{doi:10.7312/gumb92958}

\bibitem[{{Hainline} {et~al.}(2024){Hainline}, {Helton}, {Johnson}, {Sun}, {Topping}, {Leisenring}, {Baker}, {Eisenstein}, {Hausen}, {Hviding}, {Lyu}, {Robertson}, {Tacchella}, {Williams}, {Willmer}, \& {Roellig}}]{Hainline2024}
{Hainline}, K.~N., {Helton}, J.~M., {Johnson}, B.~D., {et~al.} 2024, \apj, 964, 66, \dodoi{10.3847/1538-4357/ad20d1}

\bibitem[{{Harikane} {et~al.}(2024){Harikane}, {Nakajima}, {Ouchi}, {Umeda}, {Isobe}, {Ono}, {Xu}, \& {Zhang}}]{Harikane2024}
{Harikane}, Y., {Nakajima}, K., {Ouchi}, M., {et~al.} 2024, \apj, 960, 56, \dodoi{10.3847/1538-4357/ad0b7e}

\bibitem[{{Harikane} {et~al.}(2023){Harikane}, {Ouchi}, {Oguri}, {Ono}, {Nakajima}, {Isobe}, {Umeda}, {Mawatari}, \& {Zhang}}]{Harikane2023}
{Harikane}, Y., {Ouchi}, M., {Oguri}, M., {et~al.} 2023, \apjs, 265, 5, \dodoi{10.3847/1538-4365/acaaa9}

\bibitem[{{Harikane} {et~al.}(2025){Harikane}, {Inoue}, {Ellis}, {Ouchi}, {Nakazato}, {Yoshida}, {Ono}, {Sun}, {Sato}, {Ferrami}, {Fujimoto}, {Kashikawa}, {McLeod}, {P{\'e}rez-Gonz{\'a}lez}, {Sawicki}, {Sugahara}, {Xu}, {Yamanaka}, {Carnall}, {Cullen}, {Dunlop}, {Egami}, {Grogin}, {Isobe}, {Koekemoer}, {Laporte}, {Lee}, {Magee}, {Matsuo}, {Matsuoka}, {Mawatari}, {Nakajima}, {Nakane}, {Tamura}, {Umeda}, \& {Yanagisawa}}]{Harikane2025}
{Harikane}, Y., {Inoue}, A.~K., {Ellis}, R.~S., {et~al.} 2025, \apj, 980, 138, \dodoi{10.3847/1538-4357/ad9b2c}

\bibitem[{{Harish} {et~al.}(2025){Harish}, {Kartaltepe}, {Liu}, {Koekemoer}, {Casey}, {Franco}, {Akins}, {Ilbert}, {Shuntov}, {Drakos}, {Engesser}, {Faisst}, {Gozaliasl}, {Martin}, {Hirschmann}, {Kokorev}, {Lambrides}, {McCracken}, {McKinney}, {Paquereau}, {Rhodes}, \& {Robertson}}]{Harish2025}
{Harish}, S., {Kartaltepe}, J.~S., {Liu}, D., {et~al.} 2025, arXiv e-prints, arXiv:2506.03306.
\newblock \doarXiv{2506.03306}

\bibitem[{{Harrison} \& {Coles}(2011)}]{harrison11}
{Harrison}, I., \& {Coles}, P. 2011, \mnras, 418, L20, \dodoi{10.1111/j.1745-3933.2011.01134.x}

\bibitem[{{Haslbauer} {et~al.}(2022){Haslbauer}, {Kroupa}, {Zonoozi}, \& {Haghi}}]{Haslbauer2022}
{Haslbauer}, M., {Kroupa}, P., {Zonoozi}, A.~H., \& {Haghi}, H. 2022, \apjl, 939, L31, \dodoi{10.3847/2041-8213/ac9a50}

\bibitem[{{Heintz} {et~al.}(2024){Heintz}, {Watson}, {Brammer}, {Vejlgaard}, {Hutter}, {Strait}, {Matthee}, {Oesch}, {Jakobsson}, {Tanvir}, {Laursen}, {Naidu}, {Mason}, {Killi}, {Jung}, {Hsiao}, {Abdurro'uf}, {Coe}, {Arrabal Haro}, {Finkelstein}, \& {Toft}}]{Heintz2024}
{Heintz}, K.~E., {Watson}, D., {Brammer}, G., {et~al.} 2024, Science, 384, 890, \dodoi{10.1126/science.adj0343}

\bibitem[{{Hutter} {et~al.}(2024){Hutter}, {Cueto}, {Dayal}, {Gottl{\"o}ber}, {Trebitsch}, \& {Yepes}}]{Hutter2024}
{Hutter}, A., {Cueto}, E.~R., {Dayal}, P., {et~al.} 2024, arXiv e-prints, arXiv:2410.00730, \dodoi{10.48550/arXiv.2410.00730}

\bibitem[{{Ilbert} {et~al.}(2006){Ilbert}, {Arnouts}, {McCracken}, {Bolzonella}, {Bertin}, {Le F{\`e}vre}, {Mellier}, {Zamorani}, {Pell{\`o}}, {Iovino}, {Tresse}, {Le Brun}, {Bottini}, {Garilli}, {Maccagni}, {Picat}, {Scaramella}, {Scodeggio}, {Vettolani}, {Zanichelli}, {Adami}, {Bardelli}, {Cappi}, {Charlot}, {Ciliegi}, {Contini}, {Cucciati}, {Foucaud}, {Franzetti}, {Gavignaud}, {Guzzo}, {Marano}, {Marinoni}, {Mazure}, {Meneux}, {Merighi}, {Paltani}, {Pollo}, {Pozzetti}, {Radovich}, {Zucca}, {Bondi}, {Bongiorno}, {Busarello}, {de La Torre}, {Gregorini}, {Lamareille}, {Mathez}, {Merluzzi}, {Ripepi}, {Rizzo}, \& {Vergani}}]{Ilbert2006}
{Ilbert}, O., {Arnouts}, S., {McCracken}, H.~J., {et~al.} 2006, \aap, 457, 841, \dodoi{10.1051/0004-6361:20065138}

\bibitem[{{Inayoshi} {et~al.}(2022){Inayoshi}, {Harikane}, {Inoue}, {Li}, \& {Ho}}]{Inayoshi2022c}
{Inayoshi}, K., {Harikane}, Y., {Inoue}, A.~K., {Li}, W., \& {Ho}, L.~C. 2022, \apjl, 938, L10, \dodoi{10.3847/2041-8213/ac9310}

\bibitem[{{Jeong} {et~al.}(2025){Jeong}, {Jeon}, {Song}, \& {Bromm}}]{Jeong2025}
{Jeong}, T.~B., {Jeon}, M., {Song}, H., \& {Bromm}, V. 2025, \apj, 980, 10, \dodoi{10.3847/1538-4357/ada27d}

\bibitem[{{Kakiichi} {et~al.}(2024){Kakiichi}, {Egami}, {Fan}, {Lyu}, {Wang}, {Yang}, {Bechtel}, {Behroozi}, {Bosman}, {Cai}, {Champagne}, {Davies}, {De Rosa}, {Decarli}, {Eilers}, {Ellis}, {Endsley}, {Farina}, {Finkelstein}, {Fujimoto}, {Hennawi}, {Inoue}, {Jiang}, {Jin}, {Khusanova}, {Kirkpatrick}, {Kocevski}, {Kulkarni}, {Lee}, {Liu}, {Meyer}, {Ono}, {Onoue}, {Ouchi}, {Papovich}, {Satyavolu}, {Schindler}, {Sun}, {Tee}, {Vestergaard}, {Zhang}, \& {Zou}}]{Kakiichi2024}
{Kakiichi}, K., {Egami}, E., {Fan}, X., {et~al.} 2024, {COSMOS-3D: A Legacy Spectroscopic/Imaging Survey of the Early Universe}, JWST Proposal. Cycle 3, ID. \#5893

\bibitem[{Kannan {et~al.}(2021)Kannan, Garaldi, Smith, Pakmor, Springel, Vogelsberger, \& Hernquist}]{kannan22a}
Kannan, R., Garaldi, E., Smith, A., {et~al.} 2021, Monthly Notices of the Royal Astronomical Society, 511, 4005, \dodoi{10.1093/mnras/stab3710}

\bibitem[{{Kannan} {et~al.}(2022){Kannan}, {Smith}, {Garaldi}, {Shen}, {Vogelsberger}, {Pakmor}, {Springel}, \& {Hernquist}}]{kannan22}
{Kannan}, R., {Smith}, A., {Garaldi}, E., {et~al.} 2022, \mnras, 514, 3857, \dodoi{10.1093/mnras/stac1557}

\bibitem[{{Kimm} {et~al.}(2017){Kimm}, {Katz}, {Haehnelt}, {Rosdahl}, {Devriendt}, \& {Slyz}}]{Kimm2017}
{Kimm}, T., {Katz}, H., {Haehnelt}, M., {et~al.} 2017, \mnras, 466, 4826, \dodoi{10.1093/mnras/stx052}

\bibitem[{{Kocevski} {et~al.}(2023){Kocevski}, {Onoue}, {Inayoshi}, {Trump}, {Arrabal Haro}, {Grazian}, {Dickinson}, {Finkelstein}, {Kartaltepe}, {Hirschmann}, {Aird}, {Holwerda}, {Fujimoto}, {Juneau}, {Amor{\'\i}n}, {Backhaus}, {Bagley}, {Barro}, {Bell}, {Bisigello}, {Calabr{\`o}}, {Cleri}, {Cooper}, {Ding}, {Grogin}, {Ho}, {Hutchison}, {Inoue}, {Jiang}, {Jones}, {Koekemoer}, {Li}, {Li}, {McGrath}, {Molina}, {Papovich}, {P{\'e}rez-Gonz{\'a}lez}, {Pirzkal}, {Wilkins}, {Yang}, \& {Yung}}]{Kocevski2023}
{Kocevski}, D.~D., {Onoue}, M., {Inayoshi}, K., {et~al.} 2023, \apjl, 954, L4, \dodoi{10.3847/2041-8213/ace5a0}

\bibitem[{{Koekemoer} {et~al.}(2007){Koekemoer}, {Aussel}, {Calzetti}, {Capak}, {Giavalisco}, {Kneib}, {Leauthaud}, {Le F{\`e}vre}, {McCracken}, {Massey}, {Mobasher}, {Rhodes}, {Scoville}, \& {Shopbell}}]{Koekemoer2007}
{Koekemoer}, A.~M., {Aussel}, H., {Calzetti}, D., {et~al.} 2007, \apjs, 172, 196, \dodoi{10.1086/520086}

\bibitem[{{Kokorev} {et~al.}(2025{\natexlab{a}}){Kokorev}, {Ch{\'a}vez Ortiz}, {Taylor}, {Finkelstein}, {Arrabal Haro}, {Dickinson}, {Chisholm}, {Fujimoto}, {Mu{\~n}oz}, {Endsley}, {Hu}, {Napolitano}, {Wilkins}, {Akins}, {Amori{\'\i}n}, {Casey}, {Cheng}, {Cleri}, {Cole}, {Cullen}, {Daddi}, {Davis}, {Donnan}, {Dunlop}, {Fern{\'a}ndez}, {Giavalisco}, {Grogin}, {Hathi}, {Hirschmann}, {Kartaltepe}, {Koekemoer}, {Leung}, {Lucas}, {McLeod}, {Papovich}, {Pentericci}, {P{\'e}rez-Gonz{\'a}lez}, {Somerville}, {Wang}, {Yung}, \& {Zavala}}]{Kokorev2025b}
{Kokorev}, V., {Ch{\'a}vez Ortiz}, {\'O}.~A., {Taylor}, A.~J., {et~al.} 2025{\natexlab{a}}, arXiv e-prints, arXiv:2504.12504, \dodoi{10.48550/arXiv.2504.12504}

\bibitem[{{Kokorev} {et~al.}(2025{\natexlab{b}}){Kokorev}, {Atek}, {Chisholm}, {Endsley}, {Chemerynska}, {Mu{\~n}oz}, {Furtak}, {Pan}, {Berg}, {Fujimoto}, {Oesch}, {Weibel}, {Adamo}, {Blaizot}, {Bouwens}, {Dessauges-Zavadsky}, {Khullar}, {Korber}, {Goovaerts}, {Jecmen}, {Labb{\'e}}, {Leclercq}, {Marques-Chaves}, {Mason}, {McQuinn}, {Naidu}, {Natarajan}, {Nelson}, {Rosdahl}, {Saldana-Lopez}, {Schaerer}, {Trebitsch}, {Volonteri}, \& {Zitrin}}]{Kokorev2025}
{Kokorev}, V., {Atek}, H., {Chisholm}, J., {et~al.} 2025{\natexlab{b}}, \apjl, 983, L22, \dodoi{10.3847/2041-8213/adc458}

\bibitem[{Kotz \& Nadarajah(2000)}]{kotz00}
Kotz, S., \& Nadarajah, S. 2000, Extreme Value Distributions (PUBLISHED BY IMPERIAL COLLEGE PRESS AND DISTRIBUTED BY WORLD SCIENTIFIC PUBLISHING CO.), \dodoi{10.1142/p191}

\bibitem[{{Kravtsov} \& {Belokurov}(2024)}]{Kravtsov2024}
{Kravtsov}, A., \& {Belokurov}, V. 2024, arXiv e-prints, arXiv:2405.04578, \dodoi{10.48550/arXiv.2405.04578}

\bibitem[{{K{\"u}mmel} {et~al.}(2020){K{\"u}mmel}, {Bertin}, {Schefer}, {Apostolakos}, {{\'A}lvarez-Ayll{\'o}n}, \& {Dubath}}]{Kummel2020}
{K{\"u}mmel}, M., {Bertin}, E., {Schefer}, M., {et~al.} 2020, in Astronomical Society of the Pacific Conference Series, Vol. 527, Astronomical Data Analysis Software and Systems XXIX, ed. R.~{Pizzo}, E.~R. {Deul}, J.~D. {Mol}, J.~{de Plaa}, \& H.~{Verkouter}, 29

\bibitem[{{Kümmel} {et~al.}(2022){Kümmel}, {{\'A}lvarez-Ayll{\'o}n}, {Bertin}, {Dubath}, {Gavazzi}, {Hartley}, \& {Schefer}}]{Kummel2022}
{Kümmel}, M., {{\'A}lvarez-Ayll{\'o}n}, A., {Bertin}, E., {et~al.} 2022, arXiv e-prints, arXiv:2212.02428, \dodoi{10.48550/arXiv.2212.02428}

\bibitem[{Labbé {et~al.}(2023)Labbé, Van~Dokkum, Nelson, Bezanson, Suess, Leja, Brammer, Whitaker, Mathews, Stefanon, \& Wang}]{Labbe2023}
Labbé, I., Van~Dokkum, P., Nelson, E., {et~al.} 2023, Nature, 616, 266, \dodoi{10.1038/s41586-023-05786-2}

\bibitem[{{Langeroodi} \& {Hjorth}(2023)}]{Langeroodi2023}
{Langeroodi}, D., \& {Hjorth}, J. 2023, \apjl, 957, L27, \dodoi{10.3847/2041-8213/acfeec}

\bibitem[{{Leethochawalit} {et~al.}(2023){Leethochawalit}, {Roberts-Borsani}, {Morishita}, {Trenti}, \& {Treu}}]{Leethochawalit2023}
{Leethochawalit}, N., {Roberts-Borsani}, G., {Morishita}, T., {Trenti}, M., \& {Treu}, T. 2023, \mnras, 524, 5454, \dodoi{10.1093/mnras/stad2202}

\bibitem[{{Leung} {et~al.}(2023){Leung}, {Bagley}, {Finkelstein}, {Ferguson}, {Koekemoer}, {P{\'e}rez-Gonz{\'a}lez}, {Morales}, {Kocevski}, {Yang}, {Somerville}, {Wilkins}, {Yung}, {Fujimoto}, {Larson}, {Papovich}, {Pirzkal}, {Berg}, {Lotz}, {Castellano}, {Ch{\'a}vez Ortiz}, {Cheng}, {Dickinson}, {Giavalisco}, {Hathi}, {Hutchison}, {Jung}, {Kartaltepe}, {Natarajan}, \& {Rothberg}}]{Leung2023}
{Leung}, G. C.~K., {Bagley}, M.~B., {Finkelstein}, S.~L., {et~al.} 2023, \apjl, 954, L46, \dodoi{10.3847/2041-8213/acf365}

\bibitem[{{Li} {et~al.}(2024){Li}, {Dekel}, {Sarkar}, {Aung}, {Giavalisco}, {Mandelker}, \& {Tacchella}}]{Li2024}
{Li}, Z., {Dekel}, A., {Sarkar}, K.~C., {et~al.} 2024, \aap, 690, A108, \dodoi{10.1051/0004-6361/202348727}

\bibitem[{{Livermore} {et~al.}(2017){Livermore}, {Finkelstein}, \& {Lotz}}]{Livermore2017}
{Livermore}, R.~C., {Finkelstein}, S.~L., \& {Lotz}, J.~M. 2017, \apj, 835, 113, \dodoi{10.3847/1538-4357/835/2/113}

\bibitem[{{Lovell} {et~al.}(2023){Lovell}, {Harrison}, {Harikane}, {Tacchella}, \& {Wilkins}}]{Lovell2023}
{Lovell}, C.~C., {Harrison}, I., {Harikane}, Y., {Tacchella}, S., \& {Wilkins}, S.~M. 2023, \mnras, 518, 2511, \dodoi{10.1093/mnras/stac3224}

\bibitem[{{Lovell} {et~al.}(2021){Lovell}, {Vijayan}, {Thomas}, {Wilkins}, {Barnes}, {Irodotou}, \& {Roper}}]{Lovell2021}
{Lovell}, C.~C., {Vijayan}, A.~P., {Thomas}, P.~A., {et~al.} 2021, \mnras, 500, 2127, \dodoi{10.1093/mnras/staa3360}

\bibitem[{{Ma} {et~al.}(2020){Ma}, {Quataert}, {Wetzel}, {Hopkins}, {Faucher-Gigu{\`e}re}, \& {Kere{\v{s}}}}]{Ma2020}
{Ma}, X., {Quataert}, E., {Wetzel}, A., {et~al.} 2020, \mnras, 498, 2001, \dodoi{10.1093/mnras/staa2404}

\bibitem[{{Madau} \& {Dickinson}(2014)}]{Madau2014}
{Madau}, P., \& {Dickinson}, M. 2014, \araa, 52, 415, \dodoi{10.1146/annurev-astro-081811-125615}

\bibitem[{{Maraston} {et~al.}(2006){Maraston}, {Daddi}, {Renzini}, {Cimatti}, {Dickinson}, {Papovich}, {Pasquali}, \& {Pirzkal}}]{Maraston2006}
{Maraston}, C., {Daddi}, E., {Renzini}, A., {et~al.} 2006, \apj, 652, 85, \dodoi{10.1086/508143}

\bibitem[{{Maraston} {et~al.}(2010){Maraston}, {Pforr}, {Renzini}, {Daddi}, {Dickinson}, {Cimatti}, \& {Tonini}}]{Maraston2010}
{Maraston}, C., {Pforr}, J., {Renzini}, A., {et~al.} 2010, \mnras, 407, 830, \dodoi{10.1111/j.1365-2966.2010.16973.x}

\bibitem[{{Marconi} {et~al.}(2004){Marconi}, {Risaliti}, {Gilli}, {Hunt}, {Maiolino}, \& {Salvati}}]{Marconi2004}
{Marconi}, A., {Risaliti}, G., {Gilli}, R., {et~al.} 2004, \mnras, 351, 169, \dodoi{10.1111/j.1365-2966.2004.07765.x}

\bibitem[{{Mascia} {et~al.}(2024){Mascia}, {Pentericci}, {Calabr{\`o}}, {Santini}, {Napolitano}, {Arrabal Haro}, {Castellano}, {Dickinson}, {Ocvirk}, {Lewis}, {Amor{\'\i}n}, {Bagley}, {Bhatawdekar}, {Cleri}, {Costantin}, {Dekel}, {Finkelstein}, {Fontana}, {Giavalisco}, {Grogin}, {Hathi}, {Hirschmann}, {Holwerda}, {Jung}, {Kartaltepe}, {Koekemoer}, {Lucas}, {Papovich}, {P{\'e}rez-Gonz{\'a}lez}, {Pirzkal}, {Trump}, {Wilkins}, \& {Yung}}]{Mascia2024}
{Mascia}, S., {Pentericci}, L., {Calabr{\`o}}, A., {et~al.} 2024, \aap, 685, A3, \dodoi{10.1051/0004-6361/202347884}

\bibitem[{{Mason} {et~al.}(2015){Mason}, {Trenti}, \& {Treu}}]{Mason2015}
{Mason}, C.~A., {Trenti}, M., \& {Treu}, T. 2015, \apj, 813, 21, \dodoi{10.1088/0004-637X/813/1/21}

\bibitem[{{Mason} {et~al.}(2023){Mason}, {Trenti}, \& {Treu}}]{Mason2023}
---. 2023, \mnras, 521, 497, \dodoi{10.1093/mnras/stad035}

\bibitem[{Matthee {et~al.}(2024)Matthee, Naidu, Brammer, Chisholm, Eilers, Goulding, Greene, Kashino, Labbe, Lilly, Mackenzie, Oesch, Weibel, Wuyts, Xiao, Bordoloi, Bouwens, van Dokkum, Illingworth, Kramarenko, Maseda, Mason, Meyer, Nelson, Reddy, Shivaei, Simcoe, \& Yue}]{Matthee2024}
Matthee, J., Naidu, R.~P., Brammer, G., {et~al.} 2024, The Astrophysical Journal, 963, 129, \dodoi{10.3847/1538-4357/ad2345}

\bibitem[{{McCracken} {et~al.}(2012){McCracken}, {Milvang-Jensen}, {Dunlop}, {Franx}, {Fynbo}, {Le F{\`e}vre}, {Holt}, {Caputi}, {Goranova}, {Buitrago}, {Emerson}, {Freudling}, {Hudelot}, {L{\'o}pez-Sanjuan}, {Magnard}, {Mellier}, {M{\o}ller}, {Nilsson}, {Sutherland}, {Tasca}, \& {Zabl}}]{McCracken2012}
{McCracken}, H.~J., {Milvang-Jensen}, B., {Dunlop}, J., {et~al.} 2012, \aap, 544, A156, \dodoi{10.1051/0004-6361/201219507}

\bibitem[{{McLeod} {et~al.}(2024){McLeod}, {Donnan}, {McLure}, {Dunlop}, {Magee}, {Begley}, {Carnall}, {Cullen}, {Ellis}, {Hamadouche}, \& {Stanton}}]{McLeod2024}
{McLeod}, D.~J., {Donnan}, C.~T., {McLure}, R.~J., {et~al.} 2024, \mnras, 527, 5004, \dodoi{10.1093/mnras/stad3471}

\bibitem[{{Menci} {et~al.}(2024){Menci}, {Sen}, \& {Castellano}}]{Menci2024}
{Menci}, N., {Sen}, A.~A., \& {Castellano}, M. 2024, \apj, 976, 227, \dodoi{10.3847/1538-4357/ad8d5b}

\bibitem[{{Meurer} {et~al.}(1999){Meurer}, {Heckman}, \& {Calzetti}}]{meurer99}
{Meurer}, G.~R., {Heckman}, T.~M., \& {Calzetti}, D. 1999, \apj, 521, 64, \dodoi{10.1086/307523}

\bibitem[{{Mirocha} \& {Furlanetto}(2023)}]{Mirocha2023}
{Mirocha}, J., \& {Furlanetto}, S.~R. 2023, \mnras, 519, 843, \dodoi{10.1093/mnras/stac3578}

\bibitem[{{Morales} {et~al.}(2025){Morales}, {Finkelstein}, {Arrabal Haro}, {Bagley}, {Calabr{\`o}}, {Ch{\'a}vez Ortiz}, {Davis}, {Dickinson}, {Gawiser}, {Giavalisco}, {Hathi}, {Hirschmann}, {Kartaltepe}, {Koekemoer}, {Long}, {Lucas}, {Pacucci}, {Papovich}, {Pautasso}, {Pirzkal}, {Taylor}, {de la Vega}, {Wilkins}, \& {Yung}}]{Morales2025}
{Morales}, A.~M., {Finkelstein}, S.~L., {Arrabal Haro}, P., {et~al.} 2025, arXiv e-prints, arXiv:2507.03118, \dodoi{10.48550/arXiv.2507.03118}

\bibitem[{{Mu{\~n}oz} {et~al.}(2024){Mu{\~n}oz}, {Mirocha}, {Chisholm}, {Furlanetto}, \& {Mason}}]{Munoz2024}
{Mu{\~n}oz}, J.~B., {Mirocha}, J., {Chisholm}, J., {Furlanetto}, S.~R., \& {Mason}, C. 2024, \mnras, 535, L37, \dodoi{10.1093/mnrasl/slae086}

\bibitem[{{Mu{\~n}oz} {et~al.}(2023){Mu{\~n}oz}, {Mirocha}, {Furlanetto}, \& {Sabti}}]{Munoz2023}
{Mu{\~n}oz}, J.~B., {Mirocha}, J., {Furlanetto}, S., \& {Sabti}, N. 2023, \mnras, 526, L47, \dodoi{10.1093/mnrasl/slad115}

\bibitem[{{Naidu} {et~al.}(2022){Naidu}, {Oesch}, {van Dokkum}, {Nelson}, {Suess}, {Brammer}, {Whitaker}, {Illingworth}, {Bouwens}, {Tacchella}, {Matthee}, {Allen}, {Bezanson}, {Conroy}, {Labbe}, {Leja}, {Leonova}, {Magee}, {Price}, {Setton}, {Strait}, {Stefanon}, {Toft}, {Weaver}, \& {Weibel}}]{Naidu2022}
{Naidu}, R.~P., {Oesch}, P.~A., {van Dokkum}, P., {et~al.} 2022, \apjl, 940, L14, \dodoi{10.3847/2041-8213/ac9b22}

\bibitem[{{Naidu} {et~al.}(2025){Naidu}, {Oesch}, {Brammer}, {Weibel}, {Li}, {Matthee}, {Chisholm}, {Pollock}, {Heintz}, {Johnson}, {Shen}, {Hviding}, {Leja}, {Tacchella}, {Ganguly}, {Witten}, {Atek}, {Belli}, {Bose}, {Bouwens}, {Dayal}, {Decarli}, {de Graaff}, {Fudamoto}, {Giovinazzo}, {Greene}, {Illingworth}, {Inoue}, {Kane}, {Labbe}, {Leonova}, {Marques-Chaves}, {Meyer}, {Nelson}, {Roberts-Borsani}, {Schaerer}, {Simcoe}, {Stefanon}, {Sugahara}, {Toft}, {van der Wel}, {van Dokkum}, {Walter}, {Watson}, {Weaver}, \& {Whitaker}}]{Naidu2025}
{Naidu}, R.~P., {Oesch}, P.~A., {Brammer}, G., {et~al.} 2025, arXiv e-prints, arXiv:2505.11263, \dodoi{10.48550/arXiv.2505.11263}

\bibitem[{{Nozawa} {et~al.}(2007){Nozawa}, {Kozasa}, {Habe}, {Dwek}, {Umeda}, {Tominaga}, {Maeda}, \& {Nomoto}}]{Nozawa2007}
{Nozawa}, T., {Kozasa}, T., {Habe}, A., {et~al.} 2007, \apj, 666, 955, \dodoi{10.1086/520621}

\bibitem[{{Oesch} {et~al.}(2016){Oesch}, {Brammer}, {van Dokkum}, {Illingworth}, {Bouwens}, {Labb{\'e}}, {Franx}, {Momcheva}, {Ashby}, {Fazio}, {Gonzalez}, {Holden}, {Magee}, {Skelton}, {Smit}, {Spitler}, {Trenti}, \& {Willner}}]{Oesch2016}
{Oesch}, P.~A., {Brammer}, G., {van Dokkum}, P.~G., {et~al.} 2016, \apj, 819, 129, \dodoi{10.3847/0004-637X/819/2/129}

\bibitem[{{Oke} \& {Gunn}(1983)}]{Oke1983}
{Oke}, J.~B., \& {Gunn}, J.~E. 1983, \apj, 266, 713, \dodoi{10.1086/160817}

\bibitem[{{Ono} {et~al.}(2023){Ono}, {Harikane}, {Ouchi}, {Yajima}, {Abe}, {Isobe}, {Shibuya}, {Wise}, {Zhang}, {Nakajima}, \& {Umeda}}]{Ono2023}
{Ono}, Y., {Harikane}, Y., {Ouchi}, M., {et~al.} 2023, \apj, 951, 72, \dodoi{10.3847/1538-4357/acd44a}

\bibitem[{{Pallottini} \& {Ferrara}(2023)}]{Pallottini2023}
{Pallottini}, A., \& {Ferrara}, A. 2023, \aap, 677, L4, \dodoi{10.1051/0004-6361/202347384}

\bibitem[{{Papovich} {et~al.}(2023){Papovich}, {Cole}, {Yang}, {Finkelstein}, {Barro}, {Buat}, {Burgarella}, {P{\'e}rez-Gonz{\'a}lez}, {Santini}, {Seill{\'e}}, {Shen}, {Arrabal Haro}, {Bagley}, {Bell}, {Bisigello}, {Calabr{\`o}}, {Casey}, {Castellano}, {Chworowsky}, {Cleri}, {Costantin}, {Cooper}, {Dickinson}, {Ferguson}, {Fontana}, {Giavalisco}, {Grazian}, {Grogin}, {Hathi}, {Holwerda}, {Hutchison}, {Kartaltepe}, {Kewley}, {Kirkpatrick}, {Kocevski}, {Koekemoer}, {Larson}, {Long}, {Lucas}, {Pentericci}, {Pirzkal}, {Ravindranath}, {Somerville}, {Trump}, {Urbano Stawinski}, {Weiner}, {Wilkins}, {Yung}, \& {Zavala}}]{Papovich2023}
{Papovich}, C., {Cole}, J.~W., {Yang}, G., {et~al.} 2023, \apjl, 949, L18, \dodoi{10.3847/2041-8213/acc948}

\bibitem[{{Paquereau} {et~al.}(2025){Paquereau}, {Laigle}, {McCracken}, {Shuntov}, {Ilbert}, {Akins}, {Allen}, {Arango- Togo}, {Berman}, {Bethermin}, {Casey}, {McCleary}, {Dubois}, {Drakos}, {Faisst}, {Franco}, {Harish}, {Jespersen}, {Kartaltepe}, {Koekemoer}, {Kokorev}, {Lambrides}, {Larson}, {Liu}, {Le Borgne}, {Lewis}, {McKinney}, {Mercier}, {Rhodes}, {Robertson}, {Toft}, {Trebitsch}, {Tresse}, \& {Weaver}}]{Paquereau2025}
{Paquereau}, L., {Laigle}, C., {McCracken}, H.~J., {et~al.} 2025, arXiv e-prints, arXiv:2501.11674, \dodoi{10.48550/arXiv.2501.11674}

\bibitem[{{P{\'e}rez-Gonz{\'a}lez} {et~al.}(2023){P{\'e}rez-Gonz{\'a}lez}, {Costantin}, {Langeroodi}, {Rinaldi}, {Annunziatella}, {Ilbert}, {Colina}, {N{\o}rgaard-Nielsen}, {Greve}, {{\"O}stlin}, {Wright}, {Alonso-Herrero}, {{\'A}lvarez-M{\'a}rquez}, {Caputi}, {Eckart}, {Le F{\`e}vre}, {Labiano}, {Garc{\'\i}a-Mar{\'\i}n}, {Hjorth}, {Kendrew}, {Pye}, {Tikkanen}, {van der Werf}, {Walter}, {Ward}, {Bik}, {Boogaard}, {Bosman}, {G{\'o}mez}, {Gillman}, {Iani}, {Jermann}, {Melinder}, {Meyer}, {Moutard}, {van Dishoek}, {Henning}, {Lagage}, {Guedel}, {Peissker}, {Ray}, {Vandenbussche}, {Garc{\'\i}a-Argum{\'a}nez}, \& {Mar{\'\i}a M{\'e}rida}}]{Perez-Gonzalez2023}
{P{\'e}rez-Gonz{\'a}lez}, P.~G., {Costantin}, L., {Langeroodi}, D., {et~al.} 2023, \apjl, 951, L1, \dodoi{10.3847/2041-8213/acd9d0}

\bibitem[{{P{\'e}rez-Gonz{\'a}lez} {et~al.}(2025){P{\'e}rez-Gonz{\'a}lez}, {{\"O}stlin}, {Costantin}, {Melinder}, {Finkelstein}, {Somerville}, {Annunziatella}, {{\'A}lvarez-M{\'a}rquez}, {Colina}, {Dekel}, {Ferguson}, {Li}, {Yung}, {Bagley}, {Boogard}, {Burgarella}, {Calabr{\`o}}, {Caputi}, {Cheng}, {Eckart}, {Giavalisco}, {Gillman}, {Greve}, {Hathi}, {Hjorth}, {Huertas-Company}, {Kartaltepe}, {Koekemoer}, {Kokorev}, {Labiano}, {Langeroodi}, {Leung}, {Natarajan}, {Papovich}, {Peissker}, {Pentericci}, {Pirzkal}, {Rinaldi}, {van der Werf}, \& {Walter}}]{Perez-Gonzalez2025}
{P{\'e}rez-Gonz{\'a}lez}, P.~G., {{\"O}stlin}, G., {Costantin}, L., {et~al.} 2025, arXiv e-prints, arXiv:2503.15594, \dodoi{10.48550/arXiv.2503.15594}

\bibitem[{{Pforr} {et~al.}(2012){Pforr}, {Maraston}, \& {Tonini}}]{Pforr2012}
{Pforr}, J., {Maraston}, C., \& {Tonini}, C. 2012, \mnras, 422, 3285, \dodoi{10.1111/j.1365-2966.2012.20848.x}

\bibitem[{{Planck Collaboration} {et~al.}(2016){Planck Collaboration}, {Ade}, {Aghanim}, {Arnaud}, {Ashdown}, {Aumont}, {Baccigalupi}, {Banday}, {Barreiro}, {Bartlett}, {Bartolo}, {Battaner}, {Battye}, {Benabed}, {Beno{\^\i}t}, {Benoit-L{\'e}vy}, {Bernard}, {Bersanelli}, {Bielewicz}, {Bock}, {Bonaldi}, {Bonavera}, {Bond}, {Borrill}, {Bouchet}, {Boulanger}, {Bucher}, {Burigana}, {Butler}, {Calabrese}, {Cardoso}, {Catalano}, {Challinor}, {Chamballu}, {Chary}, {Chiang}, {Chluba}, {Christensen}, {Church}, {Clements}, {Colombi}, {Colombo}, {Combet}, {Coulais}, {Crill}, {Curto}, {Cuttaia}, {Danese}, {Davies}, {Davis}, {de Bernardis}, {de Rosa}, {de Zotti}, {Delabrouille}, {D{\'e}sert}, {Di Valentino}, {Dickinson}, {Diego}, {Dolag}, {Dole}, {Donzelli}, {Dor{\'e}}, {Douspis}, {Ducout}, {Dunkley}, {Dupac}, {Efstathiou}, {Elsner}, {En{\ss}lin}, {Eriksen}, {Farhang}, {Fergusson}, {Finelli}, {Forni}, {Frailis}, {Fraisse}, {Franceschi}, {Frejsel}, {Galeotta}, {Galli}, {Ganga}, {Gauthier}, {Gerbino}, {Ghosh}, {Giard},
  {Giraud-H{\'e}raud}, {Giusarma}, {Gjerl{\o}w}, {Gonz{\'a}lez-Nuevo}, {G{\'o}rski}, {Gratton}, {Gregorio}, {Gruppuso}, {Gudmundsson}, {Hamann}, {Hansen}, {Hanson}, {Harrison}, {Helou}, {Henrot-Versill{\'e}}, {Hern{\'a}ndez-Monteagudo}, {Herranz}, {Hildebrandt}, {Hivon}, {Hobson}, {Holmes}, {Hornstrup}, {Hovest}, {Huang}, {Huffenberger}, {Hurier}, {Jaffe}, {Jaffe}, {Jones}, {Juvela}, {Keih{\"a}nen}, {Keskitalo}, {Kisner}, {Kneissl}, {Knoche}, {Knox}, {Kunz}, {Kurki-Suonio}, {Lagache}, {L{\"a}hteenm{\"a}ki}, {Lamarre}, {Lasenby}, {Lattanzi}, {Lawrence}, {Leahy}, {Leonardi}, {Lesgourgues}, {Levrier}, {Lewis}, {Liguori}, {Lilje}, {Linden-V{\o}rnle}, {L{\'o}pez-Caniego}, {Lubin}, {Mac{\'\i}as-P{\'e}rez}, {Maggio}, {Maino}, {Mandolesi}, {Mangilli}, {Marchini}, {Maris}, {Martin}, {Martinelli}, {Mart{\'\i}nez-Gonz{\'a}lez}, {Masi}, {Matarrese}, {McGehee}, {Meinhold}, {Melchiorri}, {Melin}, {Mendes}, {Mennella}, {Migliaccio}, {Millea}, {Mitra}, {Miville-Desch{\^e}nes}, {Moneti}, {Montier}, {Morgante}, {Mortlock},
  {Moss}, {Munshi}, {Murphy}, {Naselsky}, {Nati}, {Natoli}, {Netterfield}, {N{\o}rgaard-Nielsen}, {Noviello}, {Novikov}, {Novikov}, {Oxborrow}, {Paci}, {Pagano}, {Pajot}, {Paladini}, {Paoletti}, {Partridge}, {Pasian}, {Patanchon}, {Pearson}, {Perdereau}, {Perotto}, {Perrotta}, {Pettorino}, {Piacentini}, {Piat}, {Pierpaoli}, {Pietrobon}, {Plaszczynski}, {Pointecouteau}, {Polenta}, {Popa}, {Pratt}, \& {Pr{\'e}zeau}}]{Planck2016}
{Planck Collaboration}, {Ade}, P.~A.~R., {Aghanim}, N., {et~al.} 2016, \aap, 594, A13, \dodoi{10.1051/0004-6361/201525830}

\bibitem[{{Pontoppidan} {et~al.}(2022){Pontoppidan}, {Barrientes}, {Blome}, {Braun}, {Brown}, {Carruthers}, {Coe}, {DePasquale}, {Espinoza}, {Marin}, {Gordon}, {Henry}, {Hustak}, {James}, {Jenkins}, {Koekemoer}, {LaMassa}, {Law}, {Lockwood}, {Moro-Martin}, {Mullally}, {Pagan}, {Player}, {Proffitt}, {Pulliam}, {Ramsay}, {Ravindranath}, {Reid}, {Robberto}, {Sabbi}, {Ubeda}, {Balogh}, {Flanagan}, {Gardner}, {Hasan}, {Meinke}, \& {Nota}}]{Pontoppidan2022}
{Pontoppidan}, K.~M., {Barrientes}, J., {Blome}, C., {et~al.} 2022, \apjl, 936, L14, \dodoi{10.3847/2041-8213/ac8a4e}

\bibitem[{{Quilley} {et~al.}(2025){Quilley}, {de Lapparent}, {Baes}, {Bolzonella}, {Damjanov}, {H{\"a}u{\ss}ler}, {Marleau}, {Nersesian}, {Saifollahi}, {Scott}, {Sorce}, {Tortora}, {Urbano}, {Aghanim}, {Altieri}, {Amara}, {Andreon}, {Auricchio}, {Baccigalupi}, {Baldi}, {Balestra}, {Bardelli}, {Basset}, {Battaglia}, {Biviano}, {Bonchi}, {Bonino}, {Branchini}, {Brescia}, {Brinchmann}, {Caillat}, {Camera}, {Capobianco}, {Carbone}, {Carretero}, {Casas}, {Castellano}, {Castignani}, {Cavuoti}, {Cimatti}, {Colodro-Conde}, {Congedo}, {Conselice}, {Conversi}, {Copin}, {Courbin}, {Courtois}, {Cropper}, {Cuillandre}, {Da Silva}, {Degaudenzi}, {De Lucia}, {Di Giorgio}, {Dinis}, {Dubath}, {Duncan}, {Dupac}, {Dusini}, {Ealet}, {Farina}, {Farrens}, {Faustini}, {Ferriol}, {Fotopoulou}, {Frailis}, {Franceschi}, {Fumana}, {Galeotta}, {George}, {Gillis}, {Giocoli}, {G{\'o}mez-Alvarez}, {Grazian}, {Grupp}, {Haugan}, {Hoar}, {Holmes}, {Hormuth}, {Hornstrup}, {Hudelot}, {Jahnke}, {Jhabvala}, {Keih{\"a}nen}, {Kermiche},
  {Kiessling}, {Kilbinger}, {Kubik}, {Kuijken}, {K{\"u}mmel}, {Kunz}, {Kurki-Suonio}, {Laureijs}, {Le Mignant}, {Ligori}, {Lilje}, {Lindholm}, {Lloro}, {Mainetti}, {Maino}, {Maiorano}, {Mansutti}, {Marggraf}, {Markovic}, {Martinelli}, {Martinet}, {Marulli}, {Massey}, {Medinaceli}, {Mei}, {Melchior}, {Mellier}, {Meneghetti}, {Merlin}, {Meylan}, {Mora}, {Moresco}, {Moscardini}, {Nakajima}, {Neissner}, {Nichol}, {Niemi}, {Padilla}, {Paltani}, {Pasian}, {Pedersen}, {Percival}, {Pettorino}, {Pires}, {Polenta}, {Poncet}, {Popa}, {Pozzetti}, {Raison}, {Rebolo}, {Renzi}, {Rhodes}, {Riccio}, {Romelli}, {Roncarelli}, {Rossetti}, {Saglia}, {Sakr}, {Sapone}, {Sartoris}, {Schirmer}, {Schneider}, {Schrabback}, {Secroun}, {Sefusatti}, {Seidel}, {Serrano}, {Sirignano}, {Sirri}, {Stanco}, {Steinwagner}, {Tallada-Cresp{\'\i}}, {Taylor}, {Tereno}, {Toledo-Moreo}, {Torradeflot}, {Tutusaus}, {Valenziano}, {Vassallo}, {Verdoes Kleijn}, {Veropalumbo}, {Wang}, {Weller}, {Zamorani}, {Zucca}, {Burigana}, \& {Scottez}}]{Quilley2025}
{Quilley}, L., {de Lapparent}, V., {Baes}, M., {et~al.} 2025, arXiv e-prints, arXiv:2502.15581, \dodoi{10.48550/arXiv.2502.15581}

\bibitem[{{Rest} {et~al.}(2023){Rest}, {Pierel}, {Correnti}, {Canipe}, {Hilbert}, {Engesser}, {Sunnquist}, \& {Fox}}]{Rest2023}
{Rest}, A., {Pierel}, J., {Correnti}, M., {et~al.} 2023, {arminrest/jhat: The JWST HST Alignment Tool (JHAT)}, v2,  Zenodo, \dodoi{10.5281/zenodo.7892935}

\bibitem[{{Ricotti}(2002)}]{Ricotti2002}
{Ricotti}, M. 2002, \mnras, 336, L33, \dodoi{10.1046/j.1365-8711.2002.05990.x}

\bibitem[{{Rieke} {et~al.}(2005){Rieke}, {Kelly}, \& {Horner}}]{rieke05}
{Rieke}, M.~J., {Kelly}, D., \& {Horner}, S. 2005, in Society of Photo-Optical Instrumentation Engineers (SPIE) Conference Series, Vol. 5904, Cryogenic Optical Systems and Instruments XI, ed. J.~B. {Heaney} \& L.~G. {Burriesci}, 1--8, \dodoi{10.1117/12.615554}

\bibitem[{{Rieke} {et~al.}(2003){Rieke}, {Baum}, {Beichman}, {Crampton}, {Doyon}, {Eisenstein}, {Greene}, {Hodapp}, {Horner}, {Johnstone}, {Lesyna}, {Lilly}, {Meyer}, {Martin}, {McCarthy}, {Rieke}, {Roellig}, {Stauffer}, {Trauger}, \& {Young}}]{rieke03}
{Rieke}, M.~J., {Baum}, S.~A., {Beichman}, C.~A., {et~al.} 2003, in Society of Photo-Optical Instrumentation Engineers (SPIE) Conference Series, Vol. 4850, IR Space Telescopes and Instruments, ed. J.~C. {Mather}, 478--485, \dodoi{10.1117/12.489103}

\bibitem[{{Rieke} {et~al.}(2023){Rieke}, {Kelly}, {Misselt}, {Stansberry}, {Boyer}, {Beatty}, {Egami}, {Florian}, {Greene}, {Hainline}, {Leisenring}, {Roellig}, {Schlawin}, {Sun}, {Tinnin}, {Williams}, {Willmer}, {Wilson}, {Clark}, {Rohrbach}, {Brooks}, {Canipe}, {Correnti}, {DiFelice}, {Gennaro}, {Girard}, {Hartig}, {Hilbert}, {Koekemoer}, {Nikolov}, {Pirzkal}, {Rest}, {Robberto}, {Sunnquist}, {Telfer}, {Wu}, {Ferry}, {Lewis}, {Baum}, {Beichman}, {Doyon}, {Dressler}, {Eisenstein}, {Ferrarese}, {Hodapp}, {Horner}, {Jaffe}, {Johnstone}, {Krist}, {Martin}, {McCarthy}, {Meyer}, {Rieke}, {Trauger}, \& {Young}}]{rieke23}
{Rieke}, M.~J., {Kelly}, D.~M., {Misselt}, K., {et~al.} 2023, \pasp, 135, 028001, \dodoi{10.1088/1538-3873/acac53}

\bibitem[{{Roberts-Borsani} {et~al.}(2024){Roberts-Borsani}, {Treu}, {Shapley}, {Fontana}, {Pentericci}, {Castellano}, {Morishita}, {Bergamini}, \& {Rosati}}]{Roberts-Borsani2024}
{Roberts-Borsani}, G., {Treu}, T., {Shapley}, A., {et~al.} 2024, \apj, 976, 193, \dodoi{10.3847/1538-4357/ad85d3}

\bibitem[{{Roberts-Borsani} {et~al.}(2025){Roberts-Borsani}, {Bagley}, {Rojas-Ruiz}, {Treu}, {Morishita}, {Finkelstein}, {Trenti}, {Arrabal Haro}, {Ba{\~n}ados}, {Ch{\'a}vez Ortiz}, {Chworowsky}, {Hutchison}, {Larson}, {Leethochawalit}, {Leung}, {Mason}, {Somerville}, {Stiavelli}, {Yung}, {Kassin}, \& {Soto}}]{Roberts-Borsani2025}
{Roberts-Borsani}, G., {Bagley}, M., {Rojas-Ruiz}, S., {et~al.} 2025, \apj, 983, 18, \dodoi{10.3847/1538-4357/adba60}

\bibitem[{{Robertson}(2022)}]{Robertson2022}
{Robertson}, B.~E. 2022, \araa, 60, 121, \dodoi{10.1146/annurev-astro-120221-044656}

\bibitem[{{Robertson} {et~al.}(2015){Robertson}, {Ellis}, {Furlanetto}, \& {Dunlop}}]{Robertson2015}
{Robertson}, B.~E., {Ellis}, R.~S., {Furlanetto}, S.~R., \& {Dunlop}, J.~S. 2015, \apjl, 802, L19, \dodoi{10.1088/2041-8205/802/2/L19}

\bibitem[{{Rojas-Ruiz} {et~al.}(2024){Rojas-Ruiz}, {Bagley}, {Roberts-Borsani}, {Treu}, {Finkelstein}, {Morishita}, {Leethochawalit}, {Mason}, {Ba{\~n}ados}, {Trenti}, {Stiavelli}, {Yung}, {Arrabal Haro}, {Somerville}, \& {Soto}}]{Rojas-Ruiz2024}
{Rojas-Ruiz}, S., {Bagley}, M.~B., {Roberts-Borsani}, G., {et~al.} 2024, arXiv e-prints, arXiv:2408.00843, \dodoi{10.48550/arXiv.2408.00843}

\bibitem[{{Rowe} {et~al.}(2015){Rowe}, {Jarvis}, {Mandelbaum}, {Bernstein}, {Bosch}, {Simet}, {Meyers}, {Kacprzak}, {Nakajima}, {Zuntz}, {Miyatake}, {Dietrich}, {Armstrong}, {Melchior}, \& {Gill}}]{Rowe2015}
{Rowe}, B.~T.~P., {Jarvis}, M., {Mandelbaum}, R., {et~al.} 2015, Astronomy and Computing, 10, 121, \dodoi{10.1016/j.ascom.2015.02.002}

\bibitem[{{Schaye} {et~al.}(2015){Schaye}, {Crain}, {Bower}, {Furlong}, {Schaller}, {Theuns}, {Dalla Vecchia}, {Frenk}, {McCarthy}, {Helly}, {Jenkins}, {Rosas-Guevara}, {White}, {Baes}, {Booth}, {Camps}, {Navarro}, {Qu}, {Rahmati}, {Sawala}, {Thomas}, \& {Trayford}}]{Schaye2015}
{Schaye}, J., {Crain}, R.~A., {Bower}, R.~G., {et~al.} 2015, \mnras, 446, 521, \dodoi{10.1093/mnras/stu2058}

\bibitem[{{Schechter}(1976)}]{Schechter1976}
{Schechter}, P. 1976, \apj, 203, 297, \dodoi{10.1086/154079}

\bibitem[{{Schmidt}(1968)}]{Schmidt1968}
{Schmidt}, M. 1968, \apj, 151, 393, \dodoi{10.1086/149446}

\bibitem[{{Schouws} {et~al.}(2024){Schouws}, {Bouwens}, {Ormerod}, {Smit}, {Algera}, {Sommovigo}, {Hodge}, {Ferrara}, {Oesch}, {Rowland}, {van Leeuwen}, {Stefanon}, {Herard-Demanche}, {Fudamoto}, {R{\"o}ttgering}, \& {van der Werf}}]{Schouws2024}
{Schouws}, S., {Bouwens}, R.~J., {Ormerod}, K., {et~al.} 2024, arXiv e-prints, arXiv:2409.20549, \dodoi{10.48550/arXiv.2409.20549}

\bibitem[{{Scoville} {et~al.}(2007){Scoville}, {Aussel}, {Brusa}, {Capak}, {Carollo}, {Elvis}, {Giavalisco}, {Guzzo}, {Hasinger}, {Impey}, {Kneib}, {LeFevre}, {Lilly}, {Mobasher}, {Renzini}, {Rich}, {Sanders}, {Schinnerer}, {Schminovich}, {Shopbell}, {Taniguchi}, \& {Tyson}}]{scoville07a}
{Scoville}, N., {Aussel}, H., {Brusa}, M., {et~al.} 2007, \apjs, 172, 1, \dodoi{10.1086/516585}

\bibitem[{{Shen} {et~al.}(2023){Shen}, {Vogelsberger}, {Boylan-Kolchin}, {Tacchella}, \& {Kannan}}]{Shen2023}
{Shen}, X., {Vogelsberger}, M., {Boylan-Kolchin}, M., {Tacchella}, S., \& {Kannan}, R. 2023, \mnras, 525, 3254, \dodoi{10.1093/mnras/stad2508}

\bibitem[{{Shuntov} {et~al.}(2025{\natexlab{a}}){Shuntov}, {Akins}, {Paquereau}, {Casey}, {Ilbert}, {Arango-Toro}, {McCracken}, {Franco}, {Harish}, {Kartaltepe}, {Koekemoer}, {Yang}, {Huertas-Company}, {Berman}, {McCleary}, {Toft}, {Gavazzi}, {Achenbach}, {Bertin}, {Brinch}, {Champagne}, {Chartab}, {Drakos}, {Egami}, {Endsley}, {Faisst}, {Fan}, {Flayhart}, {Hartley}, {Hatamnia}, {Gozaliasl}, {Gentile}, {Jermann}, {Jin}, {Kakiichi}, {Khostovan}, {K{\"u}mmel}, {Laigle}, {Laishram}, {Lambrides}, {Liu}, {Lyu}, {Magdis}, {Mobasher}, {Moutard}, {Renzini}, {Robertson}, {Schefer}, {Scognamiglio}, {Scoville}, {Sattari}, {Sanders}, {Taamoli}, {Trakhtenbrot}, {Valentino}, {Wang}, {Weaver}, \& {Yang}}]{Shuntov2025}
{Shuntov}, M., {Akins}, H.~B., {Paquereau}, L., {et~al.} 2025{\natexlab{a}}, arXiv e-prints, arXiv:2506.03243.
\newblock \doarXiv{2506.03243}

\bibitem[{{Shuntov} {et~al.}(2025{\natexlab{b}}){Shuntov}, {Oesch}, {Toft}, {Meyer}, {Covelo-Paz}, {Paquereau}, {Bouwens}, {Brammer}, {Gelli}, {Giovinazzo}, {Herard-Demanche}, {Illingworth}, {Mason}, {Naidu}, {Weibel}, \& {Xiao}}]{Shuntov2025b}
{Shuntov}, M., {Oesch}, P.~A., {Toft}, S., {et~al.} 2025{\natexlab{b}}, arXiv e-prints, arXiv:2503.14280, \dodoi{10.48550/arXiv.2503.14280}

\bibitem[{{Smith} {et~al.}(2022){Smith}, {Kannan}, {Garaldi}, {Vogelsberger}, {Pakmor}, {Springel}, \& {Hernquist}}]{smith22}
{Smith}, A., {Kannan}, R., {Garaldi}, E., {et~al.} 2022, \mnras, 512, 3243, \dodoi{10.1093/mnras/stac713}

\bibitem[{{Somerville} {et~al.}(2025){Somerville}, {Yung}, {Lancaster}, {Menon}, {Sommovigo}, \& {Finkelstein}}]{Somerville2025}
{Somerville}, R.~S., {Yung}, L.~Y.~A., {Lancaster}, L., {et~al.} 2025, arXiv e-prints, arXiv:2505.05442, \dodoi{10.48550/arXiv.2505.05442}

\bibitem[{{Song} {et~al.}(2023){Song}, {Fang}, {Lin}, {Gu}, \& {Kong}}]{Song2023}
{Song}, J., {Fang}, G., {Lin}, Z., {Gu}, Y., \& {Kong}, X. 2023, \apj, 958, 82, \dodoi{10.3847/1538-4357/ad0365}

\bibitem[{{Stark}(2016)}]{Stark2016}
{Stark}, D.~P. 2016, \araa, 54, 761, \dodoi{10.1146/annurev-astro-081915-023417}

\bibitem[{{Stefanon} {et~al.}(2022){Stefanon}, {Bouwens}, {Illingworth}, {Labb{\'e}}, {Oesch}, \& {Gonzalez}}]{Stefanon2022}
{Stefanon}, M., {Bouwens}, R.~J., {Illingworth}, G.~D., {et~al.} 2022, \apj, 935, 94, \dodoi{10.3847/1538-4357/ac7e44}

\bibitem[{{Stefanon} {et~al.}(2021){Stefanon}, {Bouwens}, {Labb{\'e}}, {Illingworth}, {Gonzalez}, \& {Oesch}}]{Stefanon2021}
{Stefanon}, M., {Bouwens}, R.~J., {Labb{\'e}}, I., {et~al.} 2021, \apj, 922, 29, \dodoi{10.3847/1538-4357/ac1bb6}

\bibitem[{{Steidel} {et~al.}(1996){Steidel}, {Giavalisco}, {Pettini}, {Dickinson}, \& {Adelberger}}]{Steidel1996}
{Steidel}, C.~C., {Giavalisco}, M., {Pettini}, M., {Dickinson}, M., \& {Adelberger}, K.~L. 1996, \apjl, 462, L17, \dodoi{10.1086/310029}

\bibitem[{{Tacchella} {et~al.}(2022){Tacchella}, {Finkelstein}, {Bagley}, {Dickinson}, {Ferguson}, {Giavalisco}, {Graziani}, {Grogin}, {Hathi}, {Hutchison}, {Jung}, {Koekemoer}, {Larson}, {Papovich}, {Pirzkal}, {Rojas-Ruiz}, {Song}, {Schneider}, {Somerville}, {Wilkins}, \& {Yung}}]{Tacchella2022}
{Tacchella}, S., {Finkelstein}, S.~L., {Bagley}, M., {et~al.} 2022, \apj, 927, 170, \dodoi{10.3847/1538-4357/ac4cad}

\bibitem[{{Taffoni} {et~al.}(2020){Taffoni}, {Becciani}, {Garilli}, {Maggio}, {Pasian}, {Umana}, {Smareglia}, \& {Vitello}}]{Taffoni_etal_2020}
{Taffoni}, G., {Becciani}, U., {Garilli}, B., {et~al.} 2020, in Astronomical Society of the Pacific Conference Series, Vol. 527, Astronomical Data Analysis Software and Systems XXIX, ed. R.~{Pizzo}, E.~R. {Deul}, J.~D. {Mol}, J.~{de Plaa}, \& H.~{Verkouter}, 307, \dodoi{10.48550/arXiv.2002.01283}

\bibitem[{{Taylor} {et~al.}(2025){Taylor}, {Kokorev}, {Kocevski}, {Akins}, {Cullen}, {Dickinson}, {Finkelstein}, {Arrabal Haro}, {Bromm}, {Giavalisco}, {Inayoshi}, {Juneau}, {Leung}, {Perez-Gonzalez}, {Somerville}, {Trump}, {Amorin}, {Barro}, {Burgarella}, {Brooks}, {Carnall}, {Casey}, {Cheng}, {Chisholm}, {Chworowsky}, {Davis}, {Donnan}, {Dunlop}, {Ellis}, {Fernandez}, {Fujimoto}, {Grogin}, {Gupta}, {Hathi}, {Jung}, {Hirschmann}, {Kartaltepe}, {Koekemoer}, {Larson}, {Leung}, {Llerena}, {Lucas}, {McLeod}, {McLure}, {Napolitano}, {Papovich}, {Stanton}, {Tripodi}, {Wang}, {Wilkins}, {Yung}, \& {Zavala}}]{Taylor2025}
{Taylor}, A.~J., {Kokorev}, V., {Kocevski}, D.~D., {et~al.} 2025, arXiv e-prints, arXiv:2505.04609, \dodoi{10.48550/arXiv.2505.04609}

\bibitem[{{Topping} {et~al.}(2022){Topping}, {Stark}, {Endsley}, {Plat}, {Whitler}, {Chen}, \& {Charlot}}]{Topping2022}
{Topping}, M.~W., {Stark}, D.~P., {Endsley}, R., {et~al.} 2022, \apj, 941, 153, \dodoi{10.3847/1538-4357/aca522}

\bibitem[{{Topping} {et~al.}(2024){Topping}, {Stark}, {Endsley}, {Whitler}, {Hainline}, {Johnson}, {Robertson}, {Tacchella}, {Chen}, {Alberts}, {Baker}, {Bunker}, {Carniani}, {Charlot}, {Chevallard}, {Curtis-Lake}, {DeCoursey}, {Egami}, {Eisenstein}, {Ji}, {Maiolino}, {Williams}, {Willmer}, {Willott}, \& {Witstok}}]{Topping2024}
---. 2024, \mnras, 529, 4087, \dodoi{10.1093/mnras/stae800}

\bibitem[{{Trapp} \& {Furlanetto}(2020)}]{Trapp2020}
{Trapp}, A.~C., \& {Furlanetto}, S.~R. 2020, \mnras, 499, 2401, \dodoi{10.1093/mnras/staa2828}

\bibitem[{{Trenti} \& {Stiavelli}(2008)}]{Trenti2008}
{Trenti}, M., \& {Stiavelli}, M. 2008, \apj, 676, 767, \dodoi{10.1086/528674}

\bibitem[{{Trinca} {et~al.}(2024){Trinca}, {Schneider}, {Valiante}, {Graziani}, {Ferrotti}, {Omukai}, \& {Chon}}]{Trinca2024}
{Trinca}, A., {Schneider}, R., {Valiante}, R., {et~al.} 2024, \mnras, 529, 3563, \dodoi{10.1093/mnras/stae651}

\bibitem[{{Ucci} {et~al.}(2021){Ucci}, {Dayal}, {Hutter}, {Yepes}, {Gottl{\"o}ber}, {Legrand}, {Pentericci}, {Castellano}, \& {Choudhury}}]{Ucci2021}
{Ucci}, G., {Dayal}, P., {Hutter}, A., {et~al.} 2021, \mnras, 506, 202, \dodoi{10.1093/mnras/stab1229}

\bibitem[{{Varadaraj} {et~al.}(2023){Varadaraj}, {Bowler}, {Jarvis}, {Adams}, \& {H{\"a}u{\ss}ler}}]{Varadaraj2023}
{Varadaraj}, R.~G., {Bowler}, R.~A.~A., {Jarvis}, M.~J., {Adams}, N.~J., \& {H{\"a}u{\ss}ler}, B. 2023, \mnras, 524, 4586, \dodoi{10.1093/mnras/stad2081}

\bibitem[{{Vijayan} {et~al.}(2021){Vijayan}, {Lovell}, {Wilkins}, {Thomas}, {Barnes}, {Irodotou}, {Kuusisto}, \& {Roper}}]{Vijayan2021}
{Vijayan}, A.~P., {Lovell}, C.~C., {Wilkins}, S.~M., {et~al.} 2021, \mnras, 501, 3289, \dodoi{10.1093/mnras/staa3715}

\bibitem[{{Volonteri} \& {Gnedin}(2009)}]{Volonteri2009}
{Volonteri}, M., \& {Gnedin}, N.~Y. 2009, \apj, 703, 2113, \dodoi{10.1088/0004-637X/703/2/2113}

\bibitem[{{Wang} {et~al.}(2024){Wang}, {Sun}, {Zhou}, {Xu}, {Cheng}, {Li}, {Chen}, {Mo}, {Dekel}, {Yang}, {Wang}, {Zheng}, {Cai}, {Elbaz}, {Dai}, \& {Huang}}]{Wang2024}
{Wang}, T., {Sun}, H., {Zhou}, L., {et~al.} 2024, arXiv e-prints, arXiv:2403.02399, \dodoi{10.48550/arXiv.2403.02399}

\bibitem[{{Weaver} {et~al.}(2022){Weaver}, {Kauffmann}, {Ilbert}, {McCracken}, {Moneti}, {Toft}, {Brammer}, {Shuntov}, {Davidzon}, {Hsieh}, {Laigle}, {Anastasiou}, {Jespersen}, {Vinther}, {Capak}, {Casey}, {McPartland}, {Milvang-Jensen}, {Mobasher}, {Sanders}, {Zalesky}, {Arnouts}, {Aussel}, {Dunlop}, {Faisst}, {Franx}, {Furtak}, {Fynbo}, {Gould}, {Greve}, {Gwyn}, {Kartaltepe}, {Kashino}, {Koekemoer}, {Kokorev}, {Le F{\`e}vre}, {Lilly}, {Masters}, {Magdis}, {Mehta}, {Peng}, {Riechers}, {Salvato}, {Sawicki}, {Scarlata}, {Scoville}, {Shirley}, {Silverman}, {Sneppen}, {Smolc̆i{\'c}}, {Steinhardt}, {Stern}, {Tanaka}, {Taniguchi}, {Teplitz}, {Vaccari}, {Wang}, \& {Zamorani}}]{Weaver2022}
{Weaver}, J.~R., {Kauffmann}, O.~B., {Ilbert}, O., {et~al.} 2022, \apjs, 258, 11, \dodoi{10.3847/1538-4365/ac3078}

\bibitem[{{Whitler} {et~al.}(2023){Whitler}, {Endsley}, {Stark}, {Topping}, {Chen}, \& {Charlot}}]{Whitler2023}
{Whitler}, L., {Endsley}, R., {Stark}, D.~P., {et~al.} 2023, \mnras, 519, 157, \dodoi{10.1093/mnras/stac3535}

\bibitem[{{Whitler} {et~al.}(2025){Whitler}, {Stark}, {Topping}, {Robertson}, {Rieke}, {Hainline}, {Endsley}, {Chen}, {Baker}, {Bhatawdekar}, {Bunker}, {Carniani}, {Charlot}, {Chevallard}, {Curtis-Lake}, {Egami}, {Eisenstein}, {Helton}, {Ji}, {Johnson}, {P{\'e}rez-Gonz{\'a}lez}, {Rinaldi}, {Tacchella}, {Williams}, {Willmer}, {Willott}, \& {Witstok}}]{Whitler2025}
{Whitler}, L., {Stark}, D.~P., {Topping}, M.~W., {et~al.} 2025, arXiv e-prints, arXiv:2501.00984, \dodoi{10.48550/arXiv.2501.00984}

\bibitem[{{Williams} {et~al.}(2018){Williams}, {Curtis-Lake}, {Hainline}, {Chevallard}, {Robertson}, {Charlot}, {Endsley}, {Stark}, {Willmer}, {Alberts}, {Amorin}, {Arribas}, {Baum}, {Bunker}, {Carniani}, {Crandall}, {Egami}, {Eisenstein}, {Ferruit}, {Husemann}, {Maseda}, {Maiolino}, {Rawle}, {Rieke}, {Smit}, {Tacchella}, \& {Willott}}]{Williams2018}
{Williams}, C.~C., {Curtis-Lake}, E., {Hainline}, K.~N., {et~al.} 2018, \apjs, 236, 33, \dodoi{10.3847/1538-4365/aabcbb}

\bibitem[{{Willott} {et~al.}(2024){Willott}, {Desprez}, {Asada}, {Sarrouh}, {Abraham}, {Brada{\v{c}}}, {Brammer}, {Estrada-Carpenter}, {Iyer}, {Martis}, {Matharu}, {Mowla}, {Muzzin}, {Noirot}, {Sawicki}, {Strait}, {Rihtar{\v{s}}i{\v{c}}}, \& {Withers}}]{Willott2024}
{Willott}, C.~J., {Desprez}, G., {Asada}, Y., {et~al.} 2024, \apj, 966, 74, \dodoi{10.3847/1538-4357/ad35bc}

\bibitem[{{Witstok} {et~al.}(2023){Witstok}, {Shivaei}, {Smit}, {Maiolino}, {Carniani}, {Curtis-Lake}, {Ferruit}, {Arribas}, {Bunker}, {Cameron}, {Charlot}, {Chevallard}, {Curti}, {de Graaff}, {D'Eugenio}, {Giardino}, {Looser}, {Rawle}, {Rodr{\'\i}guez del Pino}, {Willott}, {Alberts}, {Baker}, {Boyett}, {Egami}, {Eisenstein}, {Endsley}, {Hainline}, {Ji}, {Johnson}, {Kumari}, {Lyu}, {Nelson}, {Perna}, {Rieke}, {Robertson}, {Sandles}, {Saxena}, {Scholtz}, {Sun}, {Tacchella}, {Williams}, \& {Willmer}}]{Witstok2023}
{Witstok}, J., {Shivaei}, I., {Smit}, R., {et~al.} 2023, \nat, 621, 267, \dodoi{10.1038/s41586-023-06413-w}

\bibitem[{{Witstok} {et~al.}(2025){Witstok}, {Jakobsen}, {Maiolino}, {Helton}, {Johnson}, {Robertson}, {Tacchella}, {Cameron}, {Smit}, {Bunker}, {Saxena}, {Sun}, {Alberts}, {Arribas}, {Baker}, {Bhatawdekar}, {Boyett}, {Cargile}, {Carniani}, {Charlot}, {Chevallard}, {Curti}, {Curtis-Lake}, {D'Eugenio}, {Eisenstein}, {Hainline}, {Jones}, {Kumari}, {Maseda}, {P{\'e}rez-Gonz{\'a}lez}, {Rinaldi}, {Scholtz}, {{\"U}bler}, {Williams}, {Willmer}, {Willott}, \& {Zhu}}]{Witstok2025}
{Witstok}, J., {Jakobsen}, P., {Maiolino}, R., {et~al.} 2025, \nat, 639, 897, \dodoi{10.1038/s41586-025-08779-5}

\bibitem[{{Yan} {et~al.}(2023){Yan}, {Ma}, {Ling}, {Cheng}, \& {Huang}}]{Yan2023}
{Yan}, H., {Ma}, Z., {Ling}, C., {Cheng}, C., \& {Huang}, J.-S. 2023, \apjl, 942, L9, \dodoi{10.3847/2041-8213/aca80c}

\bibitem[{{Yang} {et~al.}(2022){Yang}, {Morishita}, {Leethochawalit}, {Castellano}, {Calabr{\`o}}, {Treu}, {Bonchi}, {Fontana}, {Mason}, {Merlin}, {Paris}, {Trenti}, {Roberts-Borsani}, {Bradac}, {Vanzella}, {Vulcani}, {Marchesini}, {Ding}, {Nanayakkara}, {Birrer}, {Glazebrook}, {Jones}, {Boyett}, {Santini}, {Strait}, \& {Wang}}]{Yang2022}
{Yang}, L., {Morishita}, T., {Leethochawalit}, N., {et~al.} 2022, \apjl, 938, L17, \dodoi{10.3847/2041-8213/ac8803}

\bibitem[{{Yang} {et~al.}(2025){Yang}, {Kartaltepe}, {Franco}, {Ding}, {Achenbach}, {Arango-Toro}, {Casey}, {Drakos}, {Faisst}, {Gillman}, {Gozaliasl}, {Huertas-Company}, {Jin}, {Liu}, {Magdis}, {Massey}, {Silverman}, {Tanaka}, {Yu}, {Akins}, {Allen}, {Ilbert}, {Koekemoer}, {McCracken}, {Paquereau}, {Rhodes}, {Robertson}, {Shuntov}, \& {Toft}}]{Yang2025}
{Yang}, L., {Kartaltepe}, J.~S., {Franco}, M., {et~al.} 2025, arXiv e-prints, arXiv:2504.07185, \dodoi{10.48550/arXiv.2504.07185}

\bibitem[{{Yung} {et~al.}(2019){Yung}, {Somerville}, {Finkelstein}, {Popping}, \& {Dav{\'e}}}]{yung19}
{Yung}, L.~Y.~A., {Somerville}, R.~S., {Finkelstein}, S.~L., {Popping}, G., \& {Dav{\'e}}, R. 2019, \mnras, 483, 2983, \dodoi{10.1093/mnras/sty3241}

\bibitem[{{Yung} {et~al.}(2024){Yung}, {Somerville}, {Finkelstein}, {Wilkins}, \& {Gardner}}]{Yung2024}
{Yung}, L.~Y.~A., {Somerville}, R.~S., {Finkelstein}, S.~L., {Wilkins}, S.~M., \& {Gardner}, J.~P. 2024, \mnras, 527, 5929, \dodoi{10.1093/mnras/stad3484}

\bibitem[{{Zavala} {et~al.}(2023){Zavala}, {Buat}, {Casey}, {Finkelstein}, {Burgarella}, {Bagley}, {Ciesla}, {Daddi}, {Dickinson}, {Ferguson}, {Franco}, {Jim{\'e}nez-Andrade}, {Kartaltepe}, {Koekemoer}, {Le Bail}, {Murphy}, {Papovich}, {Tacchella}, {Wilkins}, {Aretxaga}, {Behroozi}, {Champagne}, {Fontana}, {Giavalisco}, {Grazian}, {Grogin}, {Kewley}, {Kocevski}, {Kirkpatrick}, {Lotz}, {Pentericci}, {P{\'e}rez-Gonz{\'a}lez}, {Pirzkal}, {Ravindranath}, {Somerville}, {Trump}, {Yang}, {Yung}, {Almaini}, {Amor{\'\i}n}, {Annunziatella}, {Arrabal Haro}, {Backhaus}, {Barro}, {Bell}, {Bhatawdekar}, {Bisigello}, {Buitrago}, {Calabr{\`o}}, {Castellano}, {Ch{\'a}vez Ortiz}, {Chworowsky}, {Cleri}, {Cohen}, {Cole}, {Cooke}, {Cooper}, {Cooray}, {Costantin}, {Cox}, {Croton}, {Dav{\'e}}, {de La Vega}, {Dekel}, {Elbaz}, {Estrada-Carpenter}, {Fern{\'a}ndez}, {Finkelstein}, {Freundlich}, {Fujimoto}, {Garc{\'\i}a-Argum{\'a}nez}, {Gardner}, {Gawiser}, {G{\'o}mez-Guijarro}, {Guo}, {Hamilton}, {Hathi}, {Holwerda}, {Hirschmann},
  {Huertas-Company}, {Hutchison}, {Iyer}, {Jaskot}, {Jha}, {Jogee}, {Juneau}, {Jung}, {Kassin}, {Kurczynski}, {Larson}, {Leung}, {Long}, {Lucas}, {Magnelli}, {Mantha}, {Matharu}, {McGrath}, {McIntosh}, {Medrano}, {Merlin}, {Mobasher}, {Morales}, {Newman}, {Nicholls}, {Pandya}, {Rafelski}, {Ronayne}, {Rose}, {Ryan}, {Santini}, {Seill{\'e}}, {Shah}, {Shen}, {Simons}, {Snyder}, {Stanway}, {Straughn}, {Teplitz}, {Vanderhoof}, {Vega-Ferrero}, {Wang}, {Weiner}, {Willmer}, {Wuyts}, \& {Ceers Team}}]{Zavala2023}
{Zavala}, J.~A., {Buat}, V., {Casey}, C.~M., {et~al.} 2023, \apjl, 943, L9, \dodoi{10.3847/2041-8213/acacfe}

\bibitem[{{Ziparo} {et~al.}(2023){Ziparo}, {Ferrara}, {Sommovigo}, \& {Kohandel}}]{Ziparo2023}
{Ziparo}, F., {Ferrara}, A., {Sommovigo}, L., \& {Kohandel}, M. 2023, \mnras, 520, 2445, \dodoi{10.1093/mnras/stad125}

\end{thebibliography}

\bibliographystyle{aasjournal}

\newpage
\allauthors

\end{document}